\documentclass[12pt]{iopart}

\usepackage{iopams}
\usepackage{amssymb}
\usepackage{setstack}
\usepackage{graphicx}
\usepackage{dcolumn}
\usepackage{bm}
\usepackage{amssymb}
\usepackage[active]{srcltx}

\usepackage{color}

\definecolor{yellow}{rgb}{0.5,0.5,0}

\begin{document}
\date{\today}
\title[Active transport on networks]{\bf  Exclusion processes
on networks as models for cytoskeletal transport}

\author{Izaak Neri$^{1,2}$, Norbert Kern$^{1,2}$ and Andrea
Parmeggiani$^{1,2,3,4}$}
\address{$^{1}$ Universit\'e Montpellier 2, Laboratoire Charles Coulomb UMR
5221,
F-34095, Montpellier, France}
\address{$^{2}$ CNRS, Laboratoire Charles Coulomb UMR 5221, F-34095,
Montpellier,
France} 
\address{${^3}$ Universit\'e Montpellier 2, Laboratoire DIMNP UMR 5235, F-34095,
Montpellier, France}
\address{${^4}$ CNRS, Laboratoire DIMNP UMR 5235, F-34095, Montpellier, France}
\ead{izaak.neri@gmail.com,
norbert.kern@univ-montp2.fr,
andrea.parmeggiani@univ-montp2.fr}

\begin{abstract}
We present a study of exclusion processes on networks
as models for complex transport phenomena and in particular for active transport
of motor proteins along the cytoskeleton.  We argue that
active transport processes on networks spontaneously develop  density
heterogeneities at various scales.  These heterogeneities can be regulated
through a variety of multi-scale factors, such as the interplay of  exclusion
interactions, the non-equilibrium nature of the transport process and the
network topology.  

We show how an effective rate approach allows to
develop an understanding of the stationary state of transport processes through
complex networks from the  phase
diagram of one single segment.  For exclusion processes we rationalize that the
stationary state can be classified in three
qualitatively different regimes: a homogeneous phase as well as inhomogeneous
network and segment phases.

In particular, we present here a study of the stationary state on networks of
three paradigmatic models from non-equilibrium statistical physics:
the totally asymmetric simple exclusion process, the
partially asymmetric simple exclusion process and the totally asymmetric simple
exclusion process with Langmuir kinetics.  
With these models we can interpolate
between equilibrium (due to bi-directional motion along a network 
or infinite diffusion) and
out-of-equilibrium active directed
motion along a network. 
The study of these models sheds
further light on the emergence of density heterogeneities in active phenomena. 
 \end{abstract}

\section{Introduction} 

Over the last decades our  knowledge on
the composition and functioning of the cellular organelles has 
increased considerably \cite{Alberts}, but understanding how cells
self-organize and make their molecular components self-assemble into 
cellular compartments and structures is still a major challenge in
cellular biology \cite{Fletcher, Vignaud}.   How a cell works is undoubtedly
related to
its internal organization and the spatial distributions of its components on
different scales.  
One may guess that order at mesoscopic scales  is a consequence of the
non-equilibrium nature of cellular processes. 
How  self-organization arises spontaneously in non-equilibrium 
physics is also a 
a topic of current interest in non-equilibrium statistical physics, and is
referred to as the
study of active matter
\cite{Ram, active}.

Cells require active fluxes of matter to maintain their internal
organization. To function properly cells thus need 
to  establish  a specific spatio-temporal organization of proteins, organelles,
etc. 
Delivery of cargoes in eukaryotic cells is functional for biological
processes and mainly
realized using an active transport process based on motor proteins
moving along cytoskeletal filaments \cite{Alberts, HowardBook, Vale}.  
In general, cytoskeletal filaments in cells form  intracellular networks which
act as macromolecular highways along which motor proteins can
deliver cargoes to specific locations
in cells.    In nerve cells, for instance, proteins and membranes must be
transported from the cell body to the  synaptic terminal, a distance
which can reach several meters.  Besides their important role in
transporting cargo, motor proteins have  various other functions, e.g.~force
production leading to muscle contraction, depolymerization or
rearrangement of the
cytoskeletal filaments in cytoskeleton dynamics.  Hence, the spatial
organization of the motor proteins
is an important aspect in the understanding of the physical and
biological properties of cells.  Motor-protein transport  also has an important
impact on the health of organisms: motor-protein mutations have been shown to
lead to neurodegenerative diseases and they also play an important role in
left-right body determination, tumour suppression, etc. \cite{LS, Hir}.

The physical picture of motor protein transport inside the cell is roughly the following.  Motor proteins  consume chemical energy at a high rate, which they employ to perform active motion along the polymer filaments of the cytoskeleton, in a preferential direction set by the filament polarity.
These directed runs along the cytoskeleton typically
alternate with diffusion in the cytoplasm, as the motors stochastically 
detach and re-attach via specific binding and unbinding processes.
The distance which the motor protein typically
moves between an attachment and a detachment event is a measure for its {\it processivity} (see for instance \cite{Block}).
When motors reach a junction, where filaments branch or interconnect, they can change their direction or switch filament \cite{Ross}.

In recent years single motor protein motion has become experimentally
observable due to
progress in  
super-resolution imaging, physical manipulations of single molecules as well as
electron and light microscopy.
These have considerably boosted the ability to study
the spatio-temporal organization of proteins within the cell.  For instance, it is now
possible to determine quantitatively, both {\it in-vitro} and {\it in-cellulo}, 
 dynamic properties of single motor-proteins  \cite{Cai,
Pier, Yoo, bal}, to quantitatively analyze the collective
effects between interacting motors \cite{VargaCell,Varg2},
to observe the overall organization of proteins, organelles or membranes in
the cell \cite{Nasc}, as
well as to present a 3D mapping of the
topological structure of the cytoskeleton
\cite{Medalia, Harush, Fleischer} and to control the formation of 
actin and microtubule networks of various topologies
\cite{Rey, Por}.   It is a challenge to develop the modelling
tools to complement the insights given by these experimental studies.

From a theoretical point of view, motor-protein transport is an intriguing
example of a  stochastic transport phenomenon of molecular entities, far from
thermodynamic equilibrium and subject to mutual interactions. 
In statistical physics motor-protein transport is modelled by 
particles performing an active stochastic motion along a
one-dimensional segment \cite{Jul, Reimann, Guer, Chow}.  One class of commonly
used models are  lattice gas {\it exclusion processes}, for which the particles are constrained in their movement by their excluded volume \cite{Spitzer, Spohn,
Schutz2}. Their instantaneous velocities during the stepping process plays no
role as such, since at the nanometer scale all motion is overdamped by the 
viscous environment.
Active exclusion  processes have been proposed originally  in the context
of mRNA translation by ribosomes \cite{Mac68, Mac69} and, more recently,
have allowed to make quantitative predictions for  in-vitro
experiments on motor-protein transport along single filaments
\cite{Varg2, Nishinari, Reese}.   Thanks to extensive
studies, we have now a rather in-depth understanding of 
non-equilibrium transport in a one-dimensional setting  \cite{Chou2011}.  These
lattice gas models have stimulated a lot of fundamental research in
non-equilibrium physics \cite{Mal}, but also in more applied topics such as
modelling macromolecules which move through
capillary vessels \cite{Chou1999}, electrons hopping through a chain of
quantum-dots \cite{oppen}, vehicular  flow in traffic \cite{stad, Helbing},
translation of mRNA \cite{Luca2}.  Current
studies have mainly focused on one-dimensional models, but  they did not address
how motor proteins spatially self-organize along the cytoskeleton.

Motor protein transport along the cytoskeleton can be envisaged as a generalisation of these processes to complex networks.  Studies of transport processes on complex networks abound in the literature, starting from the seminal work of 
Kirchhoff on currents through electrical circuits \cite{Kirchoff}. Diffusion
through networks is well-studied \cite{Lovasz}, and active transport processes
through networks have been considered as models for motor protein transport
\cite{Klumpp, Goldsteinx, Kahana, Ned, Bresslof, Klann}.
To date, most such models are mainly based on non-interacting particles.
However, much interesting physics is expected when motors interact
through exclusion interactions, and work on this aspect is recent.
Greulich and Santen have studied particles moving actively on a spatially disordered network, also accounting for finite diffusion in the surrounding
reservoir \cite{Santen}.
Ezaki and Nishinari have developed an exactly solvable model of an
exclusion process on a  network respecting a balance condition \cite{Ezaki}.  
In our recent work  \cite{NeriT, NeriTT} we have shown that the stationary
state of exclusion processes on complex networks can be understood in terms of
their behaviour on a single-segment.   

The role of heterogeneities has emerged from our previous studies as an important feature: 
since exclusion models in one dimension  display  a boundary-induced 
first-order phase transition in the particle density \cite{Krug, Der}, 
transport through complex networks leads to various
regimes of density heterogeneities at different spatial scales \cite{NeriT,
NeriTT}, which also depend on the network topology.
We rationalize these phenomena starting from the transport characteristics
of a single  segment between two particle reservoirs.   In particular, we
build on the idea of effective rate diagrams  which allow to visualize the
stationary state of the network  from the single-segment phase diagram 
\cite{NeriTT}.   
Using this effective rate approach which we further develop in this article, we
have identified three stationary
regimes   in exclusion models on networks, characterized by the spatial
heterogeneities in the particle densities:
a {\it heterogeneous network regime},
a {\it heterogeneous segment regime} and a {\it homogeneous regime}.   We
link these scales of heterogeneities 
to an interplay between the topology of the network, the microscopic (molecular)
parameters for the transport process and the fraction of the network filled with particles. Our approach
can be applied to any model for which the single-segment phase diagram and
current density profiles through a  single segment have been determined. 
Our approach is straightforwardly
applicable to a large number of transport models for which the single-segment
diagram
is known  \cite{Schutz2,
Chou2011}.   

We apply our method to three different paradigmatic models: the totally
asymmetric simple exclusion process (TASEP) \cite{Mac68, Mac69, Derrida}, the
partially asymmetric simple
exclusion process (PASEP) \cite{Sandow} and the totally asymmetric simple
exclusion process
with Langmuir kinetics (TASEP-LK) \cite{Par03, parmeggiani2}. The TASEP
is a model of active particles
which hop  stochastically and {\it uni-directionally} through a network and
mutually interact with exclusion interactions.  Interestingly, in  recent work
we have shown that transport through closed disordered networks by active
particles
following TASEP rules spontaneously leads to  strong density heterogeneities
between the different segments of the network \cite{NeriT}. 
To establish a clear understanding on how these heterogeneities appear in
non-equilibrium transport processes we consider an extension of TASEP
 in two ways: we consider {\it bi-directional} motion of active
particles on networks (PASEP) and the coupling of
active
motion through a network with the {\it passive diffusion} of  particles in a
reservoir (TASEP-LK).  These extensions allow us to interpolate between a
passive transport process (observed for fully bi-directional or diffusive
motion) and
an active transport process (for uni-directional motion along the network).  In
such a way we gain insight into the emergence of
density heterogeneities  on networks (which
are absent in passive processes but appear in active processes).   
In the
perspective of biological modelling,  
these models add important molecular parameters to the TASEP description of 
motor protein transport along the cytoskeleton.

In the following section we describe our general mathematical framework of
active
particles moving along complex networks, presented here in the perspective of motor
protein transport along the cytoskeleton.  We define the random networks and
excluded volume processes
which model, respectively, the cytoskeletal architecture and 
motor protein motion along the biofilaments.  In the third section we introduce
the two main concepts which allow us to intuitively understand and characterize
the stationary state of excluded volume processes on networks: effective
rate
diagrams for the network and a classification of stationary states based on three 
regimes of particle density heterogeneities.   We introduce these concepts
first  on TASEP,  
revisiting and extending the results presented in \cite{NeriT}.   We then
show in section four how this effective rate approach can be applied to 
understand bi-directional transport on networks  and, in section five,
to understand active transport on networks coupled to a homogeneous particle
 reservoir. The latter analysis considerably extends 
the results presented in \cite{NeriTT}.  The  study of these three
models shows how the stationary states of exclusion processes can be 
classified in a
unified
way using the three stationary regimes of density heterogeneities sketched
above.  We discuss the potential implications of our findings in the context of
cytoskeletal motor protein transport using experimental values from the
literature.

In the conclusions we summarize how this classification allows to present 
in a compact way how 
spatial density heterogeneities
appear in active transport on networks through an interplay between network
topology, disorder, bi-directionality and
finite particle processivity.

\section{Modelling motor-protein  transport}
\begin{figure}[h!]
\begin{center}
\includegraphics[width=1\textwidth]{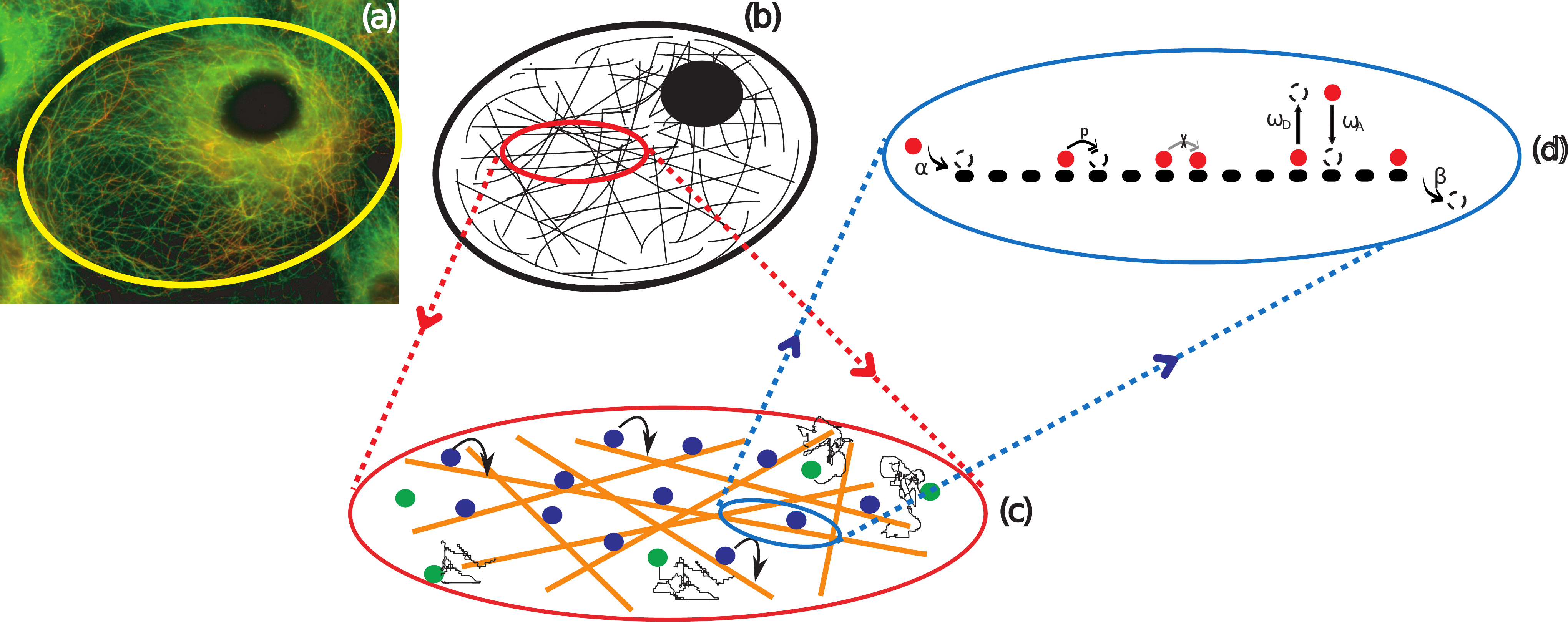}
\end{center}
\caption{(a) Microtubular network of a COS cell (courtesy of P.
Montcourrier, CRLC Val
d'Aurelle); (b) scheme of the microtubular network; (c) zoom on filaments
with motor transport, binding and unbinding; (d) TASEP-LK microscopic rules for
transport along a single filament.
}\label{fig:biotransport}
\end{figure}

In this section we present our general modelling framework to study
transport along complex networks, with a view towards motor-protein 
transport on the cytoskeleton. 
Cells are hugely complex systems consisting of a variety of  building blocks
based on macromolecular assemblies such as proteins, filaments, membranes,
organelles, etc.  \cite{Alberts, Howard, Lodisch}.
 We use minimal models to
capture some of the essential qualitative features of
motor
protein transport.  These minimal models consist of
particles (motor proteins) moving directionally along a graph (the cytoskeleton) as
represented
in figure \ref{fig:biotransport}.

\subsection{Cytoskeleton as a directed network}
\begin{figure}[h!]
\begin{center}
\includegraphics[width=1\textwidth]{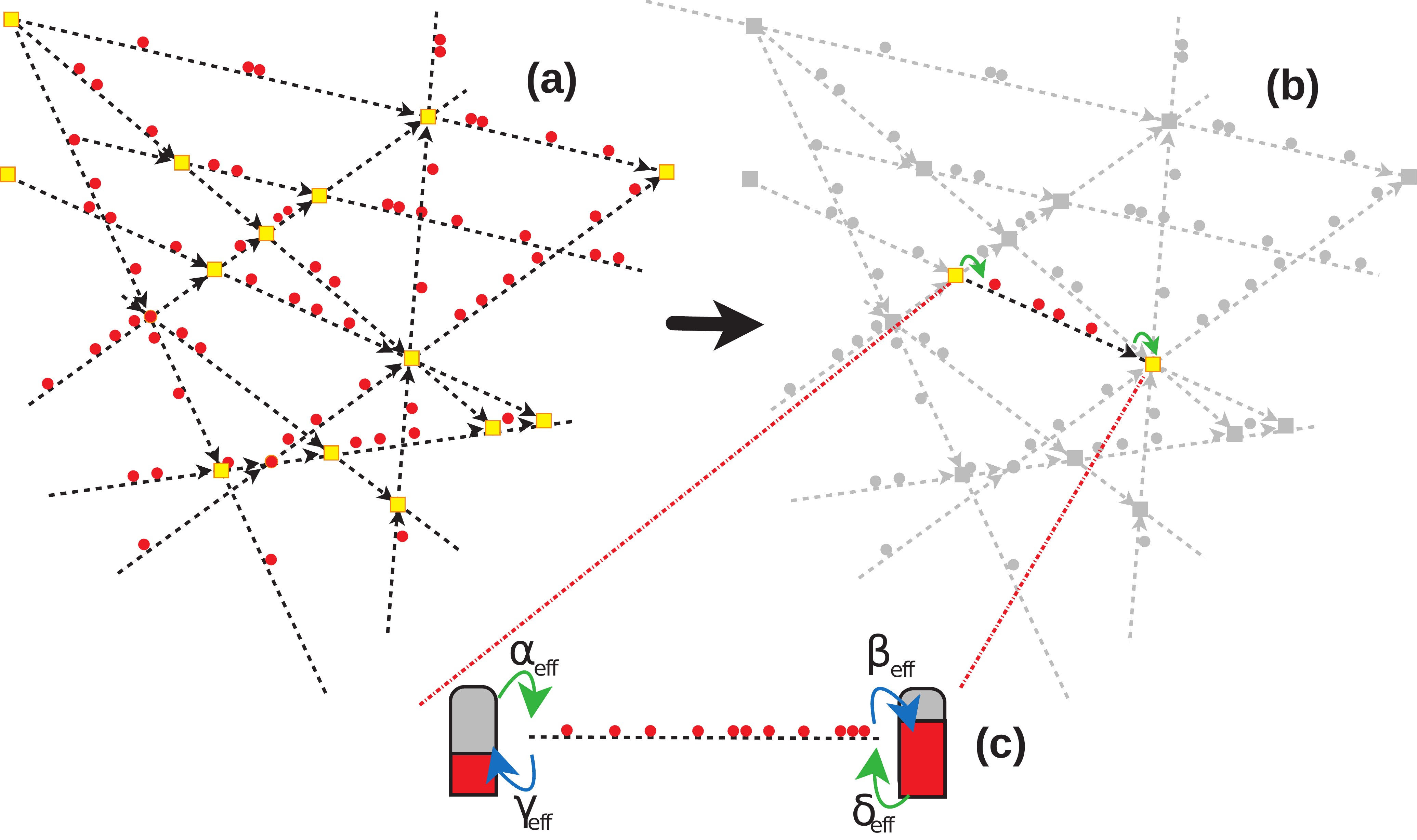}
\end{center}
\caption{(a) Transport of particles along a network as a minimal model for
cytoskeletal motor-protein driven transport;  (b)
The stationary state of the whole network can be studied by considering that
every segment is an open segment which connects two
reservoirs from the entrance of the segment towards the exit.  (c) The two
reservoirs are characterized by certain effective rates which depend on the
state of the
network, in particular the particle density at the junction nodes.
}\label{fig:transport-network}
\end{figure}

The cytoskeletal meshwork of filaments is represented as a  directed graph of
segments of length $L$ which are interconnected at junction sites or vertices.  A directed graph is a couple $G = (V,E)$ of the set of vertices $v\in V$ and 
directed edges or segments $s\in E\subset
V\times V$.  It is represented through nodes interlinked by
arrows (see figures \ref{fig:biotransport} and \ref{fig:transport-network}). 
In figure \ref{fig:transport-network} the segments are
represented as dashed lines and the junction sites  as squares.  
This representation of the cytoskeleton as a directed network of segments takes
into consideration the polarization of the cytoskeletal filaments, their discrete microscopic nature as well as the one-dimensional nature of the protofilaments.

In principle we could consider a specific topology of 
 the cytoskeletal network, as it may be known from experiments
with cryo-electron tomography  \cite{Harush, Fleischer} or from  in-vitro
reconstituted polymer networks using  micropatterning methods
\cite{Rey, Por}.
A static characterization of the cytoskeleton is however
 difficult, since the cytoskeleton is in fact a highly dynamic network due to
e.g.~(de)polymerization of filaments or (un)binding of cross-linker proteins.  Moreover, due to the intrinsic complexity of the cytoskeleton, working on any particular structure would make it more difficult to unveil
the main mechanisms leading to overall motor protein organization.  
In this perspective, theoretical studies  are useful to explore the influence of any
possible realization of a network topology on motor protein transport.  

\begin{figure}[h!]
 \begin{center}
 \includegraphics[width=1
 \textwidth]{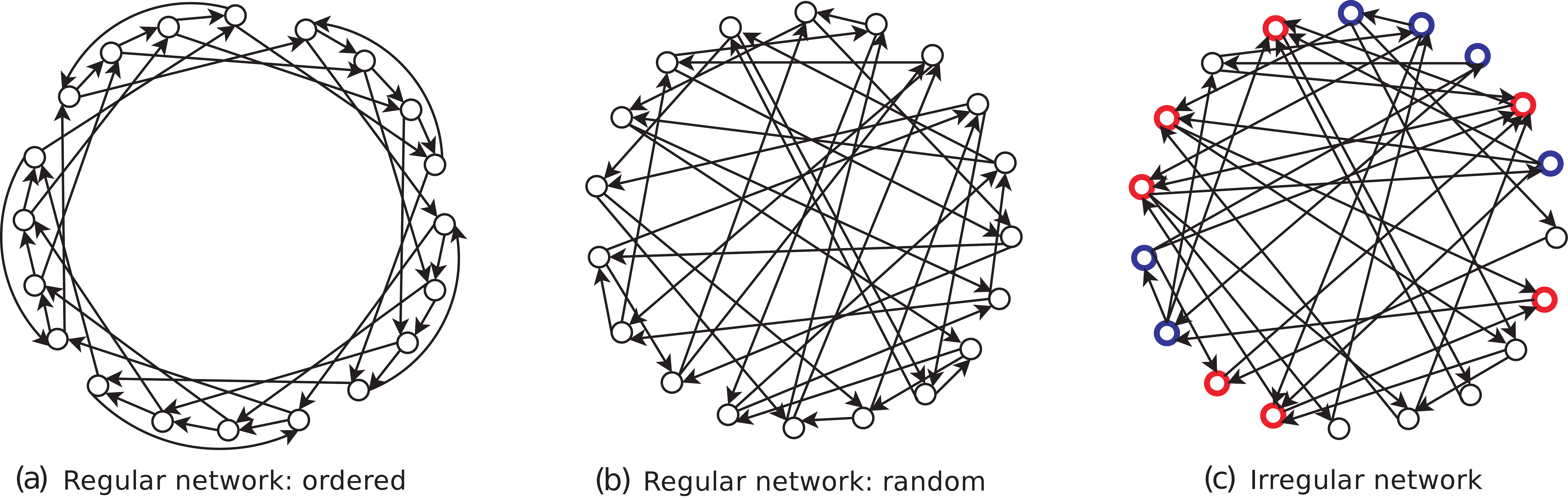}
 \caption{Three directed graphs which could serve as a model
 for the topology of a cytoskeleton.  
 Left: ordered square lattice of degree $c=2$. 
 Middle: 2-regular random graph.  Although the local connectivities of the
 vertices are the same, disorder is present in how vertices
 are connected at long distances. 
 Right: irregular random graph of mean connectivity $c=2$.  In this case, the
 local neighbourhoods of
 vertices vary from site to site.  We remark that the circles denote
 the junction sites while the crossings between segments do not represent real
 intersections.  Red vertices have more incoming segments than
 outgoing ones, blue ones
 have more outgoing segments than incoming ones,  and for black vertices the
 number of ingoing and outgoing segments is equal.}
 \label{reg:networks}
 \end{center}
 \end{figure}

Here we consider random networks in which a single
graph instance is drawn with a certain probability from an ensemble of graphs \cite{BB}.
 The randomness in the construction of the graph
 reflects to a certain extent  our lack of knowledge in the precise 
cytoskeletal
structure, as well as the complexity of the cytoskeleton.  In this work
we consider two types of graph ensembles: one without local
disorder (the $c$-regular ensemble, a.k.a.~Bethe lattice \cite{Bethe, Baxter}), and one with local
disorder (the irregular Erd\"os-R\'enyi ensemble) \cite{Gilbert, Erdo}.  
Local disorder is
defined through disorder in vertex degrees.  We define the indegree $c^{\rm in}_v$ of a
vertex $v$
as the number of segments arriving at a vertex while the outdegree $c^{\rm
out}_v$ denotes the number of segments leaving a vertex $v$.  Regular graphs are
then defined by the constraint $c^{\rm out}_v = c^{\rm in}_v = c$, such that all
vertices have equivalent local neighbourhoods.  As illustrated in figure
\ref{reg:networks}, we can consider ordered regular
graphs such as the square lattice or random regular graphs such as Bethe
lattices.  In irregular graphs the local vertex degrees of
different vertices can differ,  as in the  Erd\"os-R\'enyi ensemble mentioned above.  In this ensemble single
graph instances are constructed by
randomly
drawing  edges between the vertices: each directed segment
of the graph is present with a probability $c/|V|$ and absent with a probability
$1-c/|V|$.  Networks drawn from
the Erd\"os-R\'enyi  ensemble have 
a Poissonian degree distribution.   
We select the strongly connected component of the graph \cite{NeriT, Tar}, whenever necessary, to make sure that every junction can be reached from every other junction.  

As an
illustration,  we juxtapose in figure \ref{reg:networks} the square
lattice, as well as single graph instances drawn from the ensembles of 
 regular and  irregular graphs.  These ensembles have
been studied extensively in graph theory \cite{BB}, and  have been used to
study complex networks appearing in various sciences, e.g. the Internet, social
networks, regulatory networks in biology \cite{Barrat, Newman}.

\subsection{Motor proteins as active particles}
\begin{figure}[t]
 \includegraphics[ width =\textwidth]{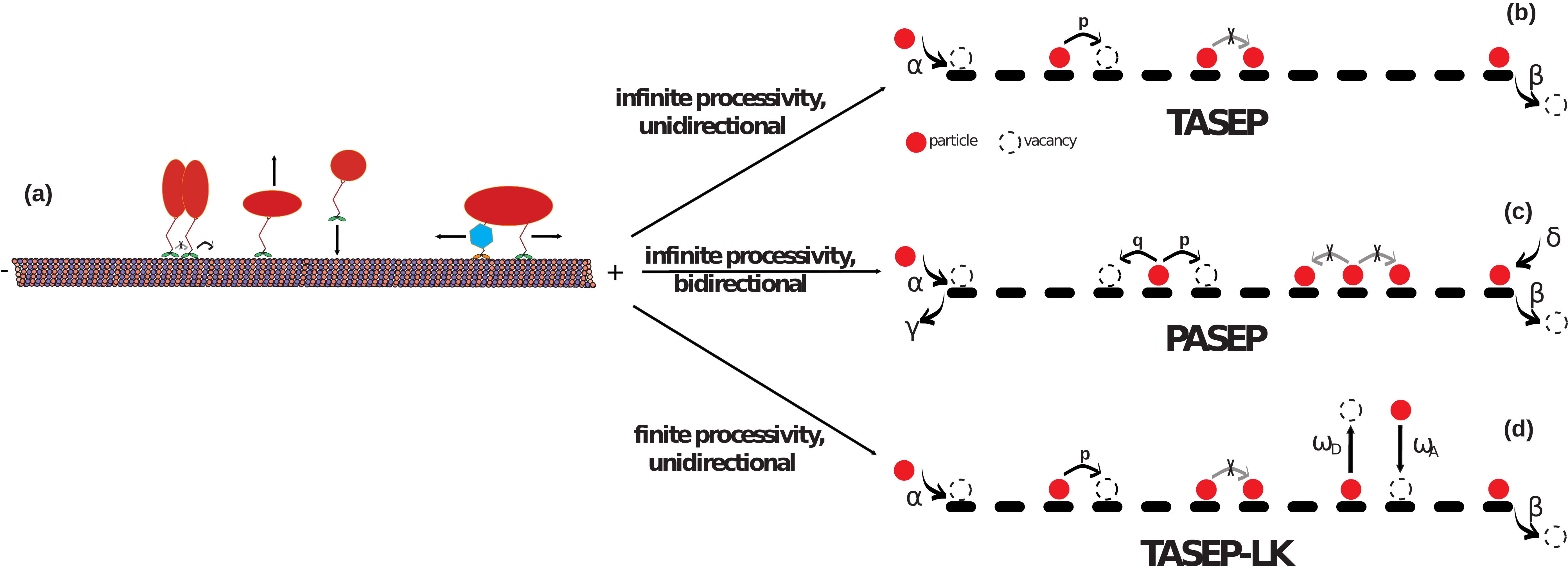}
 \begin{center}
 \caption{
   (a) Motor protein transport on a single filament: motors can
   stochastically step, without overtaking. They can bind
   and unbind in a kinetic exchange between the filament and the cytoplasm. 
   In certain cases the motors, or cargoes driven by competing motors, may
 also 
   undergo bi-directional motion.
   (b) TASEP process mimicking stochastic motion in one direction only, 
   with site-exclusion and perfect processivity of motors. 
   (c) PASEP process representing bi-directionality of
   motors or cargoes. 
   (d) TASEP-LK reproducing unidirectional motion, but also binding and
 unbinding processes of motors. 
 }\label{fig:models}
 \end{center}
 \end{figure}

Motor protein motion through one cytoskeletal filament is modelled as
a stochastic and biased motion through a single directed segment of the
network, see figure \ref{fig:models}. 
In this work we base the particle motion on one specific class of microscopic
models known as {\it exclusion processes}. For these particles
hop stochastically along the sites of the segments, with an excluded volume
condition which forbids that several particles occupy the same site.

There are
several reasons why exclusion processes are good models to study motor protein
transport on a mesoscopic level.   
First, they reduce many complex details of the stepping process to a rather
simple set of rules.   In particular                    
they allow to study collective effects in active transport due
to the interactions between the particles (which is more difficult to compute in more
elaborate models of motor protein transport \cite{Jul, Reimann}).
Second, it has been shown that exclusion processes
can describe both qualitatively and quantitatively the spatio-temporal organization of motor proteins along single biofilaments \cite{Nishinari, Reese, Varg2}.
Third,  they have been studied extensively over the last decades, in the
one-dimensional configuration of an open segment, interconnecting two particle reservoirs with injection/extraction rates $(\alpha, \delta)$ and $(\gamma, \beta)$, respectively (see figures \ref{fig:models} and
\ref{fig:transport-network}).  At this moment, analytical expressions for the
average current and density are known for numerous exclusion processes
\cite{Schutz2,Chou2011}, which will prove important when addressing generalizations to networks.

In figure \ref{fig:models} we illustrate three examples of exclusion
models which retain some of the essential characteristics of motor protein
transport.  The simplest variant is the totally asymmetric simple exclusion
process (TASEP), in which particles hop uni-directionally along a single segment
at a fixed rate $p$ \cite{Mac68, Mac69, Derrida} (see figure
\ref{fig:models}-(b)).  A simple extension of this model is given by
the partially  asymmetric simple exclusion process (PASEP)  \cite{Sandow}.  
Here the forward rate $p$ and the
backward rate $q$ differ (see figure \ref{fig:models}-(c)).  Such
a generalization of TASEP is useful for capturing the
bi-directional motion of motors, which can be due to fluctuations or to
competing motors transporting a cargo. Finally we consider
the totally asymmetric simple exclusion process with Langmuir kinetics
(TASEP-LK).   This model adds particle exchange with a reservoir,
through a binding/unbinding process obeying Langmuir kinetics, to the 
TASEP model (see figure
\ref{fig:models}-(d)).   Such exchange kinetics are important when studying
active transport of motors with finite
processivity, which can only cover a finite distance along the segment before they detach stochastically.


In the following we generalize the models presented in figure \ref{fig:models} to a
complex network and study their stationary state.  
In terms of the dynamics on the network, 
we have to complement the above rules for particle hopping in the segments by a
set of microscopic 
rules for their behavior at the junctions. A particle located at a junction
site can jump to either of the outgoing segments with equal probability, and
will then continue its one-dimensional dynamics along the new segment (see for
instance figures \ref{fig:biotransport} and \ref{fig:transport-network}). Here
we consider the simplest choice, where  jumps to all outgoing segments are
equally likely, but other choices are possible \cite{Em09,Ade}, to which our analysis could easily be adapted.

\subsection{Balancing currents at the junctions and segment properties from
effective rates}\label{sec:modelCyto}

We present now a method to deduce the stationary density profiles of
particles moving through a network from the stationary density profiles of an
individual segment \cite{NeriT}.   As 
illustrated in figure \ref{fig:transport-network}, we consider that the end
points of every segment in the network connect to reservoirs which
inject/extract particles with certain effective rates. 
The expressions of these rates are given, following
mean-field arguments,  by the occupation probabilities of the junction sites $v$, i.e. by the average density $\rho_v$ at the junctions.  This allows us to 
develop a description of transport through a complex network for which the
densities $\rho_v$ at the junction sites are
sufficient to determine all the currents and densities in the whole network
(even within the segments, since the junction densities determine the
effective rates of each segment).  

We determine the (average) densities $\rho_v$ by balancing currents at the junctions \cite{NeriT}. The continuity equations in $\rho_v$ read as: 
\begin{eqnarray}
 \frac{\partial \rho_v}{\partial t} = \sum_{s\rightarrow
v}J_{s\rightarrow v} -\sum_{s\leftarrow v}J_{v\rightarrow s}. 
\label{eq:continuity}
\end{eqnarray}
The quantity $J_{s\rightarrow v}$ is the  average current of particles
flowing from a segment $s$ into a junction site $v$, whereas $J_{s\leftarrow v}$ is
the reverse flow from the junction $v$ into the segment $s$.  Note that
the sums in equation (\ref{eq:continuity}) run over the segments of the graph and therefore
considers its specific topology. 

We assume now that the particle current in any  given segment $s = (v, v')$ 
is given by the current  between two particle reservoirs with appropriately chosen effective rates 
$(\alpha^{\rm eff}_s, \delta^{\rm eff}_s)$ 
and 
$(\beta^{\rm eff}_s, \gamma^{\rm eff}_s)$, 
see figure \ref{fig:transport-network}-(c).   
Hence the exact continuity equations (\ref{eq:continuity}) are approximated by 
the following mean-field equations
\begin{eqnarray}
 \frac{\partial \rho_v}{\partial t} = \sum_{s\rightarrow
v}J^-\left[\alpha^{\rm eff}_s, \delta^{\rm eff}_s;
\gamma^{\rm eff}_s, \beta^{\rm eff}_s\right] -\sum_{s\leftarrow
v}J^+\left[\alpha^{\rm eff}_s, \delta^{\rm eff}_s; 
\gamma^{\rm eff}_s, \beta^{\rm eff}_s\right], \label{eq:RhoL}
\end{eqnarray}
where $J^-$ and $J^+$, the current entering (leaving) the given segment, 
have the generic functional form known from a single segment: the in/out currents vary from one segment to the other {\it only} through the effective rates of the respective segments. 

In the stationary state, closure of the set of equations (\ref{eq:RhoL}) is  achieved by establishing
the expressions of the effective rates, for any segment $s= (v, v')$, by linking them
to the (average) junction densities $\rho_v$ and $\rho_{v'}$.  
The appropriate expression for the effective rates can be found  using a
mean-field approximation at the junctions \cite{Em09}.    
Note that even for those cases where the exact expression for the single-segment current is known, our 
procedure remains an approximation as segment cross-correlations at the junctions are not accounted for.

A particularly interesting aspect of our approach is that it constructs 
the description of transport at a network scale,
from transport within the single segments.  This leads to
a strong simplification of the master equations, allowing to solve them
on very large networks.  Moreover, due to this decomposition one can apply
our scheme to any model, with the sole condition of knowing a solution  $J^{\pm}_v$ for a single open segment with entrance/exit rates
$\left(\alpha, \delta\right)$ and $\left(\gamma, \beta\right)$.
Since several 
non-equilibrium models for transport have been solved exactly in this particular
one-dimensional configuration \cite{Schutz2,
Chou2011}, our approach is straightforwardly applicable to a very large number
of models. 

Let us finally define a certain number of macroscopic quantities which we will
use
throughout the article.   The total density
$\rho$ and the total current $J$ of the network are given by
\begin{eqnarray}
\rho &\equiv& \frac{L\sum_{s\in S}\rho_s +
\sum_{v\in V}\rho_v}{|S|L + |V|} \approx\frac{\sum_{s\in S}\rho_s}{|S|},  \\ 
J &\equiv&
\frac{L\sum_{s\in S}J_s +
\sum_{v\in V}J_v}{|S|L + |V|} \approx \frac{\sum_{s\in S}J_s }{|S|}. 
\end{eqnarray}  
where the approximations are valid if individual segments 
are sufficiently long $(L \gg 1)$.
The quantities $\rho_s$ and $J_s$ denote the average current and density through a given segment $s$.  When strong heterogeneities appear in the network, it is
particularly interesting to consider the distribution $W$ of segment densities $\rho_s$: 
\begin{eqnarray}
W(\rho_s) \equiv |S|^{-1}\sum_{s\in S}\delta(\rho-\rho_s).
\end{eqnarray}

\section{Analyzing TASEP on networks: effective rate diagrams and regimes of heterogeneity}
\label{sec:TASEP}

In this section we introduce two concepts which will prove central for studying  exclusion processes on networks: effective rate diagrams for the network and a classification of stationary states in terms of three regimes for heterogeneities.  The  effective rate diagrams
provide an intuitive yet quantifiable method which allows to rationalize the stationary state of transport processes on networks in terms of their single segment phase diagram. 
One of the merits of this approach is a  natural classification
of the stationary state of exclusion processes on networks in terms of three
distinct regimes: a {\it heterogeneous network} regime, a {\it heterogeneous
segment} regime and a {\it homogeneous} regime.  These regimes
correspond with the different
scales on which density heterogeneities can arise in the stationary state. 
As we will show in this work, these regimes give a unified
view on the stationary state of exclusion processes on networks and allow to appreciate the effect of the network topology and of microscopic rules for particle motion.

Here we use TASEP on a network \cite{NeriT}  to introduce these concepts. 
TASEP, illustrated in figure \ref{fig:models}-(a), is the simplest model.  
Particles follow TASEP rules in each
segment of the network.  At the
 junction sites they hop, with equal
probabilities, to one of the segments leaving the junction.  The model thus consists of a closed network on which particles, at a given overall density $\rho$, move uni-directionally,
stochastically and subject to mutual exclusion interactions.

Since the effective rate diagram builds on the single-segment phase diagram we
first  recapitulate the behavior of TASEP on a single open segment connecting two reservoirs (see the setting illustrated in figures
\ref{fig:transport-network}-(c) and \ref{fig:models}-(b)).    We use the
resulting current and density profiles to determine the spatial
stationary distribution of particles along complex  networks
and discuss the 
classification in terms of the length scales associated with heterogeneities.

\subsection{One-dimensional segment connecting two reservoirs}
We recall the exact TASEP expressions for the current and density of particles
moving through an open segment connecting two particle reservoirs, as shown
in figure \ref{fig:transport-network}-(c), in the
limit of large segments ($L\rightarrow \infty$).   
At the
entrance particles are injected
into the segment from  a reservoir (with entry rate $\alpha$), whereas at the
end of
the segment they are absorbed into a reservoir (with exit rate $\beta$). In
between particles hop uni-directionally at rate $p$ with mutual exclusion
interactions, see figure \ref{fig:models}-(b). The segment
density as a function of the reservoir
rates $(\alpha, \beta)$, is given by \cite{Derrida, Schutz, Der}: 
\begin{eqnarray}
 \rho^{\rm TASEP}\left[\alpha, \beta\right]  = \left\{\begin{array}{ccc}
\alpha/p
&
\alpha<\beta, \  \alpha/p < 1/2 &
\left({\rm LD}\right)  \\  1-\beta/p& \alpha>\beta, \  \beta/p<1/2& \left({\rm
HD}\right)  
\\ 1/2 & \alpha/p, \beta/p > 1/2 & \left({\rm MC}\right)  \end{array}\right., 
\label{eq:segmentD}
\end{eqnarray}
and the average current  $J$ obeys the parabolic profile:
\begin{eqnarray}
 J^{\rm TASEP}\left[\alpha, \beta\right]/p
&=& \: \rho^{\rm TASEP}\left[\alpha, \beta\right]\left(1-\rho^{\rm
TASEP}\left[\alpha, \beta\right]\right).
\label{eq:currentD} 
\end{eqnarray}
This average current is constant along a one-dimensional segment, since the particle number is conserved.

As one can see from equations (\ref{eq:segmentD}), TASEP leads 
to three stationary phases:  a  low-density phase (LD) for low values of 
$\alpha$, a high-density phase (HD) at low values for $\beta$ and a maximal 
current phase (MC) if both $\alpha$ and $\beta$ are large.  
In the LD and HD phases the density and current
profiles are boundary-controlled by the input or the exit rate, respectively.
The maximal current phase (MC) on the other hand is a saturated, bulk-limited
phase, invariably at density $\rho=1/2$ and with maximal current $J=p/4$.  Note that in this MC phase the bulk density and current are independent of the boundaries.

The LD and HD phases
are separated by a first order transition line at $\alpha=\beta<1/2$, where 
a  coexistence phase (LD-HD) arises.  Whereas the LD and HD phases 
correspond to homogeneous density profiles (except for localised boundary
effects),  
the LD-HD phase is heterogeneous, as a
domain wall separates the LD and HD regions which coexist on the same segment
\cite{Kol98}. 

The first order transition separating LD and HD phases is  an
important feature when discussing transport on networks.  It implies a
discontinuous variation in density at $\alpha=\beta$, from the 
LD value $\rho_{LD}=\alpha/p$ to the HD value
$\rho_{HD} = 1-\beta/p$. The corresponding jump  has size $\Delta \rho =
\rho_{HD}-\rho_{LD}= 1-2\alpha/p$, and we point out that it is maximal ($\Delta
\rho = 1$) at the origin. This observation will become important when analysing
networks in the following subsections.

\begin{figure}
  \begin{center}
    \includegraphics[ width=
      0.6\textwidth]{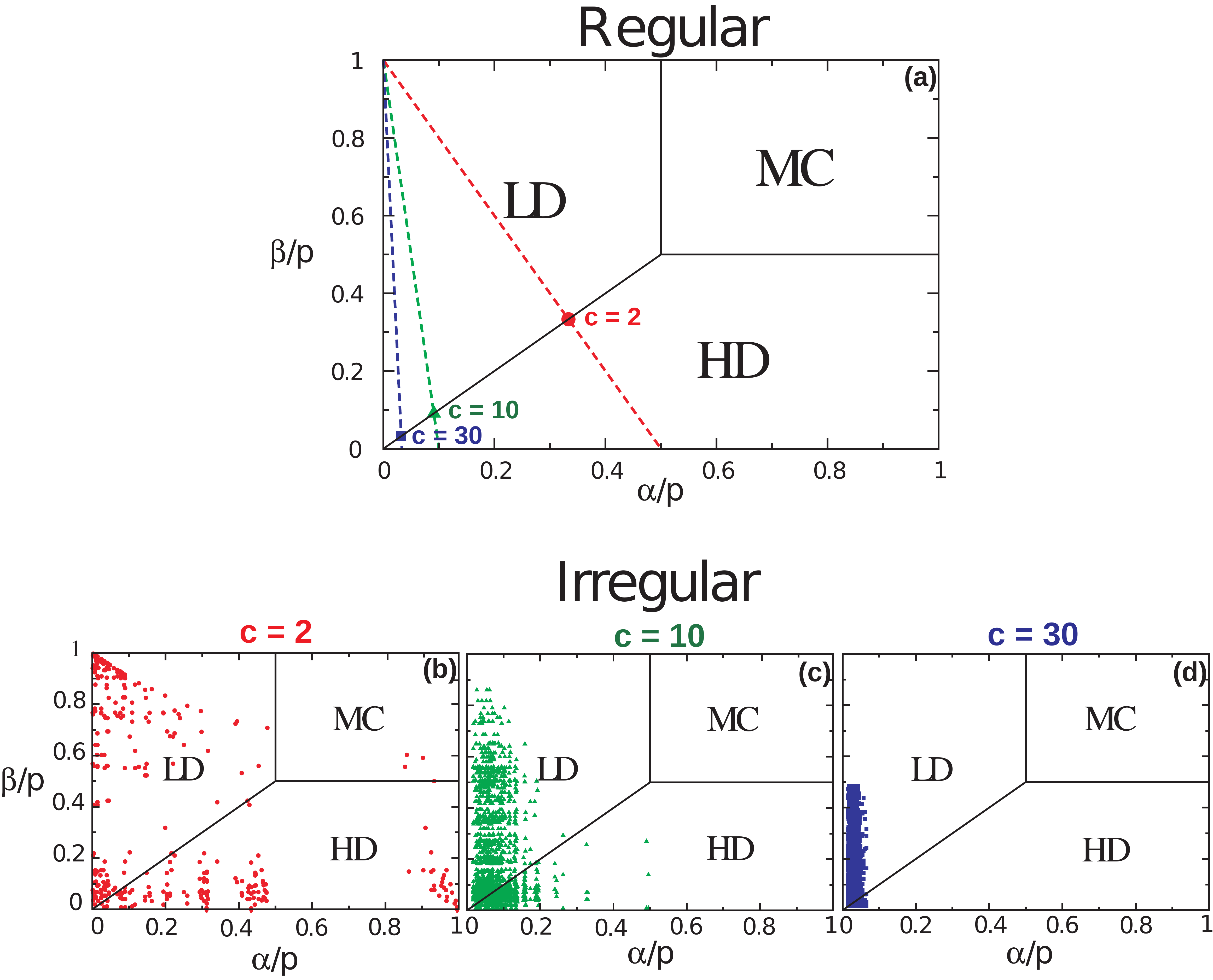}
  \end{center}
  \caption{
    Effective rate diagram, obtained by mapping the steady state effective rates 
    $(\alpha^{\rm eff}_s/p, \beta^{\rm eff}_s/p)$ for all segments $s$ in a network 
    onto the single-segment phase diagram of TASEP \cite{Der, Derrida, Schutz}.  
    The effective rates have been determined from the mean-field equations 
    (\ref{eq:MFTASEP}), and we have chosen an overall particle density $\rho=0.4$ 
    for all cases shown  
    (a): $c$-regular graph, at the given values of $c$.  
    All effective rates of all segments are equal, and coincide in one data point, shown here for  $\rho=0.4$.
   For other densities the   effective rates $(\alpha^{\rm eff}, \beta^{\rm eff})$  fall on the dashed lines.
    (b)-(d): strongly connected component of an irregular Erd\"os-R\'enyi graph, of given mean connectivity $c$. We have a scattered plot of effective rates, as
    these vary from segment to segment. 
    The number of junctions in the graphs are:
    $|V|=155$ $(c=2)$, $|V|= 200$ $(c=10)$ and $|V| = 500$ $(c=30)$.}
  \label{fig:effectiveRatesTASEP}
\end{figure}

\begin{figure}[h!]
\begin{center}
\includegraphics[width=
0.6\textwidth]{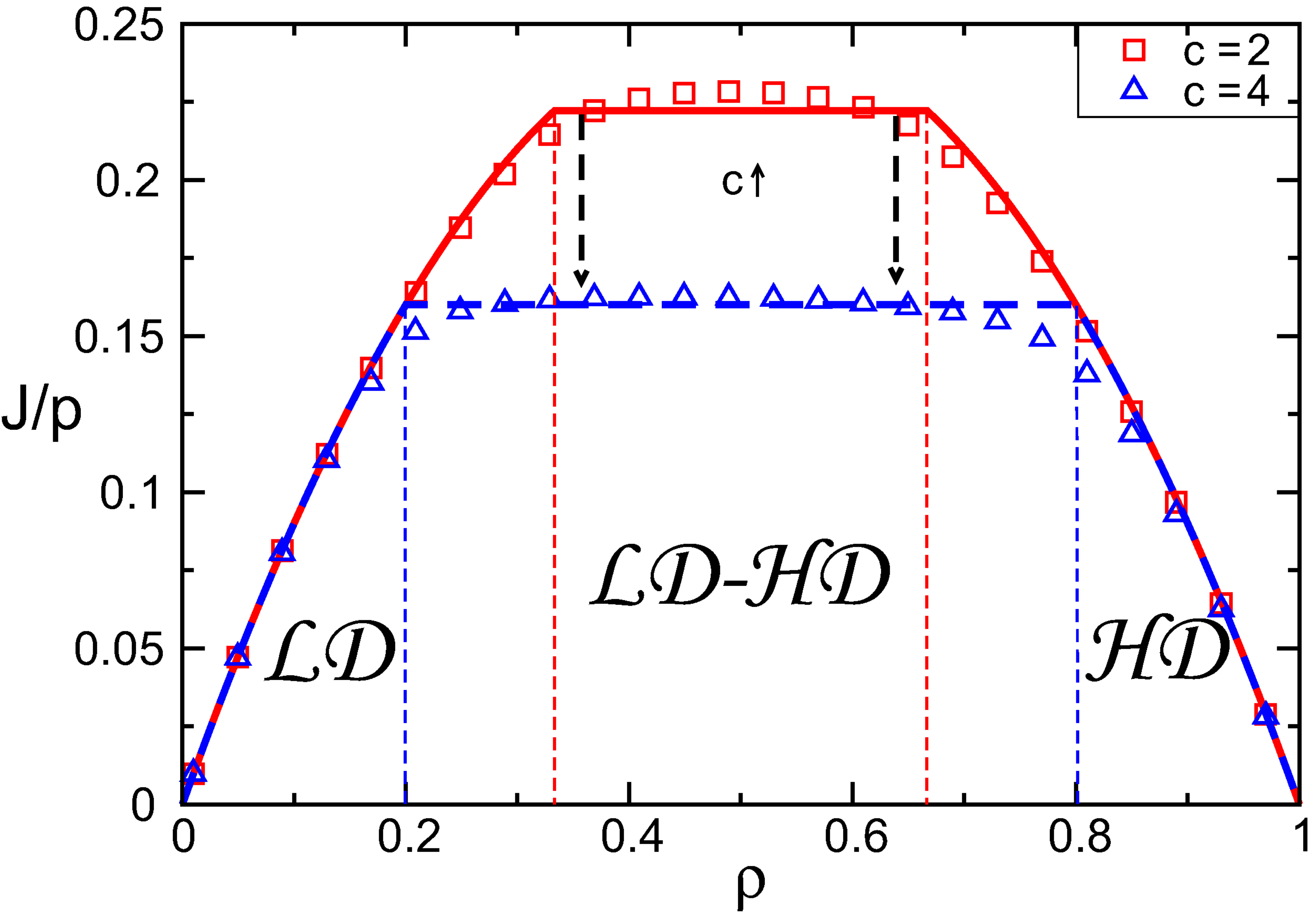}
\end{center}
\caption{Average current-density relation through regular graphs for a given
degree of connectivity $c$.   The plateau indicates the presence of a first
order transition and a coexistence phase, which dominates at increasing
connectivities.   The dashed lines indicate the transitions between the
homogeneous $\mathcal{LD}$ region, the heterogeneous $\mathcal{LD-HD}$ segment
region and the homogeneous $\mathcal{HD}$ region.  
Simulations (markers) are for segments of length $L=100$ and graphs of size $|V|
= 80$
junctions. The agreement between mean-field profiles (solid lines) 
and simulations improves further when increasing the segment length $L$.
}\label{reg:TASEPCurrentRho}
\end{figure}

\begin{figure}[h!]
\begin{center}
\includegraphics[width=
1\textwidth]{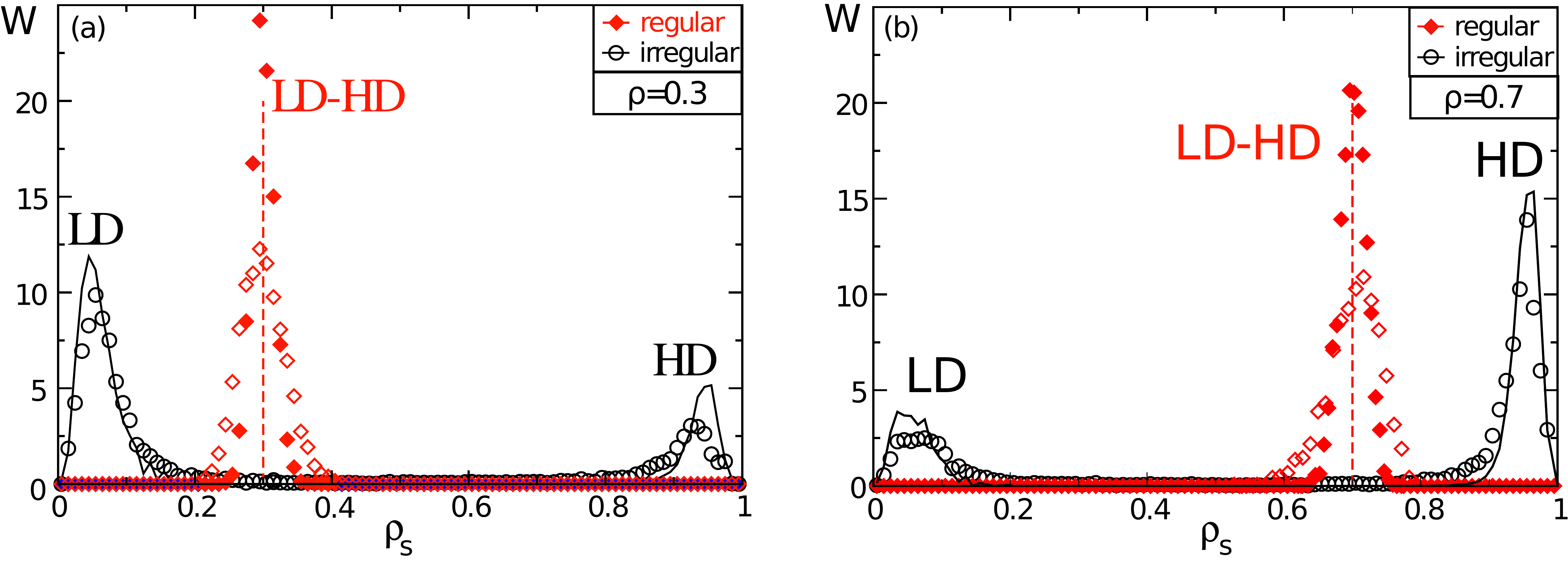}
\end{center}
\caption{The distributions $W$ of the segment densities $\rho_s$ for regular and
irregular networks are compared for two graph instances of average
connectivity $c=10$.  Note the difference between the bimodal distribution for
irregular graphs and the unimodal distribution for regular graphs.  We compare
mean-field
results (lines) with simulation  (markers). The
total density
$\rho$ equals $\rho=0.3$ (a) and $\rho=0.7$ (b).  The graphs have a size
of $|V|=1000$ junctions.   Simulations are 
run for segments of length
$L=100$.  The unimodal distribution of regular graphs predicted by mean-field
is approached gradually by simulations when increasing the runtime.
}\label{fig:bimodal}
\end{figure}

\subsection{Effective rate diagrams describing
TASEP through a network}\label{sec:TASEPNetw}
To determine the stationary distribution of particle densities for the TASEP
on networks we apply the
mean-field analysis laid out in subsection \ref{sec:modelCyto}.  The current and
density profiles  for an
individual segment $s$ in the network, $J_s$ and $\rho_s$,  are given by  $J_s =
J^{\rm
TASEP}\left[\alpha^{\rm eff}_s, \beta^{\rm
eff}_s\right]$ and $\rho_s = \rho^{\rm TASEP}\left[\alpha^{\rm eff}_s,
\beta^{\rm eff}_s\right]$. The incoming and outgoing
currents in each segment must match: $J^-[\alpha, \beta] = J^+[\alpha,
\beta] = J^{\rm TASEP}[\alpha, \beta]$,   due to current conservation along a
single segment. 
Equations (\ref{eq:RhoL}) for TASEP are now closed by relating the 
effective rates $\alpha^{\rm eff}_s[\rho_v]$  and
$\beta^{\rm eff}_s[\rho_{v'}]$ for each segment $s = (v, v')$ in the network 
to the junction densities $\left\{\rho_{v}\right\}_{v\in V}$.  
From the microscopic behaviour of the particles at the junctions, here the excluded
volume and the fact that all out-junctions are selected with equal probability, 
 mean-field
arguments lead to \cite{Em09}:
\begin{eqnarray}
  \alpha^{\rm eff}_v =p\frac{\rho_v}{c^{\rm out}_v}
  \qquad \mbox{and} \qquad
  \beta^{\rm eff}_v = p\left(1-\rho_v\right),  \label{eq:rates}
\end{eqnarray}
where $c^{\rm out}_v$ is the out-degree of junction $v$, i.e.~its number of 
outgoing segments.

Substituting the  effective rates in  equations (\ref{eq:RhoL})
leads to 
the following  set of closed equations in the density at the junctions
$\rho_v$:
\begin{eqnarray}
 \fl \frac{\partial}{\partial t} \rho_v = \sum_{v'\rightarrow
v}J^{\rm TASEP}\left[p\frac{\rho_{v'}}{c^{\rm out}_{v'}},
p\left(1-\rho_v\right)\right] -
\sum_{v''\leftarrow v}J^{\rm TASEP}\left[p\frac{\rho_v}{c^{\rm out}_v},
p\left(1-\rho_{v''}\right)\right].
\label{eq:MFTASEP}
\end{eqnarray}
Solving this set of equations, analytically or numerically, one finds 
stationary values for the densities at the junctions 
$\left\{\rho_{v}\right\}_{v\in V}$, which determine the stationary state of the
network.  These densities imply the effective rates
$(\alpha^{\rm eff}_{s}, \beta^{\rm eff}_s)$ for each segment, and thus also the average segment densities $\rho_s$ as well as the average currents $J_s$ through
each segment $s\in S$ (see
equations (\ref{eq:segmentD})-(\ref{eq:currentD})).

A very useful way to visualize and understand the transport characteristics of a
network is achieved by mapping the effective rates of the segments in the network onto the
$(\alpha/p, \beta/p)$-phase diagram of a one-dimensional open segment, see
figure \ref{fig:effectiveRatesTASEP}.  The state of the whole network can thus
be visualized by an effective rate diagram, represented in figure
\ref{fig:effectiveRatesTASEP}, from which one can read off which  phases are occupied by the segments in the network.  We notice a striking difference
between the stationary state of regular networks, for which all segments fall into the same phase, and irregular networks, for which the effective rates are
scattered over the single-segment phase diagram.  We discuss in the next two
paragraphs how this sets apart regular from irregular networks.

\subsection{Regular networks}

For regular networks all junctions are topologically identical, which also makes
all segments identical. Consequently all transport is governed by one pair of
effective rates, and ultimately (from equation (\ref{eq:rates})) by  a single
value for the junction occupancy $\rho_v$. The latter depends on the overall
density $\rho$, and 
the dashed lines in Figure \ref{fig:effectiveRatesTASEP}-(a) show how the effective rates evolve as the overall density $\rho$ varies. The labeled points correspond to an overall density of $\rho=0.4$, and here they fall onto the first order phase transition line, corresponding to a junction occupancy of $\rho_v  = c/(c+1)$. 
Note that, due to the presence of phase coexistence, this value of
the junction occupancy corresponds to a whole range of {\it overall} densities:
$\rho\in [\rho^\ast, 1-\rho^\ast]$, with
$\rho^\ast =(c+1)^{-1}$. This degeneracy is also reflected in the
current-density relation $J(\rho)$ \cite{Em09}, which is identical to that
 of each segment in the network (see Fig.
\ref{reg:TASEPCurrentRho}).  The density
heterogeneities associated with the LD-HD coexistence phase are directly linked
to a drop in transport capacity, which leads to the current plateau.


In terms of the network, the conclusion is that at low densities
($\rho\leq \rho^\ast$) all segments are in the LD phase and the particles
distribute homogeneously throughout the whole network.  
Similarly, at high densities ($\rho\geq 1-\rho^\ast$)
all segments are in the HD phase.
On the network level we refer to these as the {\it homogeneous ($\mathcal{LD}$
or $\mathcal{HD}$) regimes}, according to which phase dominates the behavior of
the network. We reserve calligraphic letters to refer to the regimes of the
network while the phases of the individual segments are 
 denoted by regular letters.  
In contrast, at intermediate densities all segments are in the 
LD-HD coexistence phase, and thus  heterogeneities are
present within each of the segments. We refer to this as a {\it heterogeneous
$\mathcal{LD-HD}$ segment regime}, to indicate that the density heterogeneities arise at the scale of single segments. 


These characteristics are further illustrated in figure \ref{fig:bimodal} 
where we present the distribution $W$ of segment densities $\rho_s$, here
superposing results from numerically solving the mean-field equations and
kinetic  Monte
Carlo simulations. There is good agreement for the irregular network.
For a regular network the above mean-field arguments would predict a delta-peak
at the overall density, here $\rho=0.3 (\rho=0.7)$, since all segments behave identically. 
The simulations corraborate this for intermediate densities, in that we indeed observe a single peak, which corresponds to segments in the LD-HD phase. 
Data for two different  simulation run lengths furthermore illustrate that the finite width of the peak reduces with the run length. Convergence to an actual Dirac distribution would require exceedingly long simulations, due to the presence of slow collective fluctuations in the coexistence phase \cite{Kol98}.

\subsection{Irregular networks}
In irregular networks the phenomenology is drastically different.  Here, as shown in figure
\ref{fig:effectiveRatesTASEP}-(b),    the variation in local connectivity at the junctions makes
the $(\alpha^{\rm eff}_s, \beta^{\rm eff}_s)$-effective rates scatter,
such that certain segments are found in a LD phase whereas others are in a HD
phase. Since the single-segment phase diagram of TASEP separates LD and HD
phases by a first-order transition line ($\alpha=\beta<1/2$), the segment
densities
on an irregular network are bimodally distributed. Such bimodality is a signature of strong heterogeneities in the way particles are distributed  on a
network scale.   
We say that irregular networks show a $\mathcal{LD/HD}$ {\it heterogeneous network regime}.

\begin{figure}[h!]
\begin{center}
\includegraphics[ width=1\textwidth]{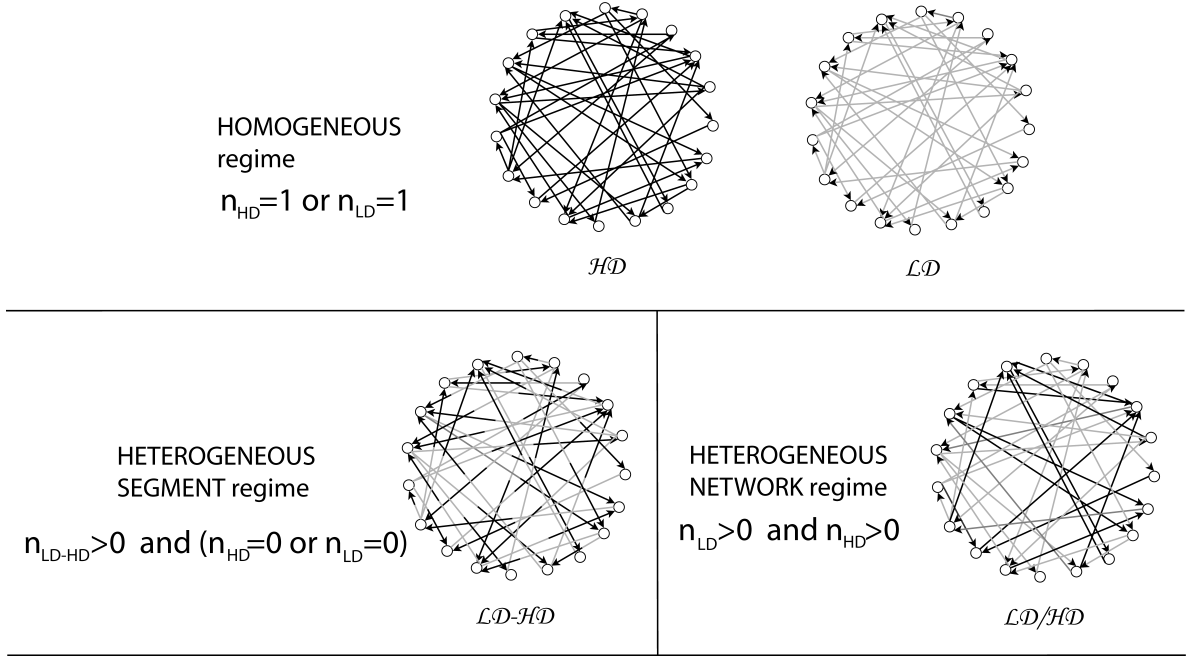}
\end{center}
\caption{The stationary states of excluded volume processes on networks are
classified in three distinct regimes.  These correspond  with the scale at which
heterogeneities in the particle densities arise throughout the network.  
They are defined in terms of the fraction $n_{\rm HD}$, $n_{\rm LD}$ and $n_{\rm
LD-HD}$  of segments which fall into the corresponding phases in the
single-segment phase diagram, see figure \ref{fig:effectiveRatesTASEP}.  The MC
phase will play a minor role in the following and therefore, for clarity, is not
considered.  }
\label{fig:networksVis}
\end{figure}

We quantify the above statements using  the
distribution of segment densities $W(\rho_s)$ in figure
\ref{fig:bimodal}.   We indeed notice the appearance of two peaks in the
distribution $W$: one peak is at low densities and accounts for the segments 
in the LD phase, whereas the other peak is at high densities and accounts for 
segments in the HD phase.  
As particles are added to the network the HD peak grows at the expense of the LD
peak, reflecting that the segments successively switch from LD to HD phases. 
This can be visualized rather intuitively in the effective rate diagram \ref{fig:effectiveRatesTASEP} as the process of certain points crossing the coexistence line.

The high connectivity limit allows particularly well to pinpoint the role of 
heterogeneities in irregular networks. Intuitively, this limit corresponds 
to the case where all junctions become bottlenecks, thus
reducing the flow of particles to almost zero. 
This is indeed apparent from the effective rates, which scale as $c$ as $(\alpha^{\rm eff}, \beta^{\rm eff}) \sim (\mathcal{O}(c^{-1}), \mathcal{O}(c^{-1}))$, see  analytical arguments  in the supplemental material of \cite{NeriT}.  
As a consequence they cluster in the $(\alpha, \beta)$-phase
plane in a zone which progressively retracts to the origin as $c$ increases,
 see figure \ref{fig:effectiveRatesTASEP}-(b).  
Since this zone  includes the first order transition lines we not only
preserve the bimodality in the high connectivity limit, but 
heterogeneities become even more pronounced, given that the density discontinuity associated with the LD to HD transition is maximum at the origin.

\subsection{Discussion: three stationary regimes two classify density heterogeneities}\label{subsec:discregimes}
Counterintuitive non-equilibrium phenomena, such as the emergence of strong
density heterogeneities in active transport on networks, can be clearly
understood using effective rate diagrams as presented in  figure
\ref{fig:effectiveRatesTASEP}.
In particular, it appears natural to classify the stationary behaviour using
three different regimes.
They are defined based on the fractions $n_{\rm LD}$, $n_{\rm HD}$, $n_{\rm
LD-HD}$  of segments which occupy the
corresponding phases in the effective rate diagrams:
\begin{itemize}
 \item 
   {\it heterogeneous network} regime ($\mathcal{LD/HD}$): 
   a finite fraction of segments occupy the LD phase 
   and a finite fraction occupies the HD phase. 
   We thus have that both  $n_{\rm LD}>0$ and $n_{\rm HD}>0$.    
   The distribution of segment densities $W(\rho_s)$ is bimodal, with 
   a LD peak at low densities and a HD peak at high densities. 
   As a consequence strong density heterogeneities
   develop on a network scale, i.e.~between individual segments.  
   At the same time LD and HD segments dominate, and the density profiles 
   within single segments remain mostly homogeneous.  
\item 
  {\it heterogeneous segment} regime ($\mathcal{LD-HD}$): the segments occupying the LD-HD phase dominate, and depending on 
  the overall density either the LD or the HD phases are 
  essentially unpopulated.   We therefore have $n_{\rm LD-HD}>0$ 
  and either $n_{\rm HD}=0$ or $n_{\rm LD}=0$.  At the network scale
  transport thus behaves rather uniformly: 
the segment density distribution $W(\rho_s)$
  in the stationary state is dominated by a single peak corresponding to 
  segments in LD-HD coexistence. The segments therefore behave similarly
  throughout the network, but strong heterogeneities are present {\it within} the 
  segments due to LD-HD phase coexistence.
\item {\it homogeneous} regime ($\mathcal{LD}$) or ($\mathcal{HD}$): all
  segments occupy either the LD, or HD phase, such that 
 either $n_{\rm LD}=1$ or $n_{\rm HD}=1$.
  Particles are distributed homogeneously throughout the network and few 
  heterogeneities appear at any scale. 
\end{itemize}
The three regimes are visualized in figure \ref{fig:networksVis}.  Again 
 calligraphic letters systematically stand for the regimes of the entire
network, rather than the phases of the segments.  

The above observations highlight how the network topology affects density  heterogeneities in TASEP transport.
Regular networks lead to a $\mathcal{LD}$ homogeneous regime (at low densities),
a $\mathcal{LD-HD}$ segment regime (at intermediate densities)
and a  $\mathcal{HD}$ homogeneous regime (at high densities).
Irregular networks on the other hand are dominated by the $\mathcal{LD/HD}$
network regime in which homogeneous LD and HD segments coexist on the network level.
This can easily be understood from the effective rate diagrams (figure
\ref{fig:effectiveRatesTASEP}).  In this picture the strong heterogeneities in
irregular networks is seen to be due to a combination of (i)
the disorder in the effective rates and (ii) the first order
transition in the single-segment phase diagram. This argument implies that 
heterogeneities must be expected to arise rather generally in excluded
volume processes on networks: whenever the junctions in the network 
are subject to some sort of local disorder, effective rates will be scattered 
on the single segment phase diagram. The presence of a first order
transition in the phase diagram then implies strong
heterogeneities, since a fraction of the effective rates will fall into the LD part 
while others fall into the HD zones of the phase diagram.  Based on these observations we
can anticipate what will happen when generalizing other models to complex
networks.  Indeed, the first order transition around the origin is also 
present in models of TASEP with extended particles \cite{Shaw, PierFrey}, TASEP
with
syncrhonuous
dynamics \cite{Tilstra}, TASEP with multiple lanes \cite{Pronx, Reich, Schiff,
Evansx}, TASEP with particles with internal states \cite{Nishinari, Luca1}, 
TASEP with directional switching \cite{Muhuri}, etc.
which suggests that they
too will lead to strong network heterogeneities on disordered graphs. 
From this argument it also becomes clear that those strong heterogeneities
do not depend on the particular choice of  Erd\"os-R\'enyi networks we
have considered here to model topological disorder.  
Rather, our results are expected to remain valid for other disordered systems such as
scale-free networks, regular networks with disorder in the hopping rates at the
junctions, etc.

\section{Bi-directional transport of infinitely processive particles}
\label{sec:ASEP}

In this work we intend to explore to which extent the microscopic motion of
motors affects the presence of density heterogeneities in transport processes on
networks.  A closer look at the motion of motor proteins and their cargoes
reveals that they can execute bi-directional moves along the  bio-filaments.
This may be due to backstepping of individual motors (known to account, for
instance, for something like $2-10\%$ of the displacements in kinesin
\cite{Schnitzer, Kojima, Visscher}), or to collective effects between motors of
opposite polarity (e.g. for the transport of organelles \cite{Welte, Gross}). 

We use PASEP to address the question of bi-directionality.
In PASEP particles move in a preferential
direction, but can also perform reverse hops, at a reduced rate, see figure \ref{fig:models}-(c).   TASEP is the special case where the rate for
backward hopping is set to zero, whereas the  opposite 'symmetric' 
limit of equal backward and forward hopping corresponds to the 
symmetric exclusion process (SEP).

On networks we generalise SEP such that the symmetric limit ensures a homogeneous equilibrium distribution of particles throughout the network. 
TASEP on the other hand has been seen to provoke strong inhomogeneities. 
In the following we use PASEP to interpolate between an active and a passive process, in order to investigate to which extent the bi-directionality of microscopic motion affects the formation of large-scale density heterogeneities in the stationary state. 

We follow the same outline as in the previous section.  First we recapitulate the
macroscopic behaviour of PASEP on a single open segment.  In  subsection \ref{subsec:PASEP:EffectiveRateDiagrams} we use the resulting
single-segment transport characteristics to determine the transport features
through a large network.  In the following subsections we explore the effect
of connectivity on the transport features through a study of PASEP
through regular  and irregular networks.  At last we discuss how heterogeneities
disappear  on networks when the PASEP process
approaches the equilibrium limit.

\subsection{Partially asymmetric exclusion process on a single segment}\label{subsec:PASEPOneSingle}
We revisit  the transport characteristics of PASEP through  a single, infinitely
long segment connecting two
reservoirs \cite{Sandow}.  Particles are injected (extracted) with rates $\alpha
(\gamma)$ on the left of the segment, and  with rates $\delta (\beta)$
on the right, see figure \ref{fig:models}-(c).   In between particles
hop forward (from left to right) at rate $p$, and they hop backwards
at rate $q$.  Just as in TASEP the particles interact through exclusion
interactions.  When we set $q=0$ we recover TASEP, while for $q=p$
this process reduces to the SEP.  

\begin{figure}
\begin{center}
\includegraphics[ width=0.6\textwidth]{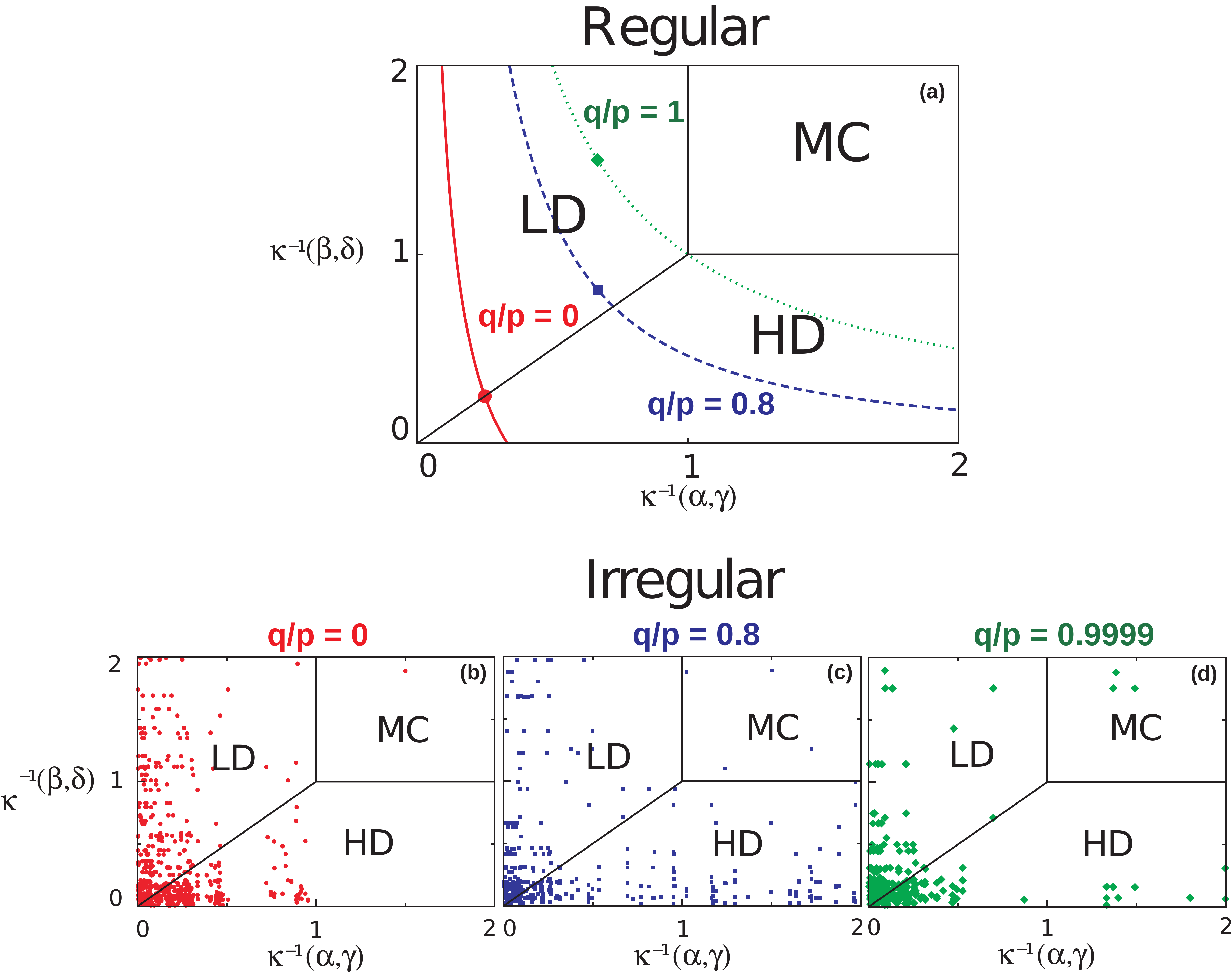}
\end{center}
\caption{
The effective rates for all segments  of a given graph instance of mean connectivity $c=4$, 
mapped onto the 
$(\kappa^{-1}\left[\beta, \delta\right], \kappa^{-1}\left[\alpha,
\gamma\right])$ phase diagram of a single open PASEP segment.   
The ratios of  backward to forward hopping rates $q/p$ are indicated.
The total density is $\rho = 0.4$ in all examples, to facilitate comparison.
(a): Effective rates  (markers) for a regular
network. The lines represent the relation between
$\kappa^{-1}\left[\beta, \delta\right]$ and $\kappa^{-1}\left[\alpha,
\gamma\right]$ when varying the total density $\rho$. 
(b)-(d): Effective rates for an irregular network of  size $|V|=193$, for
several ratios $q/p$.  The diagrams are truncated at $\kappa^{-1}=2$, but data
points are present over a larger range.
}\label{fig:EffRatePasep}
\end{figure}

The average current in
PASEP must be constant
throughout the segment, as particles can neither be destroyed nor created. 
Its current-density profile is parabolic, as is usual for exclusion
processes: 
\begin{eqnarray}
 J^{\rm PASEP}\left[\alpha,
\gamma; \delta,
\beta\right]
 = (p-q)\rho^{\rm PASEP}\left[\alpha,
\gamma; \delta,
\beta\right]\left(1-\rho^{\rm PASEP}\left[\alpha,
\gamma; \delta,
\beta\right]\right), \label{eq:JPASEP} 
\end{eqnarray}
with a homogeneous density $\rho^{\rm PASEP}$ 
\begin{eqnarray}
 \fl  \rho^{\rm PASEP}\left[\alpha, \gamma; \delta, \beta\right] = 
  \left\{
  \begin{array}{ccccc}
\frac{(p-q+\gamma+\alpha)-\sqrt{(p-q+\gamma+\alpha)^2-4(p-q)\alpha}}{2(p-q)}
&\
&\kappa[\alpha, \gamma]>\kappa[\beta, \delta], \kappa[\alpha, \gamma]>1, &\
&\mbox{(LD)} \\ 
    \frac{(p-q-\beta-\delta)+\sqrt{(p-q-\beta-\delta)^2+4(p-q)\delta}}{2(p-q)}
&\ &\kappa[\beta, \delta]>\kappa[\alpha, \gamma], \kappa[\beta, \delta]>1,
&\ &\mbox{(HD)}\\ 
    1/2     &\ & \kappa[\beta, \delta]\leq 1, \kappa[\alpha, \gamma]\leq 1.
   &\ &\mbox{(MC)}
  \end{array}\right.
  \label{eq:RMF} \nonumber \\ \label{eq:rhoASEP}
\end{eqnarray}
Here we have assumed $q<p$, without any loss of generality, and taken the limit of infinite segment length ($L \rightarrow \infty$).  The opposite case of $p<q$  is implicitly treated through particle-hole symmetry.  
The quantity  $\kappa$ is a  dimensionless function,
\begin{eqnarray}
 \kappa\left[\alpha, \gamma\right] = 
 \frac{1}{2\alpha}\left(-\alpha + \gamma + p - q + \sqrt{(-\alpha + \gamma +p -
q)^2 + 4\alpha \gamma}\right),
\end{eqnarray}
which will be central to the following discussion. We use the inverse quantities $\kappa^{-1}$, which allows us to cast the single-segment phase diagram into a familiar shape.
The phase diagram maps directly onto that for TASEP, based on the quantites$\kappa[\alpha, \gamma]$ and  $\kappa[\beta, \delta]$ introduced above, see figure \ref{fig:EffRatePasep}) \cite{Sandow}.

The PASEP has, just like TASEP,  a first-order transition between an LD
and a HD phase, along the coexistence line $\kappa\left[\alpha, \gamma\right] =
\kappa\left[\beta, \delta\right]<1$.  
Since the position of this first order transition in the phase diagram was the
key to our understanding of heterogeneities in TASEP on networks (limit $q=0$), it is now interesting to see how this picture is modified as particles are allowed to backstep ($q>0$). 
We find that the first order transition remains present,  even
in the limit of symmetric motion ($q\rightarrow p$).  The density jump is equal to
$\Delta \rho = \rho_{HD}-\rho_{LD} = 1/(1+\kappa^{-1}\left[\beta, \delta\right])
- 1/(1+\kappa\left[\alpha, \gamma\right])$. Just as for TASEP this discontinuity
increases when approaching the origin of the $(\kappa^{-1}\left[\alpha,
\gamma\right],\kappa^{-1}\left[\beta, \delta\right])$-phase plane, and reaches 
its maximal value  $\Delta\rho = 1$ at the origin.

The TASEP limit ($q \rightarrow 0$) is straightforward: equation (\ref{eq:rhoASEP}) reduces to 
$\kappa[\alpha, \gamma]= \alpha/(1+\alpha)$ and we recover the phase diagram of
TASEP.   
In contrast, the symmetric limit $q\rightarrow p$ is more subtle: the limiting process is a SEP, but we do not recover the corresponding density profile from equations (\ref{eq:rhoASEP}) when setting $q=p$: we find a  homogeneous bulk density, determined by one of the two boundaries, whereas the SEP leads to a constant density gradient, $\rho_{LD}  = \frac{\alpha}{\alpha+\gamma}$ and $\rho_{HD}  =
\frac{\delta}{\beta+\delta}$. 
The reason is that the limits $L \rightarrow \infty$ and $q/p\rightarrow 1$ do not commute. 
Since in equation (\ref{eq:rhoASEP}) we have taken the limit $L \rightarrow \infty$ before considering  $q/p\rightarrow 1$, the behaviour remains in the strongly asymmetric case \cite{Mal}. 
We will return to this point when discussing networks in the following sections.

\subsection{Effective rate diagrams for PASEP on networks} \label{subsec:PASEP:EffectiveRateDiagrams}
We base our study of PASEP transport on networks using the mean-field
algorithm given by equations (\ref{eq:RhoL}), as we did for TASEP in subsection  \ref{sec:TASEPNetw}.
The current through a segment, $J_s = J^{\rm PASEP}\left[\alpha^{\rm eff}_s,
\gamma^{\rm eff}_s; \delta^{\rm eff}_s,
\beta^{\rm eff}_s\right]$, is now given by
equations (\ref{eq:JPASEP})-(\ref{eq:rhoASEP}). 
From current conservation we still have
$J^+_s = J^-_s = J^{\rm PASEP}\left(\alpha^{\rm eff}_s,
\gamma^{\rm eff}_s; \delta^{\rm eff}_s, \beta^{\rm eff}_s\right)$.  
To determine the effective rates required to close the set of equations (\ref{eq:RhoL}) we adopt the rule that particles which leave a
junction select any of the outgoing segments with equal probability.   In this
way we obtain: 
\begin{eqnarray}
 \alpha^{\rm eff}_{(v, v')} &=& p^{{\rm junction}}_v \: \rho_v, \\ 
 \beta^{\rm eff}_{(v, v')}  &=& p \, (1-\rho_{v'}),\label{eq:eff1} \\ 
 \gamma^{\rm eff}_{(v, v')} &=& q \, (1-\rho_{v}), \label{eq:eff2}\\ 
 \delta^{\rm eff}_{(v, v')} &=& q^{{\rm junction}}_{v'} \: \rho_{v'}.
\end{eqnarray}
The microscopic rates $p^{\rm junction}_v$  and $q^{\rm junction}_{v'}$ 
denote the rates for particles at the junction site, i.e.~$p^{\rm junction}_v$ 
is the rate for a particle on a junction to step into one of its $c^{\rm out}_v$
outgoing segments, whereas $q^{\rm junction}_{v'}$ is the rate at which it
backstpdf into one of the $c^{\rm in}_{v'}$ incoming segments.  We furthermore
require the  activity of particles at the 
junctions to be equal  to the activity in  the segments, i.e. 
$p+q = c^{\rm out}_v\:p^{{\rm junction}}_v + c^{\rm in}_v q^{\rm junction}_v$. 
Moreover, for $q=0$ we wish to recover TASEP rates ($q^{{\rm
junction}}_{v'}=0$, $p^{{\rm junction}}_v = p/c^{\rm out}_v$),  whereas for $q=p$ we
impose that the dynamics fulfill detailed balance, i.e.~$p^{{\rm junction}}_v =
q^{\rm junction}_{v'}$.  Considering the above
conditions,  we are lead to: 
\begin{eqnarray}
  \alpha^{\rm eff}_{(v, v')} &=&
p\left(\frac{\rho_v}{\left(\frac{p}{p+q}\right)c^{\rm
out}_v+\left(\frac{q}{p+q}\right)c^{\rm in}_v}\right),\label{eq:eff3}\\  
\delta^{\rm eff}_{(v, v')}& =&
q\left(\frac{\rho_{v'}}{\left(\frac{p}{p+q}\right)c^{\rm
out}_{v'}+\left(\frac{q}{p+q}\right) c^{\rm
in}_{v'}}\right).\label{eq:eff4}
\end{eqnarray}
Substituting these effective rates and the current profile
equations (\ref{eq:JPASEP}) and (\ref{eq:rhoASEP}) in the expression (\ref{eq:RhoL}) leads
to the required closed set of equations in the junction densities $\rho_v$. 
Note that due to our microscopic hopping rules at the junctions  the
the symmetric limiting process (q=p, corresponding to SEP) fulfills detailed balance, which leads to 
a stationary state with a completely homogeneous distribution of 
particles over the network.

It again proves insightful to map the effective rates of the segments onto the
single-segment phase diagram of an open PASEP segment, via $\kappa^{-1}\left[\alpha^{\rm eff}_s, \gamma^{\rm
eff}_s\right]$ 
and $\kappa^{-1}\left[\beta^{\rm eff}_s, \delta^{\rm eff}_s\right]$
for each
segment $s\in S$. 
In figure \ref{fig:EffRatePasep} we compare the effective rate diagrams for a
regular network (top) and an irregular network (bottom), for various ratios
 $q/p$. Note the similarity between the PASEP effective rate diagrams 
figures \ref{fig:EffRatePasep} and the TASEP effective rate diagrams
figures \ref{fig:effectiveRatesTASEP}.

\begin{figure}[h!]
\begin{center}
\includegraphics[ width= 1\textwidth]{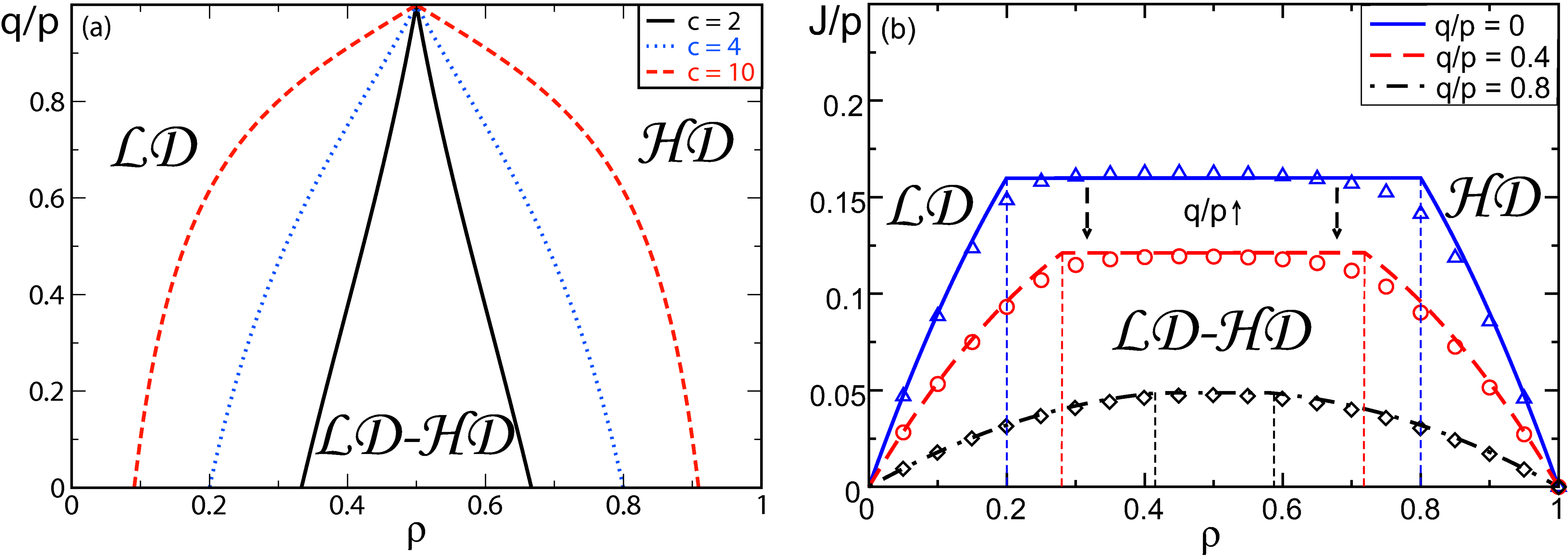}
\end{center}
\caption{PASEP through regular graphs. Left (a):   The 
transition
lines between the  $\mathcal{LD}$ and  $\mathcal{LD-HD}$
regimes, as
well as, the $\mathcal{LD-HD}$  and
$\mathcal{HD}$ regimes are presented as a function of the total
particle density $\rho$ and the
 fraction of rates
$q/p$ for regular graphs of given degree $c$.   For $q/p\rightarrow 1$ the
process becomes passive. The heterogeneous
$\mathcal{LD-HD}$ segment regime disappears in this limit. 
Right (b): Average current-density profile through regular graphs of degree
$c=4$.  The onset of a $\mathcal{LD-HD}$ regime is identified by
a plateau in the current density profile.  Mean-field results (solid lines) are
compared with simulation results
(markers)
on graphs of  $|V|=80$ junctions and  segments of length $L=400$ at different
degrees of particle
asymmetry $q/p$.
We have verified that deviations between
mean field        
and simulations decrease with increasing segment size.
}\label{reg:ASEP}
\end{figure}

\subsection{Regular networks}\label{sec:PASEPReg}

From the effective rate
diagram in figure \ref{fig:EffRatePasep} we see
that all segments have the same transport characteristics.  Indeed, 
equations (\ref{eq:RhoL}) admit  the solution $\rho_v=\rho', \ \forall v\in V$,
i.e. an identical density on all vertices $v$.  Thus all segments will have the
same average current $J_s = J$ and average density $\rho_s=\rho$.   The current-density
profile for the network is therefore the truncated parabola of a single segment and the density at the junction is $\rho_v=c/(c+1)$ on the plateau.

This shows that, all in all, PASEP through regular networks leads to a stationary 
state similar to that of TASEP.  At densities below those leading to coexistence in single segments ($\rho<\rho^\ast$) the network displays a homogeneous $\mathcal{LD}$ regime.
At intermediate densities the system is in the heterogeneous $\mathcal{LD-HD}$
segment regime, while at high densities ($\rho>1-\rho^\ast$) we have the $\mathcal{HD}$ homogeneous regime. The segment
regime corresponds to the plateau region in the current-density profile.  Just
as for TASEP the density range for the segment regime, of width $1-2\rho^\ast$, increases as a function
of the connectivity $c$.  On the other hand, the size of the segment regime
gradually decreases to zero when
approaching the symmetric $q=p$ limit corresponding to a passive process.   
Density heterogeneities thus disappear in this limit, see figure 
\ref{reg:ASEP}-(a).

We present the
analytical mean-field solution for the  current-density profile $J(\rho)$ of
the network in figure \ref{reg:ASEP}-(b). The total density value $\rho^\ast$ separating
the $\mathcal{LD-HD}$ segment
regime and the homogeneous $\mathcal{LD}$ regime is given by
\begin{eqnarray}
 \rho^\ast &=& \frac{p-q + \frac{p+q}{c+1} - \sqrt{\left(p - q +
\frac{p+q}{c+1}\right)^2 - \frac{4(p-q)p}{c+1}}}{2(p-q)}. \label{eq:lim}
\end{eqnarray}
In figure \ref{reg:ASEP}-(a) we have plotted this threshold $\rho^\ast$ as a function of $q$
and $p$ for various values of connectivity $c$.  This constitutes the phase diagram
for PASEP through regular networks.  Note how the heterogeneities gradually
disappear as the symmetric $q=p$ process is approached.

\begin{figure}
\begin{center}
\includegraphics[ width= 1\textwidth]{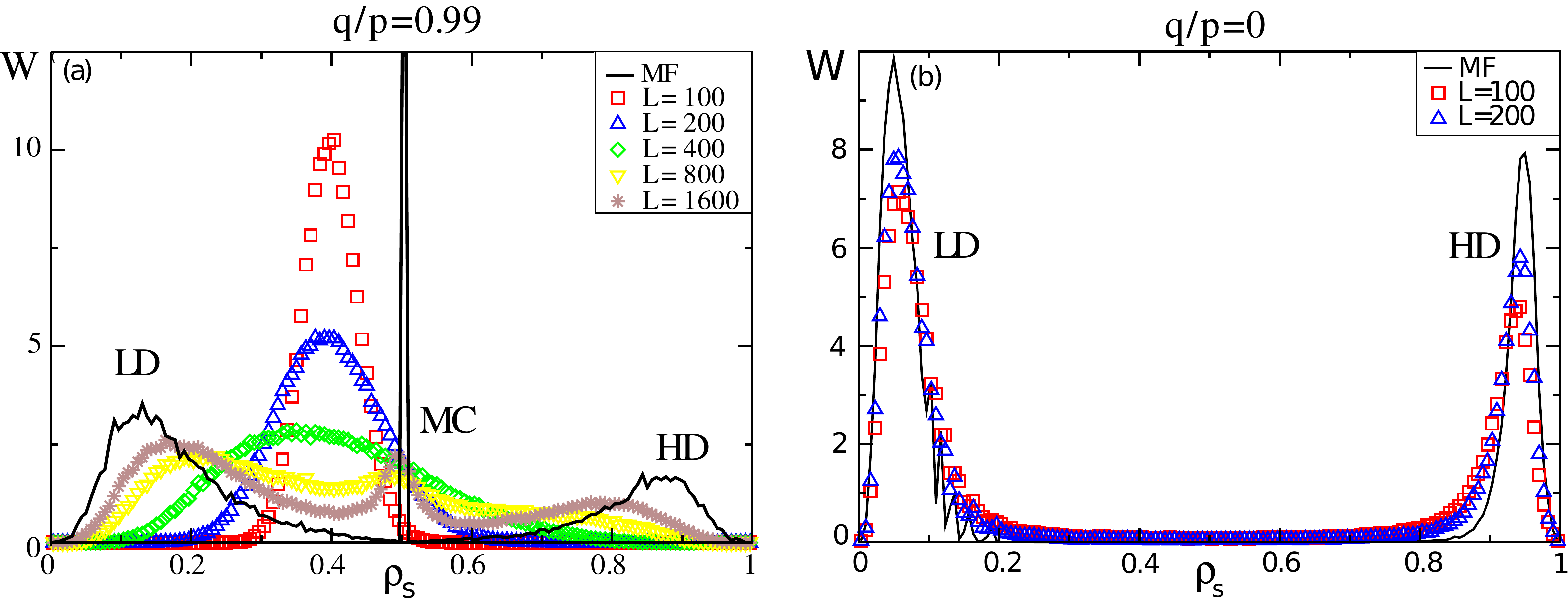}
\end{center}
\caption{The  distribution $W$ of the segment densities $\rho_s$ for a PASEP
process on irregular networks at a total density $\rho=0.4$.  We have plotted
the average
distribution over
$100$ graph instances drawn from the  Erd\"os-R\'enyi ensemble at
mean connectivity $c=10$ and size $|V|=200$. 
Solid lines denote mean-field results while  markers denote  simulation results 
for indicated segment lengths $L$. Left (a): the almost symmetric $q/p=0.99$
case. 
We                                                           
notice that for low
values of $L$ the segment distribution is unimodal, peaked at
the average value $\rho=0.4$. We are in a regime similar to passive
diffusion. For larger values of $L$ we enter  the active
regime and trimodality
arises in the segment distribution $W$  (the trimodality corresponds with the
three phases: LD, HD and MC). The
distribution seems eventually to converge to the mean-field expression. 
Right (b): We
show how $W$ depends on $L$ in the  totally
asymmetric $q=0$ case.  We see that
the bimodality is already present at  very low values of the segment lengths
$L$. 
}\label{fig:PASEPirreg}
\end{figure}

\subsection{Irregular networks} \label{sec:PASEP:irregular}
For irregular graphs the effective rate diagrams again 
show scatter plots, see figure \ref{fig:EffRatePasep}.  
As a consequence, a finite fraction of
segments will occupy the LD, HD and MC phases.
Accordingly, the segment density distribution now displays three peaks 
corresponding
to these three phases, see figure
\ref{fig:PASEPirreg}-(a).  Note also that the peak associated with the MC phase 
gradually disappears in the TASEP limit ($q/p \rightarrow 0$).  We expect this feature to change when considering a non-uniform hopping rule at the junctions.

\begin{figure}
\begin{center}
\includegraphics[width= 0.7\textwidth]{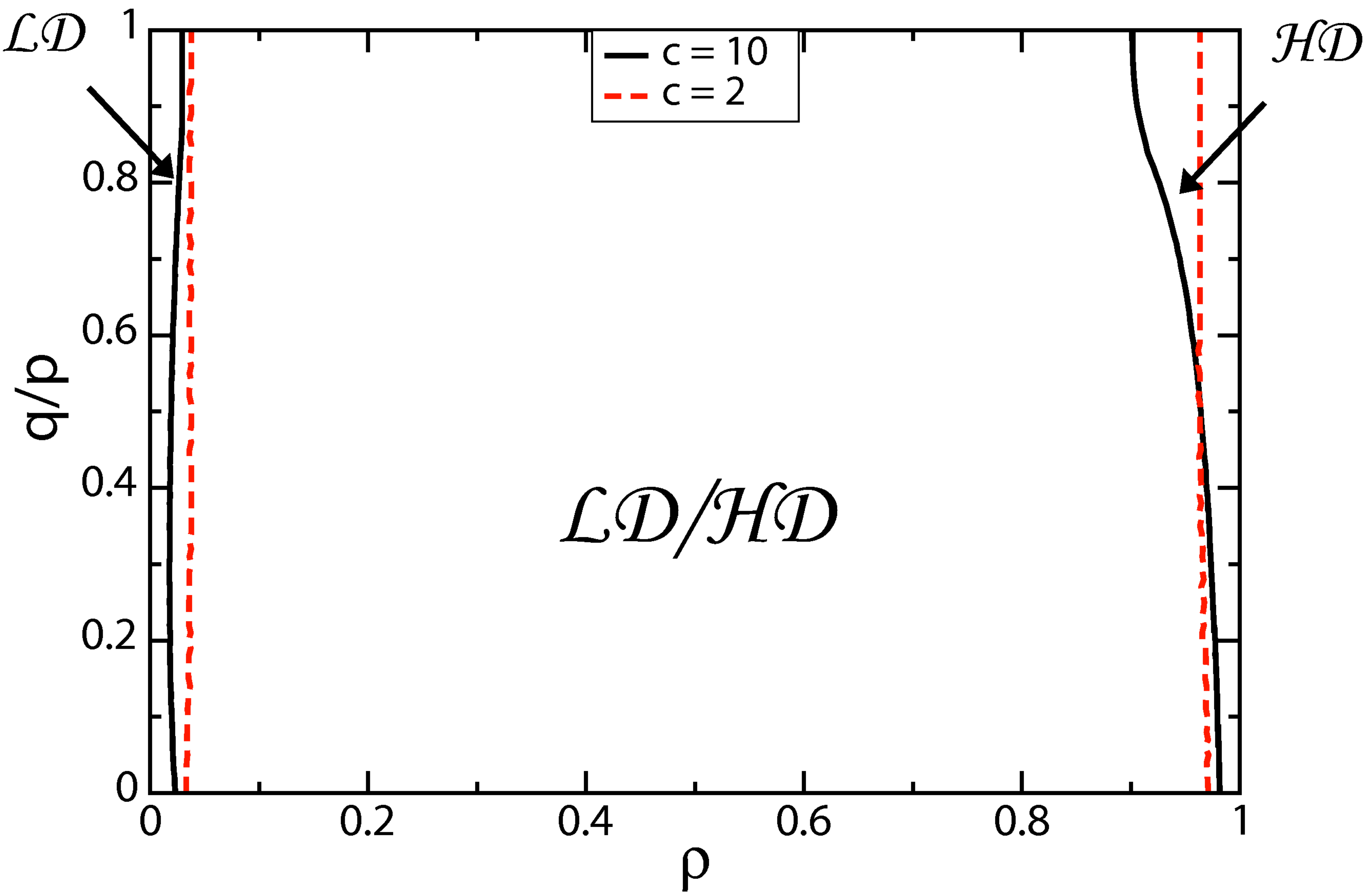}
\end{center}
\caption{The $(q/p, \rho)$-phase diagram for PASEP through irregular networks. 
We plot the 
boundaries between different regimes for single graph instances with $c=10$,
$|V| = 200$ and $c=2, |V| =607$ at $L=\infty$.  The  $\mathcal{LD}$/
$\mathcal{HD}$ regime remains present in the limit $q/p \rightarrow 1$.
}\label{fig:PASEPirregPhaseDiagram}
\end{figure}

Just as for regular graphs, it is possible to determine the overall network
phase diagram for a PASEP process through an irregular graph (see figure
\ref{fig:PASEPirregPhaseDiagram}).  We use the definitions in figure 
\ref{fig:networksVis} and in the discussion \ref{subsec:discregimes}  to find the transition lines between the $\mathcal{LD/HD}$
and $\mathcal{LD}$ or $\mathcal{HD}$ phases on the network.  Remarkably, we
notice that the $\mathcal{LD/HD}$ network regime remains prominent, even for
the symmetric limit $q\rightarrow p$. 

This symmetric limit is a point worth discussing in the
context of irregular networks. In principle we expect density heterogeneities to
disappear in this limit, since we reach a passive equilibrium process.  
However, rather surprisingly,
one can see from figure \ref{fig:PASEPirreg} that  our mean-field results 
show a pronounced trimodality at values very close to this limit (here
$q=0.99\:p$), suggesting that the trimodality persists even for  $q=p$. 
This too can be understood from the effective rate diagram, since the first
order transition in this diagram remains present in the symmetric limit (see
figure
\ref{fig:EffRatePasep}).   At first sight this result seems to contradict the
fact that passive processes must lead to a  homogeneous distribution  of particles along the network.
The reason for this apparent contradiction is the subtle interplay between the two
limits $q\rightarrow p$ and $L\rightarrow \infty$ which do not commute, as mentioned in subsection \ref{subsec:PASEPOneSingle}.

To discuss the role of the limits $q\rightarrow p$ and $L\rightarrow \infty$ 
more carefully we have performed simulations for increasing systems sizes,  
at the value $q=0.99\:p$.  These results are presented in figure 
\ref{fig:PASEPirreg}.  
At small segment lengths $L$ the density distribution is unimodal. In
contrast, this distribution is trimodal for sufficiently long segments, as
predicted by 
mean-field arguments, which assumes infinitely long segments. 
Hence, the segment length can strongly influence the
density heterogeneities on the network.
For $q \approx p$ particles distribute homogeneously over the network for small segments $L$, in correspondence with a passive process.  
For sufficiently long segments, on the other hand, particles distribute
heterogeneously in correspondence with an active process.  
It would be interesting to explore how this cross-over
from a trimodal to a unimodal distribution scales as  a function of the
asymmetry in the hopping rates $(q-p)/p$ as well as the filament length
$L$, and one might suspect that it is related to a change of universality class
of transport processes in the limit $q\rightarrow p$ \cite{Mal}.

\subsection{Discussion}
We  conclude that the stationary state of bi-directional transport
on networks has similar characteristics to that of uni-directional
transport. 
Indeed, PASEP on networks leads to stationary regimes which have their direct equivalent in TASEP: 
regular networks feature $\mathcal{LD}$, $\mathcal{LD-HD}$ and $\mathcal{HD}$
regimes, while irregular networks are
dominated by the $\mathcal{LD/HD}$ regime.    From these results it follows that the degree of asymmetry 
does not change the qualitative picture for the emergence of density heterogeneities in cytoskeletal motor protein transport.

\begin{figure}
\begin{center}
\includegraphics[ width= 1\textwidth]{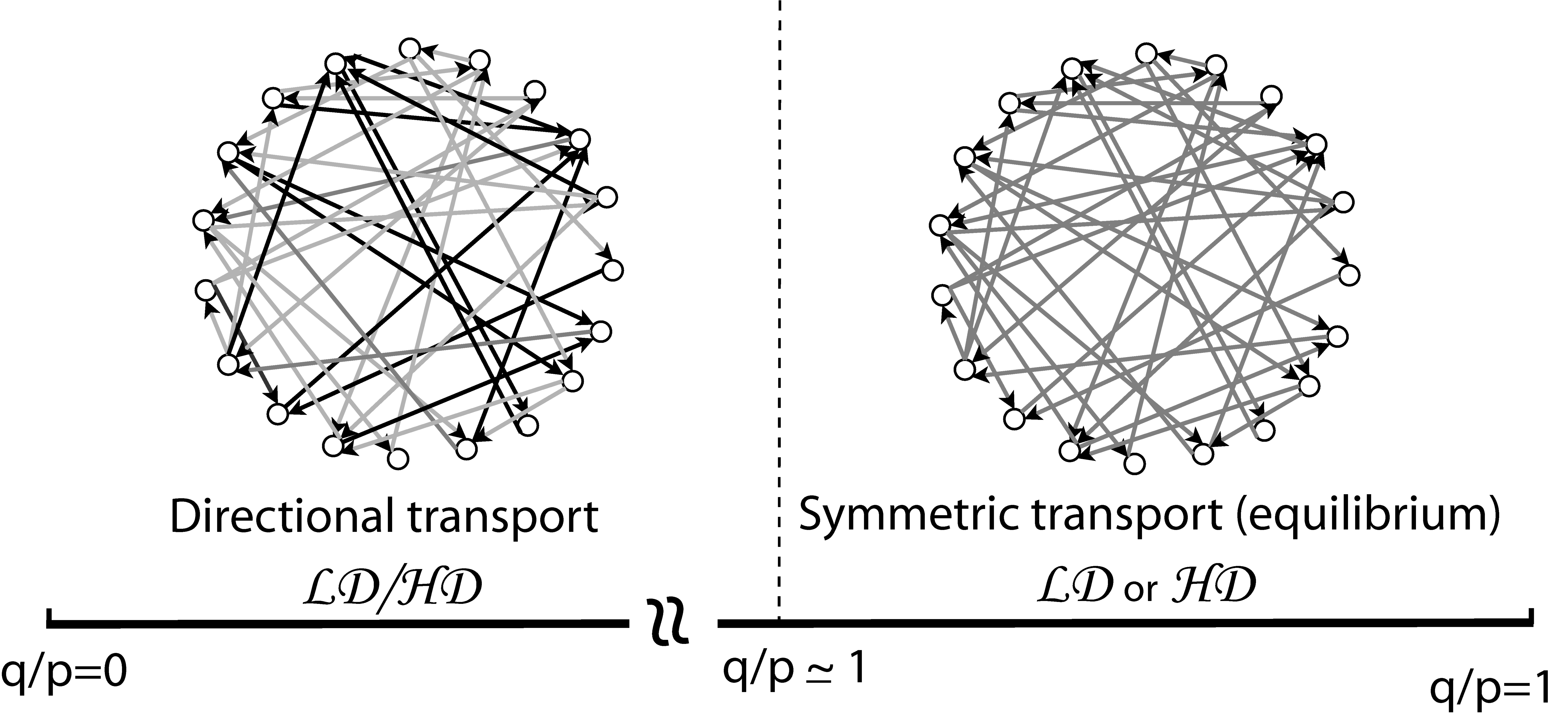}
\end{center}
\caption{Visualization of the stationary state of PASEP through irregular
networks as a function of the fraction between the hopping rates $p/q$.  When
$p/q=0$ we find a stationary state in the $\mathcal{LD/HD}$ regime corresponding
to
TASEP through irregular networks, while for $p/q=1$ we recover the homogeneous
equilibrium distribution.   The crossover from a heterogeneous to a
homogeneous distribution of particles happens at $p/q\approx 1$, and at $p/q=1$ for $L\rightarrow \infty$.
}\label{fig:visPASEP}
\end{figure}

Another interesting point is that the network topology affects how the stationary state reaches the passive equilibrium case of $q\approx p$.  
In regular networks density heterogeneities disappear gradually: the 
heterogeneous $\mathcal{LD-HD}$ segment regime
reduces gradually in size to eventually disappear at $q=p$, see figure
\ref{reg:ASEP}-(a). In the symmetric limit particles
therefore spread completely homogeneously for all particle densities. 
In irregular networks
the phenomenology is very different: density  
heterogeneities do not disappear gradually when reaching the symmetric limit,
see figure \ref{fig:PASEPirregPhaseDiagram}. 
The heterogeneous $\mathcal{LD/HD}$ network regime remains present for all
values  $q<p$ when $L$ is large enough, leading to strong density
heterogeneities on the network. This situation is visualized in figure
\ref{fig:visPASEP}. 

The simulation results also show that, in irregular networks, finite size effects play a role close to the symmetric case ($p \approx q$).
To identify the crossover from an active  to a passive process
on irregular networks we have studied the distribution of
particle densities as a function of the  segment length
$L$ at values $q\approx p$.  Then
particles are seen to be distributed homogeneously along the
network for short segments, whereas they are distributed
heterogeneously over the network for longer segments.  This can be
quantified with the distribution of segment densities (figure
\ref{fig:PASEPirreg}): this distribution is
unimodal for small $L$ and becomes trimodal at large $L$.    
Interestingly this implies that for a close to symmetric process ($q \approx p$) on
irregular networks there is a crossover from a passive to an active process in terms of the segment length $L$. 

\section{Particles with finite processivity}
\label{sec:TASEP-LK}

In the previous sections we have considered transport through closed networks. 
However, cytoskeletal transport poses an additional challenge, since motors only have a finite processivity: they can stochastically attach and detach at any point in the network, thereby alternating stretches of directed motion on the network with diffusive motion in the cytoplasm, see figure \ref{fig:biotransport}. 

\begin{figure}
  \includegraphics[ width =\textwidth]{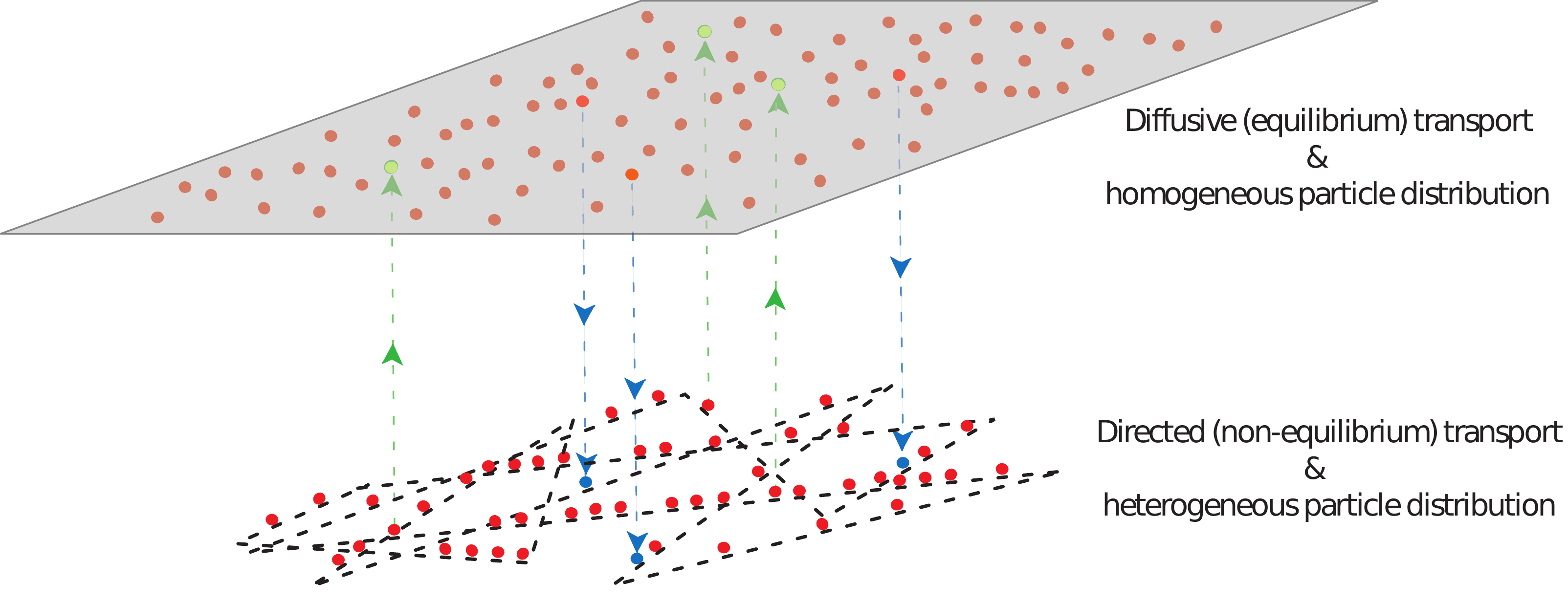}
  \caption{
    Illustration of the totally asymmetric exclusion process with Langmuir
    kinetics (TASEP-LK). Particles moving actively along a  network 
    are exchanged with a bulk reservoir, in which they diffuse.
    This leads to a competition between an active non-equilibrium transport
     process on one hand, which entails a heterogeneous distribution of
particles along the
    network, and a passive diffusion transport process, 
    which aims at a homogeneous equilibrium distribution of
    particles.
    Three parameters are relevant for this interplay:
    the topology of the network, the total particle density
    $\rho$ (equivalent to the parameter $K$), and the relative exchange rate $\Omega = \omega L/p$ (the latter 
    taking in account  the total exchange rate between reservoir and
    network $\omega$, the segment length $L$
    and the particle hopping rate along the network $p$). }
  \label{fig:illustrationLK}
\end{figure}

Here we model this behaviour based on the  TASEP-LK, which we generalize to transport along a
network \cite{NeriTT}. Particle attachment and detachment is governed by the rates $\omega_A$ and $\omega_D$, respectively. The particle reservoir is considered to be infinitely large, and diffusion is assumed to guarantee a uniform distribution in the bulk.   
In this process the total particle density along the network is seen to be set
directly by  the Langmuir exchange with the reservoir, through the ratio between
the  attachment and detachment rates. In contrast, the way in which particles
are distributed along the network requires understanding of the intricate
interplay between the effect of infinite diffusion (in the reservoir) and active
transport (along the network). This model is summarized graphically in Fig.
\ref{fig:illustrationLK}.

Indeed, the passive diffusion process in the reservoir tends to spread particles
out homogeneously, and therefore also has a homogenizing effect on the network.
Active transport on the other hand has been seen above to provoke density
heterogeneities (see sections \ref{sec:TASEP} and \ref{sec:ASEP}). Since the
bulk diffusion in the reservoir is assumed to be infinitely fast, the active
dynamics may be expected intuitively to impose heterogeneities along the network
if their motion is sufficiently processive.  We will now analyze in detail this interplay
between passive diffusion and active transport and how it regulates the
distribution of particles on the network.  These complement the results
presented briefly in \cite{NeriTT}.

We first revisit the macroscopic behaviour of TASEP-LK on a
single open segment \cite{Par03,
parmeggiani2}.  
In a second subsection we define TASEP-LK on a network, determine the
corresponding  mean-field equations and present the effective rate diagrams. 
We then present a way to establish analytical solutions to the mean-field equations
if the particle exchange with the reservoir is sufficiently strong, due to 
a decoupling of the continuity equations.  In the subsequent sections we
elaborate on the stationary state of regular and irregular networks, including
the infinitely connected limit. We end by discussing the results of  this
section and put them into the context of biophysical experiments.

\subsection{TASEP-LK on a single open segment}\label{sec:TASEPLKSegment}
TASEP-LK is similar to TASEP in that particles hop
uni-directionally along a one-dimensional segment at  rate $p$ and
are subject to exclusion interactions.   
In addition, particles
attach and detach  along the segment according to a Langmuir process
\cite{Flower}. In the single segment model, we are thus dealing with
three reservoirs: the two reservoirs at the entrance and the exit of the segment
(rates $\alpha$ and $\beta$) are now complemented by a bulk reservoir, which
allows for binding/unbinding of particles on any site along the segment with
rates $\omega_A$ and $\omega_D$, respectively.
We have illustrated this process in figure \ref{fig:models}-(d).

New behaviour emerges in TASEP-LK if there is a competition between the
directed transport and the Langmuir kinetics (LK), which is the case  in the so-called 'scaling' regime \cite{parmeggiani2},
$\omega_A = p\:\Omega_A/L$, $\omega_D = p\:\Omega_D/L$ 
(note that $\Omega_A$ and $\Omega_D$ are dimensionless). 
In this scaling regime the dynamics of the bulk is in competition with the
dynamics at the boundaries, which leads to an interesting 
$(\alpha, \beta)$-phase diagram  \cite{Par03, parmeggiani2}, represented in the appendix, figure
\ref{fig:phaseDiagramLK}.  
Indeed, the resulting density profiles in the segment interpolate between 
those of TASEP and the homogeneous profiles of a Langmuir process: on one hand the TASEP density profiles are recovered for 
small attachment/detachment rates ($\Omega_A,\Omega_D\rightarrow 0$), and on the
other hand the homogeneous Langmuir profile with an 
equilibrium density 
\begin{equation}
  \rho_l = \frac{\Omega_A}{\Omega_D+\Omega_A}
  \label{eq:LangmuirDensity}
\end{equation}
is observed for large attachment/detachment rates
($\Omega_A,\Omega_D\rightarrow \infty$). 
On a single segment these current and density profiles have been determined  
from mean-field arguments and are well corroborated by numerical simulations \cite{Par03, parmeggiani2}.    
They are conveniently expressed as a function of the rescaled position
variable along the segment, $x=i/L\in [0,1]$, with $i=1..L$.  
Since TASEP-LK is an exclusion process we
have a parabolic expression in the current-density relationship: 
\begin{equation}
  J^{\rm LK}\left[x;\alpha, \beta, \Omega_A, \Omega_D\right] = p\: \rho^{\rm
LK}\left[x;\alpha, \beta, \Omega_A, \Omega_D\right]\left(1-\rho^{\rm
LK}\left[x;\alpha, \beta, \Omega_A, \Omega_D\right]\right). \label{eq:JLK}
\end{equation}
The full expressions for $J^{\rm LK}$ and $\rho^{\rm LK}$ are discussed in
\ref{app:LK}. 
Note also that, due to particle binding/unbinding, the current is no longer constant along the segment.
Consequently we must account for the fact that the current entering a segment will generally differ from the
current leaving the same segment.   

We will require the phase diagrams of TASEP-LK on a single
segment for our analysis of the  stationary state 
on a network, using effective rate diagrams. 
We therefore discuss them briefly here; a  more elaborate discussion is
presented  in  \ref{app:LK} and in figure A1.  
It is useful to introduce the parameters $K = \Omega_A/\Omega_D$ and
$\Omega = (\Omega_A+\Omega_D)/2$. The quantity $\Omega$ characterizes the overall
exchange between the bulk reservoir and the segment, whereas the ratio in $K$
directly determines the Langmuir density imposed by the exchange with the bulk
reservoir (equation (\ref{eq:LangmuirDensity})). The phase
diagrams for TASEP-LK display several phases: 'pure' LD, HD, and MC
(also referred to as the M phase in some previous works
\cite{parmeggiani2}), as well as the combined phases LD-HD, LD-MC and MC-HD. Recall also that LD, HD and MC phases are in fact the appropriate 
generalizations of the corresponding phases in TASEP, with the difference that their density profiles are no longer constant throughout the segment.  
The density variation along the segment is continuous in all pure phases, but a discontinuous shock arises for the combined phases.
For example, in the LD-HD phase the shock corresponds to a domain wall at some position $x_w$. 
When $\Omega \rightarrow 0$ this LD-HD phase reduces to the $\alpha=\beta<p/2$
coexistence line and we recover the TASEP phase diagram as expected.  
For $\Omega \rightarrow \infty$ the LD-HD phase becomes more prominent and is
eventually present for all $\alpha/p<(K+1)^{-1}$ with $\beta/p<1/2$ when $K>1$. 
At high $\Omega$ the LD phase then disappears altogether from the 
phase diagram.  More precisely this happens 
at the threshold $\Omega_c$ given by equation (\ref{eq:critUpper}) in Appendix A.  
Similarly, the HD phase
disappears from the phase diagram for $K<1$ at a value $\Omega_c$.  
  
\begin{figure}[t]
  \includegraphics[ width =\textwidth]{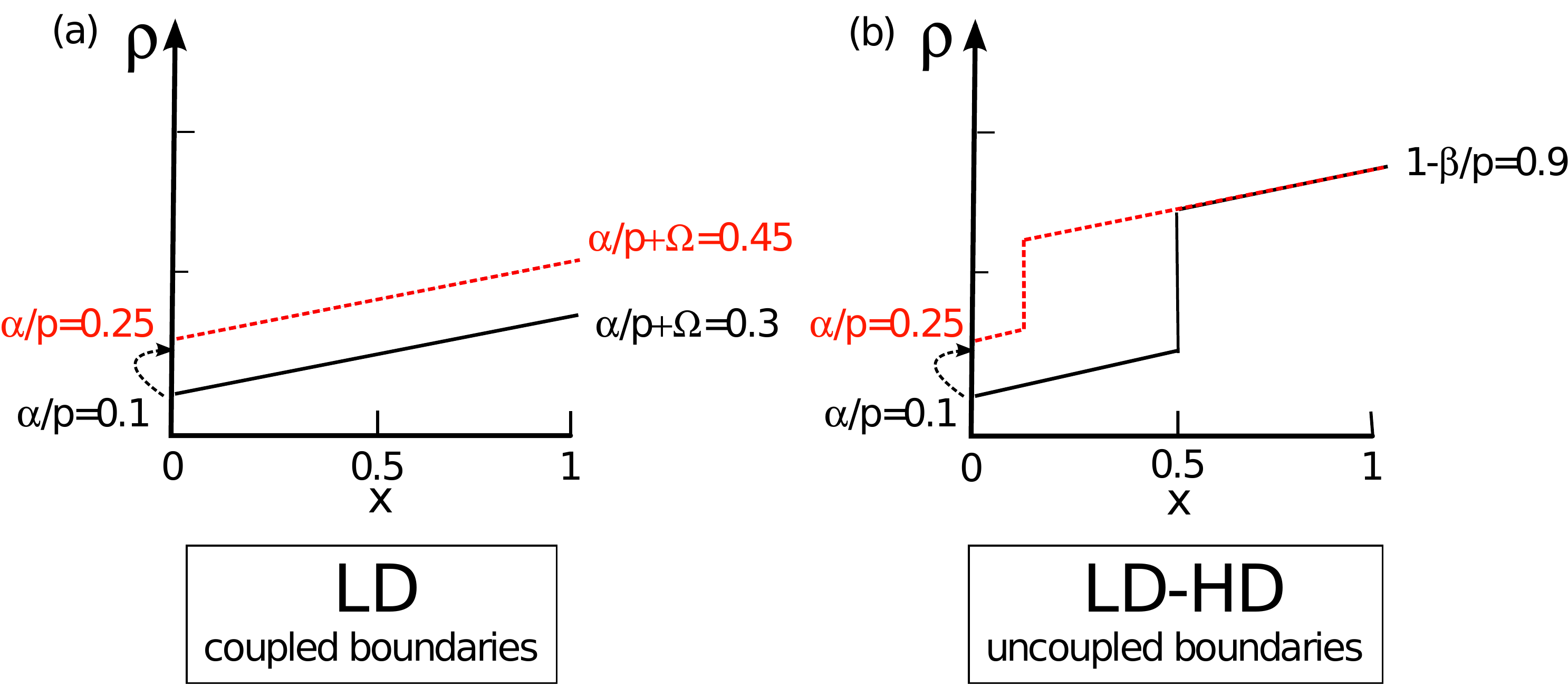}
  \begin{center}
    \caption{ 
      Density profiles $\rho_s(x)$ as a function of the rescaled position 
      $x \in [0,1]$ in the segment are presented for TASEP-LK.  A boundary 
      dependent phase (left) is compared with a boundary independent phase 
      (right).   Left (a): In the LD phase the density at the exit $x=1$ 
      changes when the entrance rate is modified: both
      boundaries are coupled.  
      Right (b): In the LD-HD coexistence phase, the density at the exit
      $x=1$ does not change with the entrance rate: hence both boundaries are
      decoupled.
    }\label{fig:profile}
  \end{center}
\end{figure}

An important notion in the study of TASEP-LK on networks is the distinction between those phases which couple boundaries and those which uncouple them.
This concept is illustrated in figure \ref{fig:profile}. 
In the LD phase (or the HD phase), changing the entrance rate $\alpha$ (or exit
rate $\beta$) modifies the bulk density and current 
throughout the whole segment, right through to the end of the segment. In this
sense the boundaries can be considered to be {\it coupled}, as is illustrated on
the left of figure \ref{fig:profile}. In the LD-HD phase on the other hand,
changing the rate $\alpha$ at the left boundary only modifies the current and
density over a finite portion of the segment, but it does not affect the density at the right boundary  
of the segment.  In that sense the boundaries are {\it uncoupled}, see the right
of figure \ref{fig:profile}.   Note that the discontinuity
in the density profile at
$x=x_w$  uncouples the LD region at the entrance from the HD region at the
exit of the segment. 
Hence boundaries are uncoupled in the LD-HD phase due to the presence of a shock. In the phases involving MC the boundaries are also decoupled.

\subsection{Effective rate diagrams describing TASEP-LK on networks}

As a first observation we point out that for TASEP-LK, although density heterogeneities arise throughout the network, the {\it overall} density $\rho$ is directly set to the Langmuir density ($\rho = \rho_\ell = K/(K+1)$). We will show that it is the exchange parameter $\Omega$ which regulates the distribution of particles on the network.

\begin{figure}
  \includegraphics[ width =\textwidth]{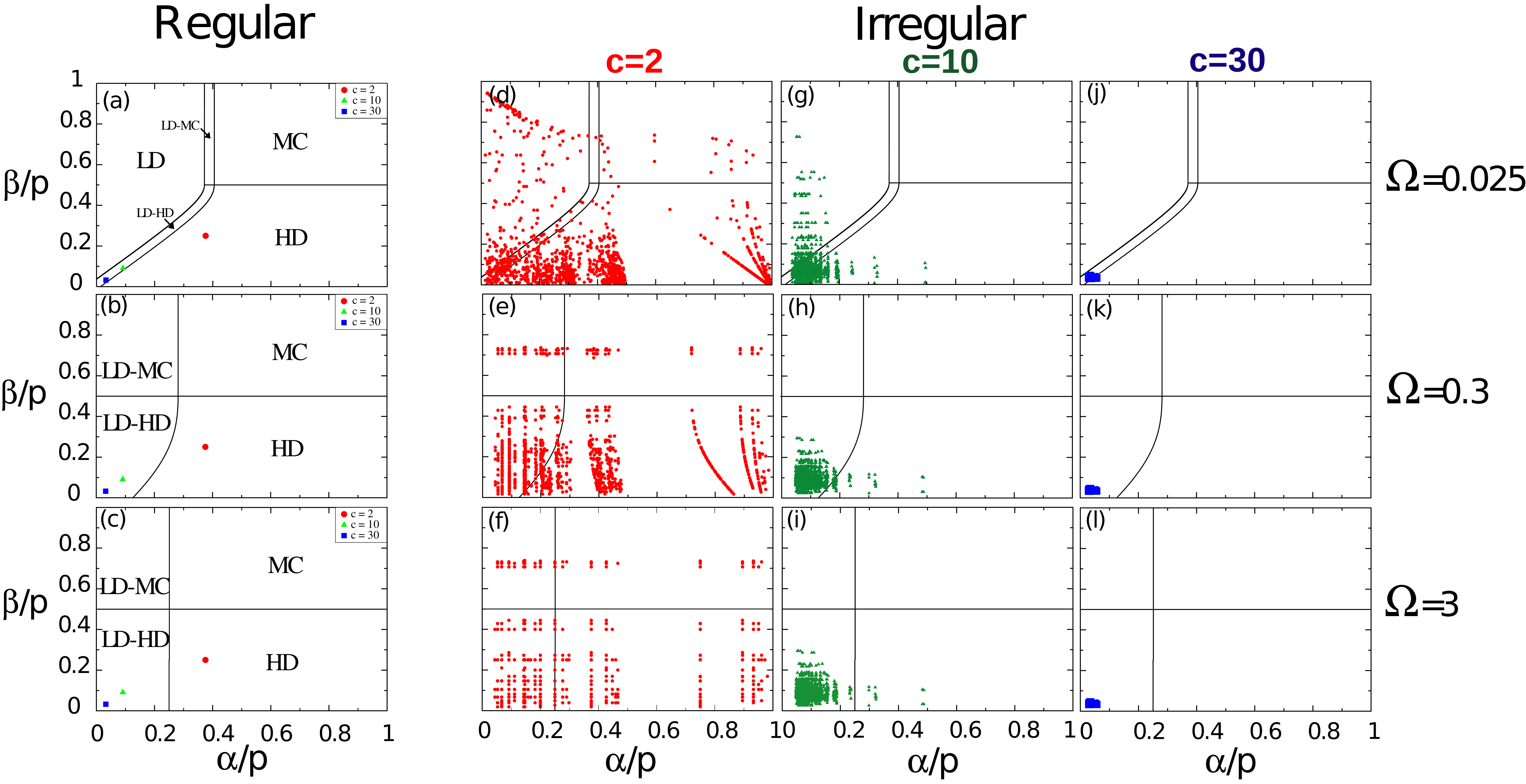}
  \caption{Effective rate diagrams for TASEP-LK through 
    regular graphs (left, (a)-(c)) and irregular graphs (right, (d)-(l)), both
filled to a 
    total density of $\rho=3/4$, are presented for the given values of 
    $\Omega$ and $c$.  We have used the same graph instances as in figure
\ref{fig:effectiveRatesTASEP}.  On regular graphs the effective rates are equal for all
segments, such that only one marker is plotted. 
  }
  \label{fig:phaseDiagram}
\end{figure}

The TASEP-LK model  on  a network can be studied using  the
general mean-field method presented in subsection \ref{sec:modelCyto}.   The
currents 
$J^-\left[\alpha_s, \beta_s\right]$ 
and 
$J^+\left[\alpha_s, \beta_s\right]$ 
entering and leaving a segment in equation (\ref{eq:RhoL})
follow,  respectively,  from the mean-field current profiles 
$J^{\rm LK}\left[x=0; \alpha_s,\beta_s,\Omega_A, \Omega_D\right]$ 
and 
$J^{\rm LK}\left[x=1; \alpha_s,\beta_s,\Omega_A, \Omega_D\right]$ 
along a single segment (see equation 
(\ref{eq:JLK}) and the expressions for the density  in the  \ref{app:LK}). 
Since particles can
attach/detach along the segment we have now $J^+\neq J^-$ in the continuity
equation (\ref{eq:RhoL}).  We establish the effective rates as
$\alpha^{\rm eff}_{(v, v')} = p\: \rho_v/c^{\rm out}_v$ 
and 
$\beta^{\rm eff}=p\:(1-\rho_v)$, corresponding to a uniform microscopic 
hopping rate at the junctions. Note that these are the same effective 
rates as presented for TASEP in equations (\ref{eq:rates}): this reflects 
the fact that attachment/detachment at the junction sites themselves is 
negligible in the scaling regime.  The resulting mean-field equations 
(\ref{eq:RhoL}) for the average junction densities of TASEP-LK are then 
given by
\begin{eqnarray}
  \fl \frac{\partial}{\partial t}\rho_v &=&
  \sum_{v'\rightarrow v}\:J^{\rm
    LK}\left[x{=}1;\frac{\rho_{v'}}{c^{\rm out}_{v'}}, 1{-}\rho_v,
    \Omega_A, \Omega_D\right] - \sum_{v'\leftarrow v} J^{\rm LK}
  \left[x{=}0;\frac{\rho_{v}}{c^{\rm
        out}_{v}}, 1{-}\rho_{v'},
    \Omega_A, \Omega_D\right]. \label{eq:rhoV}
\end{eqnarray}
We have solved the mean field equations (\ref{eq:rhoV}) and
mapped the effective rates of each of the individual segments 
of the network onto the corresponding
$(\alpha, \beta)$-phase diagrams of a single segment.   These results, 
presented in figure \ref{fig:phaseDiagram}, characterize the stationary state 
of TASEP-LK through networks. 

Several interesting features  emerge from these diagrams.  
The stationary state of TASEP-LK shares certain characteristics with TASEP and PASEP.
 For regular graphs all effective rates have the same
value and coincide in one point in the effective rate diagram.  
Irregular graphs  on the other hand lead to a scattered plot of all 
effective rates, and they cluster around the origin at high connectivities (as can be understood in terms of bottleneck formation).   However, the LD-HD coexistence increasingly widens with increasing exchange parameter $\Omega$, which has a direct impact onto the phenomenology of density heterogeneities. 
Moreover, we now see that the scatter plots 
which appear random at low exchange parameters $\Omega$ reveal a certain
regularity at high values of $\Omega$.  In this range effective rates
become  independent of $\Omega$, as well as being independent of global
topological 
features of the network (i.e.~they only depend on the local junction degrees).
This feature can be explained using the notion of
coupled and uncoupled boundaries introduced at the end of the previous
subsection.

\subsection{Uncoupling of boundaries at high $\Omega$: simplified mean-field
equations}
At weak exchange (small $\Omega$), close to the TASEP limit ($\Omega=0$), the
continuity equations (\ref{eq:rhoV}) in the junction densities are intrinsically
coupled.  Indeed, a large fraction of the segments will be in the LD or HD
phase and couple their boundaries. 
Finding a solution then typically requires a numerical procedure.  However, when
increasing $\Omega$ more and more segments switch into a LD-HD phase (see figure \ref{fig:phaseDiagram}), reducing the coupling.  For sufficiently high $\Omega$ the continuity equations (\ref{eq:rhoV}) uncouple, providing a 
route to  an exact solution to the mean-field equations.  In particular, we 
can present an exact solution of the continuity equations for any graph, based on  the 
single segment phase diagram for $\Omega\rightarrow \infty$. It will turn out that this solution
remains accurate down to rather small values of $\Omega$, which we rationalize in terms of the effective rate diagrams.

Before considering the solution in the $\Omega\rightarrow \infty$ limit, which
is valid for any graph topology, let us first  consider the simpler case
for which all segments in the network are in the
LD-HD phase. From the effective rate diagrams figure \ref{fig:phaseDiagram} we
see that this is the case for regular graphs and for irregular graphs at high
connectivities.  When all segments are in the LD-HD phase, modifying  the
density at a certain junction will not change the density of particles 
at other junctions in the network: the domain
walls in each of the segments block the
propagation of density perturbations from the adjacent junction. The
continuity equations (\ref{eq:rhoV}) then simplify into
\begin{eqnarray}
  \frac{\partial}{\partial t}\rho_v &=&  c^{\rm in}_v
  J^{-}_v -  c^{\rm out}_v J^+_v\\
  &=& 
  c^{\rm in}_{v} \rho_v(1-\rho_v)-
  \rho_v\left(1-\frac{\rho_v}{c^{\rm out}_v}\right) 
  .
  \label{eq:LKNetwork}
\end{eqnarray}
We see that the currents at the junctions $J^{\pm}_v$ only depend on the local
junction density $\rho_v$, such that the continuity equations (\ref{eq:rhoV}) 
are completely decoupled. We obtain the solution 
\begin{eqnarray}
  \rho_v &=& \frac{c^{\rm in}_v-1}{c^{\rm in}_v-\left(c^{\rm out}_v\right)^{-1}}
  .  \label{eq:simplSol} 
\end{eqnarray}
Equations (\ref{eq:simplSol}) are valid for networks at high
enough connectivities and exchange rates $\Omega$, 
and this is due to two combined effects:
increasing the connectivity effective rates cluster around the origin, whereas
increasing $\Omega$ enlarges the LD-HD region in the phase diagram (see figure 
\ref{fig:phaseDiagram}). 

We consider now the limiting case $\Omega \rightarrow \infty$, for which we can
solve the mean-field equations (\ref{eq:rhoV}) analytically using uncoupling of boundaries.   
Depending on the overall particle density $\rho =K/(K+1)$ we find the solutions:
\begin{itemize}

\item  $\rho>1/2$:
  \begin{eqnarray}
    \fl  \rho_v &=& 
    \left\{ \begin{array}{cccc}  
      c^{\rm out}_v\left(\frac{c^{\rm in}_v-1}{c^{\rm out}_vc^{\rm in}_v-1}\right) 
      && 
      & \frac{\rho_v}{c^{\rm out}_v}\leq 1-\rho\  {\rm and} \ c^{\rm in}_v\neq 1\\ 
      \frac{c^{\rm out}_v}{2} \left(1 - \sqrt{1-\left(c^{\rm out}_v\right)^{-1}}\right) &{\rm for}& &\frac{\rho_v}{c^{\rm out}_v}\leq 1-\rho \ {\rm and} \ c^{\rm in}_v=1 \\ 
      \frac{1}{2}
      \left( 1+\sqrt{1-4 \: \frac{c^{\rm out}_v}{c^{\rm in}_v} \rho(1-\rho)}\right)
      && 
      &\frac{\rho_v}{c^{\rm out}_v}\geq 1-\rho  \ {\rm and} \ \rho_v\geq 1/2
    \end{array}
    \right.\nonumber 
  \end{eqnarray}
\item  $\rho=1/2$: 
  \begin{eqnarray}
    \fl \rho_v&=&\left\{\begin{array}{cccc}c^{\rm out}_v\left(\frac{c^{\rm
        in}_v-1}{c^{\rm out}_vc^{\rm
        in}_v-1}\right)&& &c^{\rm in}_v\neq1, \: c^{\rm
      out}_v\neq 1 \\  \frac{c^{\rm out}_v}{2}\left(1-\sqrt{1-\left(c^{\rm
        out}_v\right)^{-1}}\right)&&{\rm for} &c^{\rm in}_v=1 \\ 
    \frac{1}{2}\left(1+\sqrt{1-\left(c^{\rm in}_v\right)^{-1}}\right)&& &c^{\rm
      out}_v=1 \\
    \end{array}\right. \label{eq:simple}
  \end{eqnarray}
\item  $\rho<1/2$:
  \begin{eqnarray}
    \fl  \rho_v &=& \left\{ \begin{array}{cccc}  c^{\rm out}_v\left(\frac{c^{\rm
          in}_v-1}{c^{\rm out}_vc^{\rm
          in}_v-1}\right)&& &  \rho_v\geq 1-\rho \ {\rm and} \ c^{\rm out}_v\neq 1\\ 
      \frac{1}{2}\left(1+\sqrt{1-\left(c^{\rm in}_v\right)^{-1}}\right)
      &{\rm
         for}& & \rho_v\geq 1-\rho \  {\rm and} \ c^{\rm out}_v= 1\\ 
      \frac{c^{\rm out}_v}{2}\left(1-\sqrt{1-4 \: \frac{c^{\rm in}_v}{c^{\rm
            out}_v}
        \rho(1-\rho)}\right)&& &\rho_v<1-\rho 
    \end{array}
    \right
.\nonumber 
  \end{eqnarray}
\end{itemize}
These solutions in the junction densities $\rho_v$ are valid for 
$c^{\rm in}_{v}\neq 0$ and $c^{\rm out}_v\neq 0$.  When $c^{\rm in}_v=0$ the density
will be trivially equal to $\rho_v=0$, whereas for $c^{\rm out}_v=0$ the junction
density takes the value $\rho_v=1$.  

The way to derive the above equation goes as follow: we consider the phase
diagram of TASEP-LK in the limit $\Omega\rightarrow \infty$, see bottom of
figure
\ref{fig:phaseDiagramLK}.  In this limit one notices that boundaries decouple,
i.e.~$\alpha^{\rm eff}_{v\rightarrow v'}$ is only a function of
$\rho_v/c^{\rm out}_v$ (and not of $\rho_{v'}$) 
and $\beta^{\rm eff}_{v\rightarrow v'}$ is only a function of $\rho_v'$ 
(and not of $\rho_v$).
In the end, after considering the different phases at $\Omega\rightarrow \infty$,
decoupling allows us to solve  exactly 
the full mean-field equations (\ref{eq:rhoV}) leading to equations
(\ref{eq:simple}).  Note that, when combined with the expressions for the 
segment and density profiles $J^{\rm LK}$ and $\rho^{\rm LK}$ in \ref{app:LK}, equations (\ref{eq:simple}) give us analytical expressions for the
stationary density and segment profiles of all the segments in the network.  

The simplified mean-field equations (\ref{eq:simple}) provide an explanation as to why the
effective rate diagrams for irregular graphs (figure \ref{fig:phaseDiagram}),
which contain randomly scattered effective rates at low values of $\Omega$,
become more ordered at high values of $\Omega$. 
Indeed, the decoupling leads to junction densities which only depend on the
local degrees $c^{\rm in}_v$ and $c^{\rm out}_v$, as well as the stationary 
density $\rho$. The effective rate plots at strong exchange $\Omega$ are thus 
independent of $\Omega$ and of the global random topology of the network.  

\begin{figure}
  \begin{center}
    \includegraphics[ width =1\textwidth]{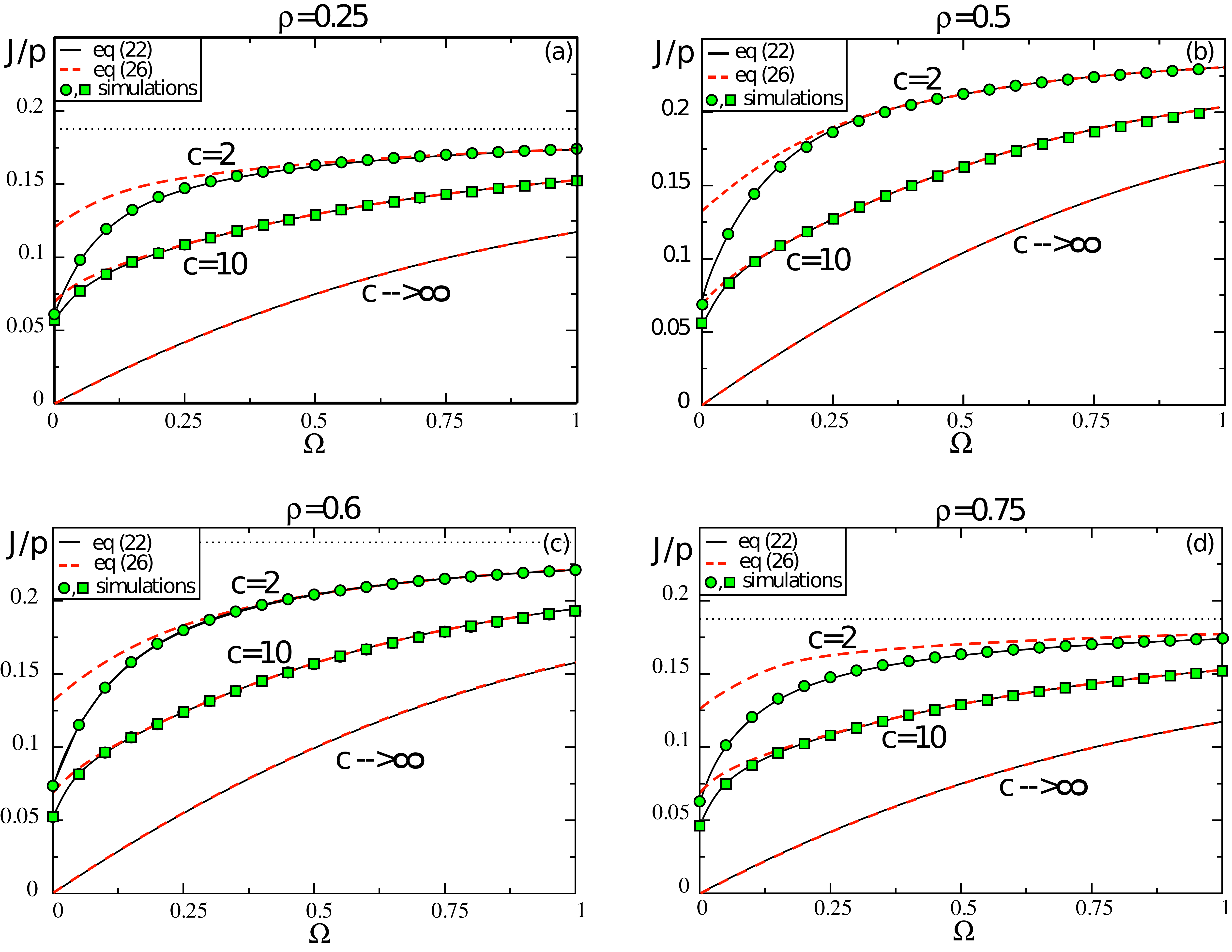}
  \end{center}
  \caption{The average current $J$  for TASEP-LK through 
    irregular networks, as a
    function of the exchange rate $\Omega$ for given particle filling $\rho$ and
    mean connectivity $c$.   Full mean-field results of equations 
    (\ref{eq:rhoV}) (solid lines) are compared with simulation data (markers) 
    and the solution to the simplified mean
    field equations (\ref{eq:simple}) (rad dashed lines).  Mean field results
are for single graph instances of mean connectivity $c=2$ and $|V| = 607$
junctions or $c = 10$ and $|V| = 200$.  Simulations are run on the same graphs
with segments of length $L=400$. Further represented are: the upper bound to the
average current given by the Langmuir expression $\rho(1-\rho)$ (dotted line)
and its lower bound, given by graphs with mean connectivity $c\rightarrow
\infty$. }\label{fig:currentprofiles}
\end{figure}

Equations (\ref{eq:simple})  can also be seen as
an approximation to the full mean-field equations (\ref{eq:rhoV}) at finite
$\Omega$.  In figure \ref{fig:currentprofiles} we compare the current profiles
of the full equations (\ref{eq:rhoV}), both with the simplified version
equation (\ref{eq:simple}) and with simulations.   Results are presented
for irregular graphs of given mean connectivity $c$ and with a given stationary
density $\rho$.  Just as expected, simulation results are in very good
agreement with those obtained by numerically solving the full mean-field equations.  But remarkably
we find that the simplified mean-field result gives already a very accurate
approximation, and this is true down to rather weak exchange parameters 
($\Omega\gtrsim0.5$).   This can be understood from the shape of the single segment phase diagram which remains similar to the one for $\Omega\rightarrow\infty$.

Moreover, the simplified equations (\ref{eq:simple}) are exact in special circumstances. This is the case at half-filling ($\rho=1/2$), for any
exchange parameter $\Omega\geq0.5$, since then the decoupling is complete (see \ref{app:C}). 
It also applies asymptotically at high connectivities $c$: as one 
can see from figures \ref{fig:phaseDiagram}, the effective rates indeed cluster around the origin in the LD-HD phase in this limit, which again leads to full decoupling. In contrast, the decoupled description by equations (\ref{eq:simple})
becomes less accurate for densities  $\rho\rightarrow 0$ or $\rho\rightarrow 1$,
since then the zone corresponding to the decoupled LD-HD phase becomes very
small in the effective rate diagrams.

To summarize, in this subsection we have shown that the mean-field description
of TASEP-LK simplifies when increasing the exchange rate with the
reservoir $\Omega$.   We have derived analytical expressions for the stationary
density and current profiles throughout the network which are very accurate at
sufficiently high exchange parameters $\Omega$ and mean connectivities $c$,  
as well as for stationary densities $\rho$ close to half filling.

\subsection{Networks of infinite connectivity} \label{sec:tasep:infconnect}
We now discuss the stationary state of networks for which the vertex degree 
of each junction is very high.  For $\Omega>0$ all 
segments fall into the LD-HD phase since the effective rates cluster around the origin.   
Thus, the network  is in the $\mathcal{LD-HD}$ regime and the simplified solution  (\ref{eq:simple}) based on the decoupling of segments is exact (within the mean field 
framework).   In this limit the stationary current and 
density profiles can be computed analytically by setting $\rho_v=1$, see \ref{app:B}.
In particular, for half filling $\rho=1/2$ we present a simple expression in \ref{app:C}.
Knowing the exact expression for the current in the
infinitely connected limit is interesting for two reasons.

First, as one can see from figure \ref{fig:currentprofiles}, the current 
reaches  in this limit ($c\rightarrow \infty$) the minimal value accessible 
for the given values of $\Omega$ and $\rho$. This is in fact intuitive, since 
all junctions become fully blocked and suppress all flow through  the junctions.
But for TASEP-LK these junction bottlenecks do not block the dynamics in the segments completely, as long as there is a non-zero exchange rate $\Omega>0$: figure \ref{fig:currentprofiles} shows that, even for small values of $\Omega$, the  current does not reduce to zero when $c\rightarrow \infty$.  
This indicates that 
even a weak exchange with a reservoir is sufficient for particles to 
circumvent bottlenecks and maintain a significant flow through the network. 

A second reason which makes the strong connectivity limit special is that all
networks behave identically, as long as $\Omega>0$.  Hence, in this limit the
stationary state of all networks is the same and we recover a universal,
topology independent stationary state.   The reason for this universality is
twofold.  First, for $c\rightarrow \infty$ the LD-HD phase dominates, such
that perturbations remain local and do not propagate throughout the network due to the continuity equations (\ref{eq:rhoV}).  Second, all junctions
become infinite bottlenecks, such that $(\alpha,\beta)\rightarrow
\mathcal{O}(c^{-1},c^{-1})$. As a consequence the currents and densities in the segments are rather insensitive to local degree fluctuations.

\subsection{TASEP on LK through regular networks}
Regular networks constitute another solvable case for TASEP-LK. 
Just as for TASEP and PASEP, the stationary state of TASEP-LK on regular
networks is given by a unique junction occupancy $\rho_v$ for all junctions.  
We find the following solution 
to the mean field equations (\ref{eq:rhoV}):
\begin{eqnarray}
  \rho_v\left[\Omega, \rho\right] 
  &=& 
  \left\{ \begin{array}{lllll}
    c\rho   && \rho<1/(c+1)         && (\mathcal{LD}) \\
    c/(c+1) && 1/(c+1)<\rho<c/(c+1) && (\mathcal{LD-HD})\\ 
    \rho    && \rho>c/(c+1)         && (\mathcal{HD})
  \end{array}\right.  
  \label{eq:rhovT}
\end{eqnarray}
Remarkably, this solution is identical to the one for TASEP \cite{NeriT}.  This
can be seen as follows.
Recall that $c^{\rm out}_v = c^{\rm in}_v =c$ at all junctions $v$  for regular
graphs, such that all junctions (and therefore all segments) are equivalent.  
Current conservation at the junctions, equation (\ref{eq:rhoV}), then implies
that we must have $J_+=J_-$ for all segments.  This points to either a solution
for which the segments have  a homogeneous density (and are thus in the LD or HD
phase) or to a solution for which the segments have a heterogeneous density with
$\alpha^{\rm eff}=\beta^{\rm eff}$ (and are in the LD-HD coexistence phase).  
The presence of a homogeneous density profile in the segments may appear
counter-intuitive, since the exchange process with the bulk typically leads to
non-homogeneous density profiles (unless $\Omega \rightarrow \infty)$. But this
proves to be correct from the solution for a single segment: equation
(\ref{eq:rightBoundarySol}), in \ref{app:LK}, shows that a homogeneous solution
 is possible, provided $\beta^{\rm eff} = (1 - \rho_\ell) = (1-\rho)$ at HD and
$\alpha^{\rm eff} = \rho_\ell = \rho$ at LD.  Using that $(\alpha^{\rm eff}
,\beta^{\rm eff} )=(\rho_{v}/c,1-\rho_v)$ leads immediatly to the solution
equation (\ref{eq:rhovT}).  In essence, these arguments show that the
homogeneous density profile
(\ref{eq:rhovT}) is the correct solution for a regular network, even for a
finite exchange parameter  $\Omega$.   In the intermediate case, when the
segments are in the
coexistence phase,  the junction densities and effective rates follow
immediately from the simplified mean field equation (\ref{eq:simple}), and are given by $\rho_v = c/(c+1)$. We thus see that, just as for TASEP, we are dealing with a
LD phase $\alpha^{\rm eff}<\beta^{\rm eff}$,  a HD phase when $\beta^{\rm
eff}<\alpha^{\rm eff}$ and a coexistence phase at $\beta^{\rm
eff}=\alpha^{\rm eff}$.

\begin{figure}[h!]
    \begin{center}
      \includegraphics[ width=
        1\textwidth]{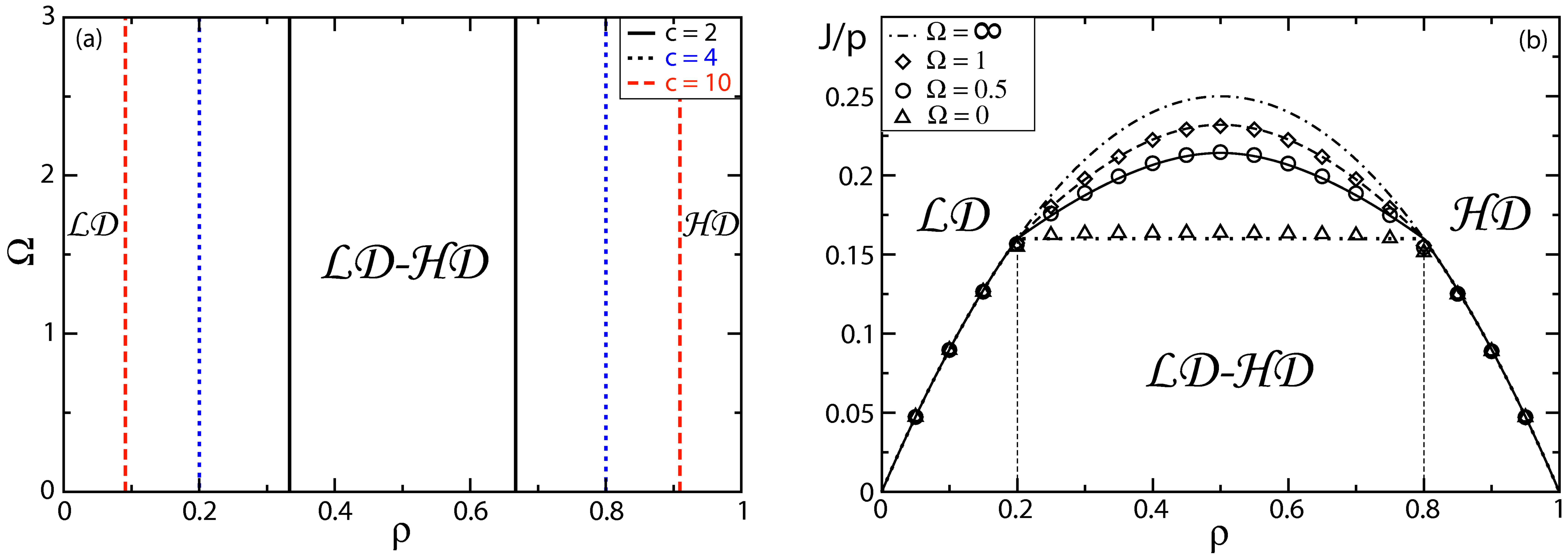}
    \end{center}
  \caption{
    TASEP-LK through regular graphs.  
    Left (a): The lines separating the $\mathcal{LD}$ and $\mathcal{LD-HD}$
    regimes, as well as the $\mathcal{LD-HD}$ and
    $\mathcal{HD}$ regimes are presented as a function of the total density
    $\rho$ and the exchange parameter $\Omega$ for regular graphs of given
degre $c$. 
    Right (b):   Average current-density profile for regular graphs of degree
    $c=4$.  Mean-field results (solid lines) are compared with simulation 
    data (markers) on graphs of  $|V|=80$ junctions and segments of 
    length $L=400$, for different values of the 
    exchange rates $\Omega$.
    For $\Omega\rightarrow \infty$ the transport process becomes
    passive.  We obtain a parabolic profile in this limit, indicating a 
    homogeneous density distribution at all scales. 
  }\label{reg:TASEP-LK}
\end{figure}

From the solution (\ref{eq:rhovT}) for the junction densities  we 
can deduce the phase diagram in figure \ref{reg:TASEP-LK}-(a) for TASEP-LK 
through regular graphs. Figure \ref{reg:TASEP-LK}-(a) shows that the same
behaviour arises as in TASEP, {\it i.e.} with increasing density we successively
observe a $\mathcal{LD}$ regime, a heterogeneous $\mathcal{LD-HD}$ segment
regime, and finally a $\mathcal{HD}$ regime. The density zone for the
$\mathcal{LD-HD}$ segment regime is identified as $1/(c+1)<\rho<c/(c+1)$, and
remarkably is altogether independent of the  exchange parameter $\Omega$. This
observation implies that the $\mathcal{LD-HD}$  does {\it not} disappear when
approaching the equilibrium process ($\Omega \rightarrow \infty$). Note that
this behaviour is 
very different from PASEP, where the symmetric limit ($p/q \rightarrow1$) makes
the segment regime disappear (compare figures \ref{reg:TASEP-LK}-(a) and
\ref{reg:ASEP}-(a)).

This equilibrium limit $\Omega \rightarrow \infty$ can be understood as follows. The inhomogeneous density profile with a domain wall, located at some position $x_w$, persists for $\Omega \rightarrow \infty$.   In fact, as $\Omega$ increases the position $x_w$  will gradually move to one of the segment ends ($x_w\rightarrow 0$ for $\rho>1/2$ or $x_w\rightarrow 1$ for $\rho<1/2$, half-filling $\rho=1/2$ is special, see \ref{app:C}). This constitutes a way of asymptotically restoring homogeneity within single segments, although the  $\mathcal{LD-HD}$ regime is maintained for the network. This analysis is further corraborated by the current-density relation $J(\rho)$, which indirectly provides a measure for the degree of heterogeneity within the segments: a parabolic profile $J=\rho(1-\rho)$ corresponds to a 
homogeneous density profile, whereas deviations from the parabola indicate a 
hinderance due to heterogeneities.
From figure \ref{reg:TASEP-LK}-(b) we see that 
for $\Omega \rightarrow \infty$ the current-density profile 
gradually reaches the parabolic Langmuir profile $J=\rho(1-\rho)$, thus 
showing that the homogeneous equilibrium profile is eventually attained. 

An analytical expression for the current density profile follows by using the 
formulas in \ref{app:LK} and the solution given by equations 
(\ref{eq:rhovT}). However, only in the case of half-filling is it simple to establish these expressions (see \ref{app:C}).

In summary, TASEP-LK through regular networks leads to the same regimes of
heterogeneities as TASEP through regular networks.  The size of the zone 
corresponding to the heterogeneous $\mathcal{LD-HD}$ regime is not affected by 
the coupling with the reservoir.  The equilibrium process is attained as the  
 LD-HD domain  walls in all segments shift towards the junction sites.  

\subsection{TASEP-LK through irregular networks}

We now determine the stationary state of TASEP-LK through irregular networks. 
It can be understood using effective rate diagrams 
(figures \ref{fig:phaseDiagram}) and the classification
in three different regimes for the particle distribution, as developed in 
section \ref{sec:TASEP} (see figure \ref{fig:networksVis}).

\begin{figure}
\includegraphics[width=1
\textwidth]{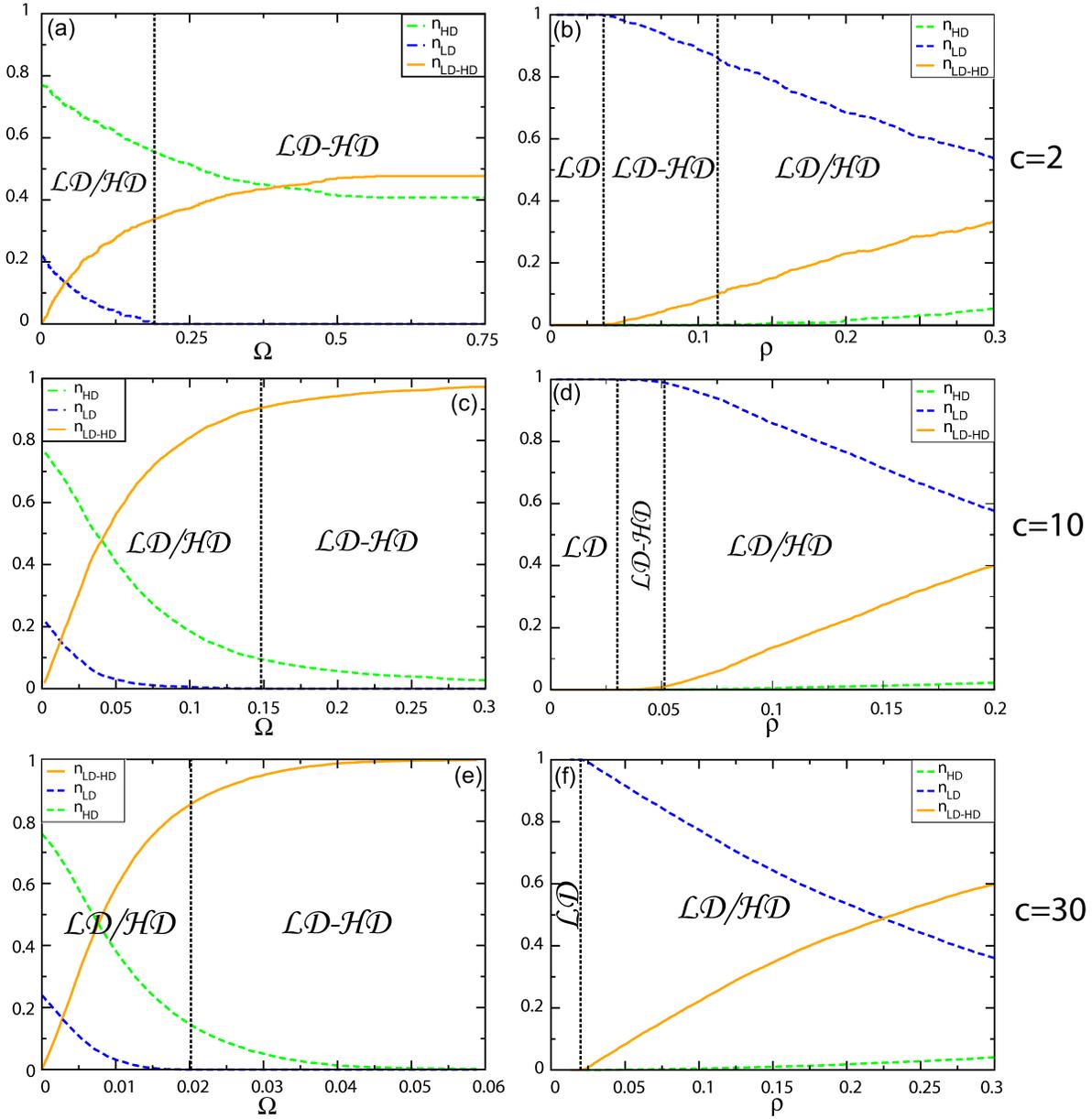}
\caption{
  The fraction of segments in the LD phase, the HD phase and the
  LD-HD phase as a function of $\Omega$ and $\rho$ 
  for TASEP-LK for single instances of irregular graphs at given
mean connectivity $c$ 
(the fractions for M, M-HD and LD-M are small and not represented).
  We have $\rho=0.75$, $c=2$ (a),  $\Omega=0.15$, $c=2$ (b), $\rho=0.75$,
$c=10$ (c),  $\Omega=0.05$, $c=10$ (d),  $\rho=0.75$,
$c=30$ (e),  $\Omega=0.01$, $c=30$ (f).   The graphs used have $|V| = 607$
junctions ($c=2$), $|V| =500$ ($c=10$) and $|V| = 500$ ($c=30$). 
  The transitions between the stationary regimes of density heterogeneities 
  are indicated: $\mathcal{LD}$ homogeneous, 
  $\mathcal{HD}$ homogeneous, $\mathcal{LD-HD}$ heterogeneous 
  and $\mathcal{LD/HD}$ heterogeneous.  
}\label{fig:irregFractions}
\end{figure}

\begin{figure}[h!]
\begin{center}
\includegraphics[width=1
\textwidth]{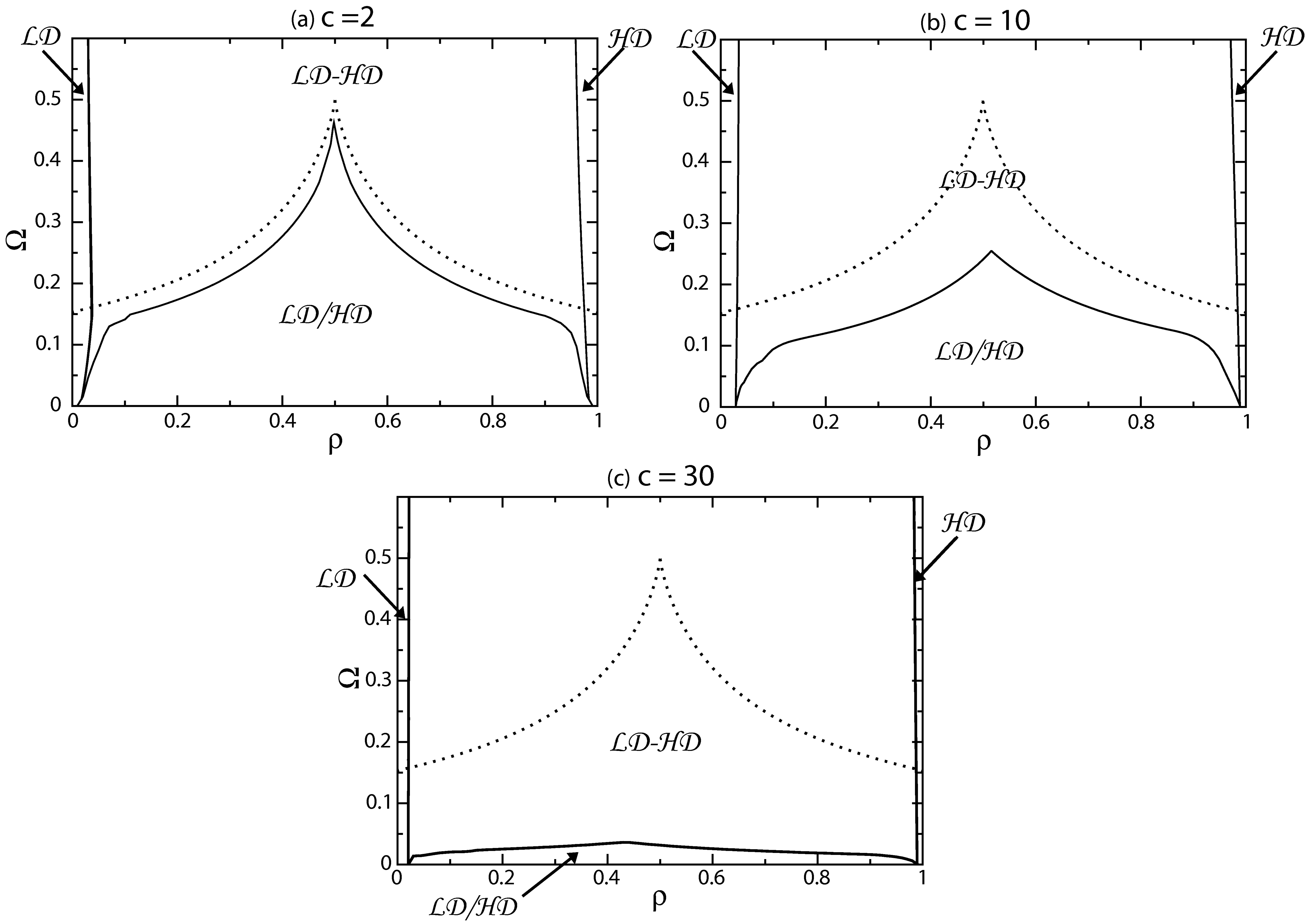}
\caption{
  The $(\Omega,\rho)$-diagram for TASEP-LK through irregular
  graphs.  The transitions (solid lines) between the different 
  regimes of the stationary state on the network are shown for the same graph
instances as in figure  \ref{fig:irregFractions}. 
The dotted line shows the upper bound to the transition
  between the  heterogeneous network $(\mathcal{LD/HD})$ and segment
  $(\mathcal{LD-HD})$ regimes (as given by equation \ref{eq:critUpperx}).
}\label{fig:irregPhaseDiagram}
\end{center}
\end{figure}

To identify the regime for the stationary state (i.e.~$\mathcal{LD}$, $\mathcal{HD}$, $\mathcal{LD-HD}$ and $\mathcal{LD/HD}$) we need to determine the fraction of
segments in the network which occupy the different phases  (i.e.~LD, HD, LD-HD)
in the effective rate diagrams, figure \ref{fig:phaseDiagram}.  
We denote these fractions by $n_{\rm LD}$, $n_{\rm HD}$, $n_{\rm LD-HD}$; the M, LD-M and M-HD phases play a minor role and will not be considered.    
In figure \ref{fig:irregFractions}  these fractions are shown as a 
function of $\Omega$ and $\rho$.  Using
the definitions of the different network regimes given in figure
\ref{fig:networksVis} we can establish the boundaries of the corresponding zones
in the $(\rho, \Omega)$ plane.  We have
indicated the resulting phase diagrams in figure \ref{fig:irregPhaseDiagram} for
irregular networks.

\begin{figure}[h!]
  \begin{center}
    \includegraphics[width=1
      \textwidth]{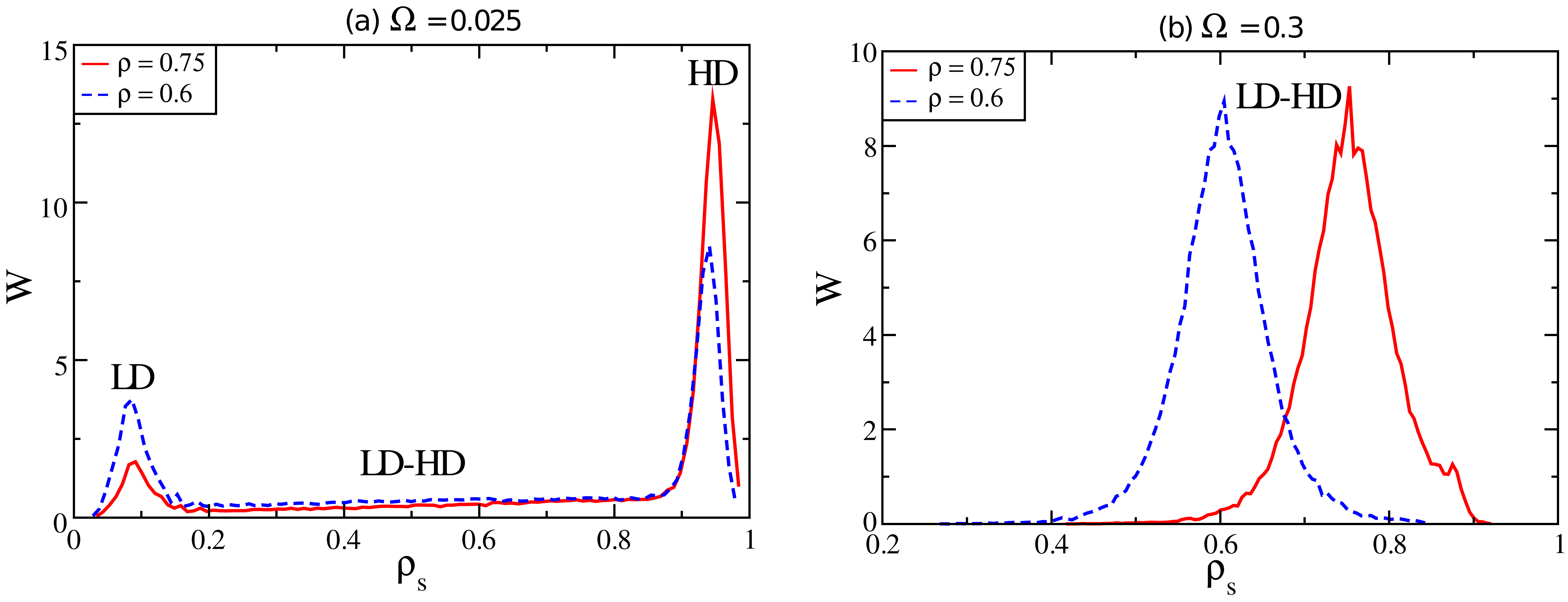}
    \caption{
      Distribution of segment densities $W(\rho_s)$ of TASEP-LK on an 
      irregular graph with mean connectivity $c=10$ at given total densities
$\rho$  and exchange rates $\Omega$.
      Mean-field results on a single graph instance of size $|V|\approx 10^4$ are 
      presented.                               
      (a): for low values of $\Omega$ the distribution is bimodal, with the two 
      peaks corresponding to segments in  the LD and the HD phases.       
      The broad intermediate ''band'' corresponds to segments in the 
      LD-HD phase. 
      The stationary state of the network is in the heterogeneous network
      regime $(\mathcal{LD/HD})$.  (b):at the value $\Omega=0.3$  segments 
      are in the LD-HD phase and the distribution is
      unimodal.  The  network is in the heterogeneous
      segment regime $(\mathcal{LD-HD})$.
    }\label{fig:distriLK}
  \end{center}
\end{figure}

We note several interesting characteristics of the stationary state
on irregular networks.  To do so let us focus on the case $\rho>1/2$
(similar arguments apply for $\rho<1/2$).  At low exchange parameters $\Omega$
the network is in the  $\mathcal{LD/HD}$  regime: a finite fraction of
segments occupy the LD and HD phases.  The strong heterogeneities in the 
particle densities are reflected in the bimodal distribution 
of segment densities, as shown in figure \ref{fig:distriLK}.   
For vanishing exchange ($\Omega \rightarrow 0$), we recover the results for 
TASEP with infinite processivity presented in section \ref{sec:TASEP}.
However, when we decrease the processivity (i.e.~for an increasing $\Omega$), the LD phase in the effective rate plane shrinks in favour of the LD-HD phase, see figure 
\ref{fig:phaseDiagram}.  Eventually the fraction of
segments in LD reduces to zero, $n_{\rm LD}=0$, and the network
enters the $\mathcal{LD-HD}$ regime.  
In this segment regime the density heterogeneities are mainly attributable to the LD-HD domain walls which separate a LD and a HD part on the same segment.
The average density between single segments however does not vary much
throughout the network in this phase.   
Indeed, the distribution of segment densities is unimodal in the 
$\mathcal{LD-HD}$ regime, see figure \ref{fig:distriLK}.    
We also note that at very high densities ($\rho\approx 1$), the
stationary state is in a homogeneous regime where $n_{\rm LD}=n_{\rm LD-HD}=0$,
and thus all segments are in the HD phase ($n_{\rm HD}=1$).  In this regime, even the 
heterogeneities within single segments have disappeared. 

The boundaries
 between the different regimes are easily understood from the
effective rate diagrams figure \ref{fig:phaseDiagram}.  Let us for instance
consider the transition from the $\mathcal{LD/HD}$ to the $\mathcal{LD-HD}$ regime.  With an increasing exchange parameter $\Omega$ the size of the 
LD phase decreases in the single-segment
$(\alpha,\beta)$-phase diagram.  At some point the LD phase is so small that no
segments occupy this phase anymore: this marks the boundary between the network
and the segment regime.
This happens  at the critical value $\Omega_c(\rho)$, at which the LD phase (or HD phase for $\rho<1/2$) in the $(\alpha,\beta)$-phase diagram (of a single segment) has completely disappeared.
This critical value is given by (see equation (\ref{eq:critUpper})):
\begin{eqnarray}
  \Omega_c = \frac{1}{2}  +
  \left|\rho-\frac{1}{2}\right|\ln\left[\frac{\left|\rho-1/2\right|}{1/2 +
      \left|\rho-1/2\right|} \right ]. \label{eq:critUpperx}
\end{eqnarray}
This expression (\ref{eq:critUpperx}) is illustrated in figure
\ref{fig:irregPhaseDiagram} through  the dotted line.  Also the transitions from the $\mathcal{LD-HD}$ segment regime to the
homogeneous $\mathcal{LD}$ regime can   be
understood intuitively from the effective rate diagram: when decreasing $\rho$ towards zero,
the LD-HD phase becomes gradually smaller and retracts to the axis $\beta=
0$, while the effective rates move towards $\beta= 1$.  A similar argument applies to  the transition from $\mathcal{LD-HD}$ to $\mathcal{HD}$, for  $\rho\rightarrow 1$.

\begin{figure}[h!] 
\begin{center}
\includegraphics[width=1
\textwidth]{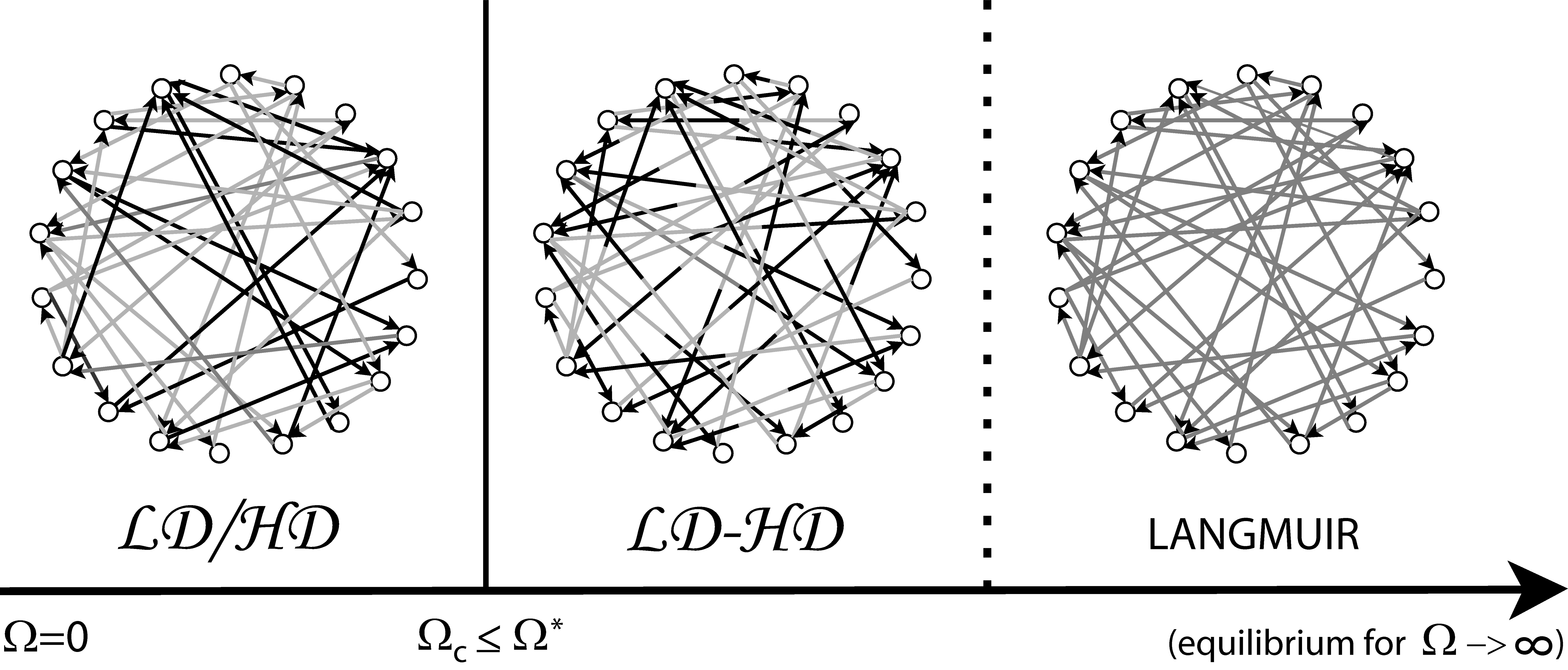}
\caption{
  Visualization of the stationary state of TASEP-LK on irregular
  networks when varying the total exchange $\Omega$ and keeping the 
  total density $\rho$ fixed.  The way particles are distributed changes 
  qualitatively when the exchange parameter $\Omega$ is varied:
  the stationary state first shifts from a $\mathcal{LD/HD}$ regime to a 
  $\mathcal{LD-HD}$ regime, indicating that the density heterogeneities
  between segments disappear and heterogeneities now arise within single 
  segments in the form of domain walls.  Increasing the exchange even 
  further, all heterogeneities disappear gradually in the network to reach 
  the Langmuir phase at $\Omega=\infty$ \cite{NeriTT}.  
}\label{fig:visTASEPLK}
\end{center}
\end{figure}

\subsection{Discussion}
We have presented a study on the interplay between active transport
through networks and passive diffusion in a bulk reservoir.  
While the active transport process leads to strong density heterogeneities 
at various scales, the passive process aims to
distribute particles homogeneously.  The competition between these two
processes is determined by three parameters: the
topology of the network, the total density $\rho$ of
particles on the network and the dimensionless parameter $\Omega = \omega L/p$.   
 Most of
the rich physical phenomenology of TASEP-LK
on networks can also be understood using effective rate diagrams and a
classification in three regimes of density heterogeneities as given in figure
\ref{fig:networksVis}.

For regular networks without local disorder, the phase diagram of TASEP-LK is
independent of the exchange rate $\Omega$.  Indeed, the $\mathcal{LD-HD}$
regime is always present at intermediate densities $\rho$ and is
unaffected by $\Omega$, see figure \ref{reg:TASEP-LK}-(a). Increasing the
coupling between network and reservoir will shift  the domain
walls, separating LD from HD phases in the segments,  towards the
junctions of the network. In this way the system will  gradually reach the
homogeneous equilibrium state for $\Omega \rightarrow \infty$.

Coupling TASEP through irregular
graphs with an infinite homogeneous reservoir leads to a rich phenomenology, as
illustrated in figure \ref{fig:visTASEPLK}. 
We have found that the heterogeneous $\mathcal{LD/HD}$ network regime 
present in TASEP 
disappears beyond some critical exchange  between network and reservoir, 
see figure \ref{fig:irregPhaseDiagram}.  For strong exchange parameters
$\Omega$ the system is in the $\mathcal{LD-HD}$ segment regime.  Increasing the
exchange even further homogenizes the particle
densities on all scales.  

The coupling $\Omega$ between reservoir and network
also affects the theoretical description of transport through the network. 
While a theoretical description of transport is
necessary on a network level when the exchange is weak (small $\Omega$), 
this is no longer the case at higher values of $\Omega$. The continuity
equations (\ref{eq:rhoV}) decouple completely at higher values of $\Omega$ in
the  $\mathcal{LD-HD}$ regime.  The stationary state in this regime can then 
be described from the local properties at the junctions (see
equations (\ref{eq:simple})). In this way, the stationary state of every 
segment can be determined, independently of the state of other segments in the
network.

\section{Exclusion processes on networks as models for motor protein transport}

In this section we put our theoretical results for PASEP and
TASEP-LK on networks into the context of motor
protein transport along the cytoskeleton.  We elaborate on the relevance of
network topology, bi-directionality and exchange with a homogeneous particle 
reservoir to the organization of motor proteins along the cytoskeleton.  As a 
model system we consider examples of motors taken from the kinesin
superfamily along a complex microtubule network.

We consider the cytoskeleton to be a disordered system, a complex meshwork
consisting of a random criss-cross of biopolymers, where disorder is
modelled using Erd\"os-R\'enyi graphs with a given mean connectivity $c$.
Although this randomness in the junction degrees cannot reflect biological
disorder in its details, this is in fact not crucial: we expect our conclusion
to remain qualitatively valid whatever the source of disorder. 
The choice of an appropriate mean connectivity $c$ is subtle,
since {\it in vivo} microtubules contain several  (typically 13) 
protofilaments.  If one takes a segment to represent an entire microtubule, then
a mean connectivity of $c=2$ appears to be most appropriate. In contrast,  when
considering segments to represent individual protofilaments, a higher mean
connectivity, of the order $c=10$, is more fitting.  The segments of the
network have a fixed length $L$ which corresponds to the typical distance
between two junctions at which microtubules interconnect.  In cells this 
distance can be set e.g.~by cross-linker proteins or branching protein complexes. 

\begin{figure}[h!]
  \begin{center}
    \includegraphics[width=1
      \textwidth]{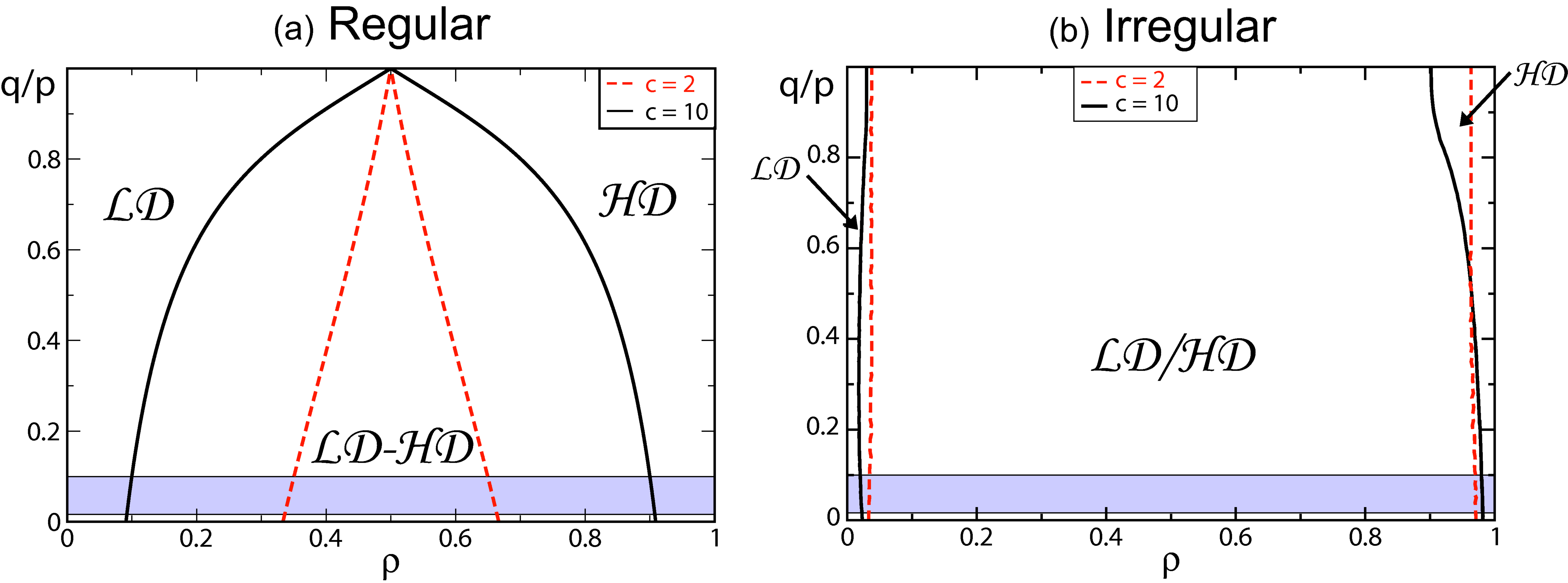}
    \caption{The stationary state of PASEP
through regular graphs (a) and irregular graphs (b) are compared.  The colour
band presents a fraction $q/p\sim 2-10\%$  as have been measured in
experiments on single kinesin-I motors \cite{Schnitzer, Kojima, Visscher}.   We
see that in general  bi-directionality of motors does not substantially alter the heterogeneities in motor densities. 
    }\label{fig:compExpa}
  \end{center}
\end{figure}

We first address the question of bi-directionality.
In figure \ref{fig:compExpa} we compare  results from figures
\ref{reg:ASEP}-(a) and \ref{fig:PASEPirregPhaseDiagram} for the stationary state
of PASEP through regular and irregular networks, indicating also the fraction $q/p$ as measured in experiments.  
As to backstepping of motors, experiments on
diluted solutions of kinesin-I have shown that in-vitro they account for
$2-10\%$ of their
displacements \cite{Schnitzer, Kojima, Visscher}.   The corresponding range $p/q\sim 0.02-0.1$, is indicated by the colour band in figure
\ref{fig:compExpa}.  We thus see that, in contrast to topology, bi-directionality has little effect on
the overall spatial organization of motors on the network.  
One also sees that the stationary state remains qualitatively unchanged 
even for much higher values of $q/p$. This suggests that also other mechanisms for bi-directionality should should not change the big picture, as for example direction switches
in cargoes transported by several motor proteins can switch 
their directionality \cite{Welte, Gross}.  A driven lattice gas model by Muhuri et al.~\cite{Muhuri} is available for this process, and the framework we have outlined would allow to study this scenario on a network.

\begin{figure}[h!]
  \begin{center}
    \includegraphics[width=1
      \textwidth]{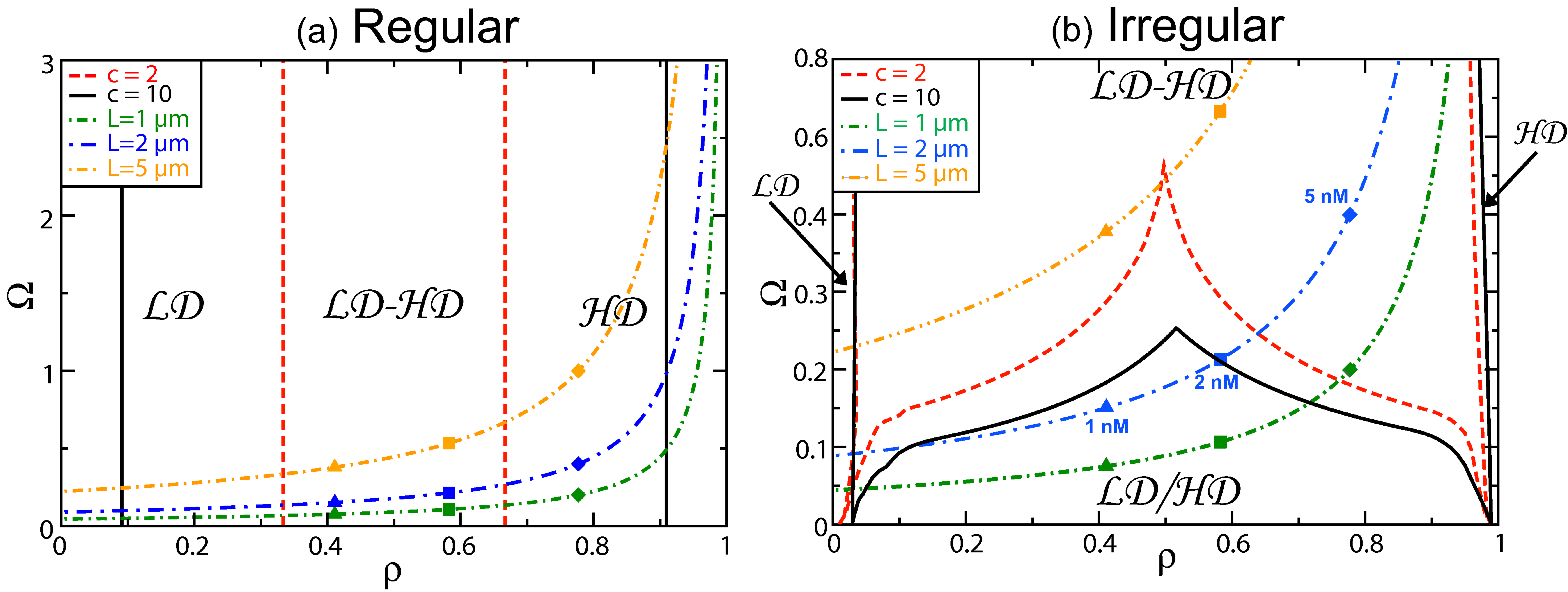}
    \caption{
      Illustration of the possible variation of the stationary state 
      for kinesin-8 motors of budding yeast moving along a network
      of microtubules. We have 
      plotted the exchange parameter $\Omega$  as a function of the total 
      density $\rho$  based on   numbers from \cite{Varg2} for given values of
the length $L$ between two microtubule intersections.  These lines are plotted
on the diagrams of TASEP-LK on regular networks (a) (from figure 
\ref{reg:TASEP-LK})  and irregular  networks  (b) (from figure 
\ref{fig:irregPhaseDiagram}).   The markers denote  the values of
$(\Omega,\rho)$ for  given values of $L$ and  the total concentration $c_m$
of motors in the solution (triangles correspond to $1nM$, squares to $2nM$ and
diamonds to $5nM$). 
    }\label{fig:compExpb}
  \end{center}
\end{figure}

Second, we turn to the question of finite motor processivity.
In figure \ref{fig:compExpb} we indicate reasonable parameter values as 
estimated from {\it in-vitro} experiments on the diagrams for the stationary
state of TASEP-LK on regular and irregular
networks, figures \ref{reg:TASEP-LK} and \ref{fig:irregPhaseDiagram}.   
To make this educated guess at the biological relevant regime of the 
parameters $\Omega$ and $\rho$, we consider the {\it in-vitro} experiments 
by Varga et al.~\cite{VargaCell} and
by Leduc et al.~\cite{Varg2}, which concern the transport of the kinesin-8 motor of budding yeast (Kip3P) along microtubules.  
Kip3P is an interesting motor in our context, since its dynamics on a single microtubule are well described by TASEP-LK \cite{Varg2}. Kip3 depolymerizes tubulin dimers at the plus end of the microtubules, and recently an extension of this model has yielded insight into the role of active transport in depolymerizing filaments \cite{MelReeseFrey, Dennis}. The microscopic parameters of Kip3P, due to this particular biological function, might be different from that of other kinesins, but this does not alter the qualitative picture of our results.  

The parameters $\Omega$ and $\rho$ can be estimated from the data
in \cite{VargaCell}, and such estimates have been
considered before in the theoretical work \cite{Reese}.  
To do so we exploit the functional relationship between $\Omega$ and $\rho$, which follows directly from the definitions as $\Omega(\rho) = \Omega_D/(2(1-\rho))$.
It therefore only depends on the appropriately scaled detachment parameter $\Omega_D = \omega_D L/p$, which can be interpreted as the fraction of a segment in the network an isolated motor typically moves before detaching.
For Kip3P we have $\omega_D/p \approx 7.5 \:10^{-4}$, and the length of
the microtubules is of the order $1-10\mu m$.  Furthermore, taking the size of tubulin dimers to be about $8.4nm$ we obtain an estimated segment length of $L\sim 100-1000$.   

In figure  \ref{fig:compExpb} we present the relation $\Omega = \Omega(\rho)$ at
fixed $\Omega_D$, corresponding to three  different segment lengths ($L=1\mu m$,
$2\mu m$ and $5\mu m$), superposing it onto the phase diagram of TASEP-LK, for 
regular and irregular networks. 
The corresponding lines indicate how the state of the network will evolve as
the total motor protein concentration $c_m$ in the solution is varied (note that
we must distinguish the total motor protein concentration $c_m$ from the
density $\rho$ of {\it bound} motors on the network: $c_m$ is the ratio of the
total number of motors to the volume of the
solution, whereas $\rho$  is the ratio of the number of bound motors to the
number of tubulin dimers on the cytoskeleton). Recall also that the overall
density of bound motors is 
$\rho = K/(K+1)$, with  $K = \tilde{\omega}_A c_m/\omega_D$, assuming an
infinitely large and homogeneous reservoir.  Using the estimates
$\tilde{\omega}_A \approx 3.3\:10^{-3} nM^{-1}s^{-1}$, $c_m = 1nM$, $2nM$ or $5
nM$ and $\omega_D = 4.7\:10^{-3} s^{-1}$, as given in \cite{VargaCell}, we can
position the state of the system at the markers in figure \ref{fig:compExpb}.


We can draw several interesting conclusions from figures \ref{fig:compExpb}.
First, we see that the biological parameters are such that all the  different
regimes of density heterogeneities  can be reached, and can in fact be targeted
for instance by varying the total motor protein density $c_m$ or by varying the
number $L$ of tubulin dimers between two junctions.  
In cells the segment length $L$ can be regulated in various ways, for instance by varying 
the concentration of crosslinker proteins, a small value of $L$ corresponding to high crosslinker concentrations.  
Second, the network topology is important for this regulation.
As the density of motor proteins on the network varies its mapping onto the 
$(\rho,\Omega)$ plane swepdf out lines, as discussed above, and which are set by
the  segment length $L$ and motor properties (rates $p$ and $\omega_D$). 
Several examples are indicated in figures \ref{fig:compExpb}. 
For regular networks, the choice of these parameters always leads 
to the same succession of transitions between regimes of heterogeneity. This is
different for irregular network topologies, for which increasing $L$ (or
$\Omega_D$) beyond a threshold value circumvents the  $\mathcal{LD/HD}$ regime
altogether.
Consequently, these estimates show that cells could indeed  exploit
heterogeneities associated with the stationary transport regimes in order to
regulate the overall organization of matter along the cytoskeleton.

\section{Conclusion}
In this work we have studied driven lattice gases through
networks.  These systems form  a class of minimal models for 
intracellular transport of motor proteins along the cytoskeleton.  Motor
proteins are considered to be active particles which move stochastically along a
complex network and the cytoskeleton is modeled as 
a network of one-dimensional lattices which interconnect at junction sites.  
Three main results on which we report in this
conclusion follow from our study.

One of the main insights of this work is that the stationary 
state of transport processes on networks can be deduced from
the  phase  diagram of a single open segment connecting two particle
reservoirs. Indeed, using mean
field arguments we can characterize the stationary state of each segment in the
network using effective entry and exit rates.  Plotting the effective rates of
all segments on the single-segment phase diagram we can represent the
stationary state of the whole network.  
 This approach leads naturally to the
classification of the stationary state of excluded volume processes in three
distinct regimes:
\begin{itemize}
 \item   {\it homogeneous} regime ($\mathcal{LD}$ or $\mathcal{HD}$): all
segments are
in the same homogeneous phase, i.e.~either
   the LD or HD phase.  As a consequence, the particles are distributed
   homogeneously along the network. TASEP, PASEP and TASEP-LK on regular 
   graphs all occupy this regime at low and high filling of particles on the 
   network.  On irregular graphs
   this regime can only appear at very low or very high particle densities. 
 \item  {\it heterogeneous segment} regime ($\mathcal{LD-HD}$): this
   regime is dominated by segments occupying the LD-HD phase.  As a consequence,
   strong density heterogeneities are present within single segments.
   TASEP, PASEP and TASEP-LK on regular graphs occupy this regime at
   intermediate particle filling.  For irregular graphs this regime appears 
   at low processivity (i.e.~high values of the exchange rate $\Omega$). 
 \item  {\it heterogeneous network} regime ($\mathcal{LD/HD}$): a
   finite fraction of segments are in the LD phase but another finite 
   fraction of segments are in the HD phase.
   Therefore, a part of the network is sparcely occupied with particles,  
   whereas another part has a very high occupancy.  Strong
   heterogeneities are present on a network scale.  This regime
   appears naturally  on irregular graphs in TASEP, PASEP and TASEP-LK with high
   processivity. 
\end{itemize}
To which of these three regimes the stationary state corresponds
thus depends on the topology of the network, on the microscopic nature of the 
transport process and on the ``molecular`` parameters of the particles. 
Having reduced the problem of studying transport along a complex network to 
the properties of one-dimensional transport is a considerable simplification 
also in practical terms: a large number of exact and approximative, established 
 over the last two decades on one-dimensional lattice gases \cite{Schutz2,
Chou2011}, can be directly exploited.  When studying transport
processes other than TASEP it might be necessary to define other
network regimes than the one presented here, which can also be constructed from
the one-dimensional phase diagram of the corresponding transport process. 

A second main insight of our work is that strong heterogeneities on a network
are a robust feature of non-equilibrium transport with
exclusion interactions.  Therefore, we expect it to be relevant in the study of
real transport processes such as motor protein transport along the
cytoskeleton (or vehicular traffic in a city, etc.).  
The occurrence of a network regime can be understood clearly from the 
effective rate diagrams as a 
consequence of two effects: the presence of a LD and a HD
phase, separated by a first order transition, on one hand, in one-dimensional transport, and the presence of {\it any} kind of 'quenched' disorder on the other hand which affects the currents at the junction sites.   
For instance,
microtubules may present local changes, due to post-translational modifications,
to the absorption of proteins on the microtubule substrate \cite{Cai} or to the
generation of microtubules by augmin protein complexes
\cite{augmin}. 
Any such disorder leads to
a scattering of the effective rates with respect to the one-dimensional phase
diagram (see figures \ref{fig:effectiveRatesTASEP}, \ref{fig:EffRatePasep} and
\ref{fig:phaseDiagram}). 
This results in a part of the segments being in a LD state and another part being in a HD state.   
These arguments can be extended to a
large number of transport problems for which the one-dimensional phase diagram
is known \cite{Chou2011}, for which we only mention a few examples relevant to motor protein transport: TASEP with extended particles \cite{Shaw}, TASEP with synchronuous dynamics \cite{Tilstra}, TASEP with multiple lanes \cite{Pronx, Reich, Schiff, Evansx}, TASEP with particles with internal states \cite{Nishinari, Luca2},  TASEP with directional switching \cite{Muhuri}, etc.
Furthermore,
our argument applies irrespectively of the origin of the disorder in the effective rates. Here we were interested in the effect of irregularity of the graph architecture, but the effective rates can also capture disorder in the way 
particles move at the junction sites, the spatial clustering of filaments, etc.
Our insights into the
appearance of network heterogeneities in the particle distributions apply
therefore far beyond the specific Erd\"os-R\'enyi
graphs we have considered here. 

A third result of our work is physical insight into how 
heterogeneities appear in equilibrium transport processes when these are
gradually driven out of equilibrium.  To resolve this question we have
interpolated between a passive process, in which particles diffuse
bi-directionally on the network, and an active process, in which they move
uni-directionally.  We have interpolated between these two limiting cases by
gradually changing the bias in the directionality of the particles.   Analyzing
the stationary state using our effective rate diagram approach has revealed 
that, for sufficiently large systems,
even a weak preference for one direction suffices to create strong density
heterogeneities.  We have also considered active transport along a network
with coupling to passive bulk diffusion  in a reservoir.   
Varying the exchange rate
we can again interpolate between an equilibrium diffusive process and an active
transport process along a network \cite{NeriTT}.  When the exchange is small,
the
active process leads to a heterogeneous network regime as in TASEP.  
Increasing the exchange rate further makes the network heterogeneities
disappear.  The stationary state corresponds then to a segment regime
with heterogeneities at the segment level.  Eventually, when the exchange rate
becomes very high, the exchange process smoothens out  all heteroeneities and
particles are distribute homogeneously over the network.

On biological grounds, since motor proteins play a key role in creating gradients within cells, but are also involved in force production and regulation as well as the control of filament length in cells.
Understanding how motor proteins organize along the cytoskeleton therefore constitutes an essential element in the
study of the microscopic statistical physics of biological cells.   
We have shown that the different
regimes of density heterogeneities of TASEP-LK through networks
could be relevant to cellular processes, as these regimes arise for
parameter values which are consistent with estimates from in-vitro
experiments on motor proteins. In particular, our results indicate that density
heterogeneities on irregular networks could be regulated 
via various parameters such as motor processivity, crosslinker density or the bulk concentration of motors.

\appendix

\section{Current and density profiles for TASEP-LK for a single segment}\label{app:LK} 
In this appendix we revisit the current and density profiles for TASEP-LK on a single segment, as they have been determined from mean-field arguments in the literature \cite{Par03,parmeggiani2}.
In the hydrodynamic limit the mean-field equation for the average density $\rho^{\rm LK}(x)$ reads: 
\begin{equation}
(2\rho-1)\partial_x\rho^{\rm LK}= \Omega_A(1-\rho^{\rm LK})-\Omega_D\rho^{\rm LK}.\label{eq:thecontinuityequation}
\end{equation}
We first consider the limiting case $K=1$ ($K=\Omega_A/\Omega_D$), which admits linear analytical solutions for the density profiles.  We then consider the general case $K\neq1$, which is more representative to real
situations  but also more technical: the density
profiles are given by the different branches of the real Lambert W function
\cite{parmeggiani2}.   We end our discussion with several analytical results with respect to the phase diagram of TASEP-LK.

\subsection{Special case of half-filling: $\Omega_A = \Omega_D$}
Let us first consider the case $K=1$, corresponding to half-filling ($\rho=1/2$), for which the continuity equation (\ref{eq:thecontinuityequation}) simplifies
considerably and leads to piecewise linear density profiles.  In order to match the boundary conditions
we must distinguish two cases, depending on the values of two
parameters $x_{\alpha}, x_{\beta} \in [0,1]$ corresponding to positions along 
the segment:
\begin{eqnarray}
  x_{\alpha} &=&  \Omega^{-1}\left(1/2-\frac{\alpha}{p}\right),\\  
  x_{\beta }&=& 1+\Omega^{-1}\left(\frac{\beta}{p}-1/2\right).
\end{eqnarray}
Depending on the relative positions $x_{\alpha}$ and $x_{\beta }$ the
density is given by
\begin{eqnarray}
  \fl \rho^{\rm LK}\left[x; \alpha, \beta,\Omega_A, \Omega_D\right]  =
  \left\{\begin{array}{cclcc} \rho_\alpha & = & \Omega\:x + \frac{\alpha}{p} &
  x<x_{\alpha}& {\rm (LD)}\\ 
  \rho_l & = & 1/2  &
  x_\beta<x<x_\alpha  & {\rm (MC)}\\ 
  \rho_\beta & = & \Omega\:(x-1) +1 -\frac{\beta}{p} & x>x_{\beta}& {\rm
    (HD)}\end{array}\right. , \label{eq:RhoLK2}
\end{eqnarray}
when $x_\alpha<x_\beta$ and by 
\begin{eqnarray}
\fl \rho^{\rm LK}\left[x;\alpha, \beta, \Omega_A, \Omega_D\right] = 
\left\{
\begin{array}{cclcc} 
\rho_\alpha & = & \Omega\:x + \frac{\alpha}{p} & \qquad
x<x_{w}& {\rm (LD)}\\   
\rho_\beta & = & \Omega\:(x-1) +1 -\frac{\beta}{p} & \qquad x>x_{w} & {\rm (HD)}
\end{array}\right., \label{eq:RhoLK1}
\end{eqnarray}
when $x_{\alpha}>x_{\beta}$.    The  variable $x_w$ denotes the position of the
domain wall in the segment and is given by:
\begin{eqnarray} 
  x_w &=& \frac{1}{2\Omega} \, \left(\frac{\beta-\alpha}{p}+\Omega\right)
  .
\end{eqnarray}

\begin{figure}[t]
\begin{center}
\includegraphics[ width =0.6\textwidth]{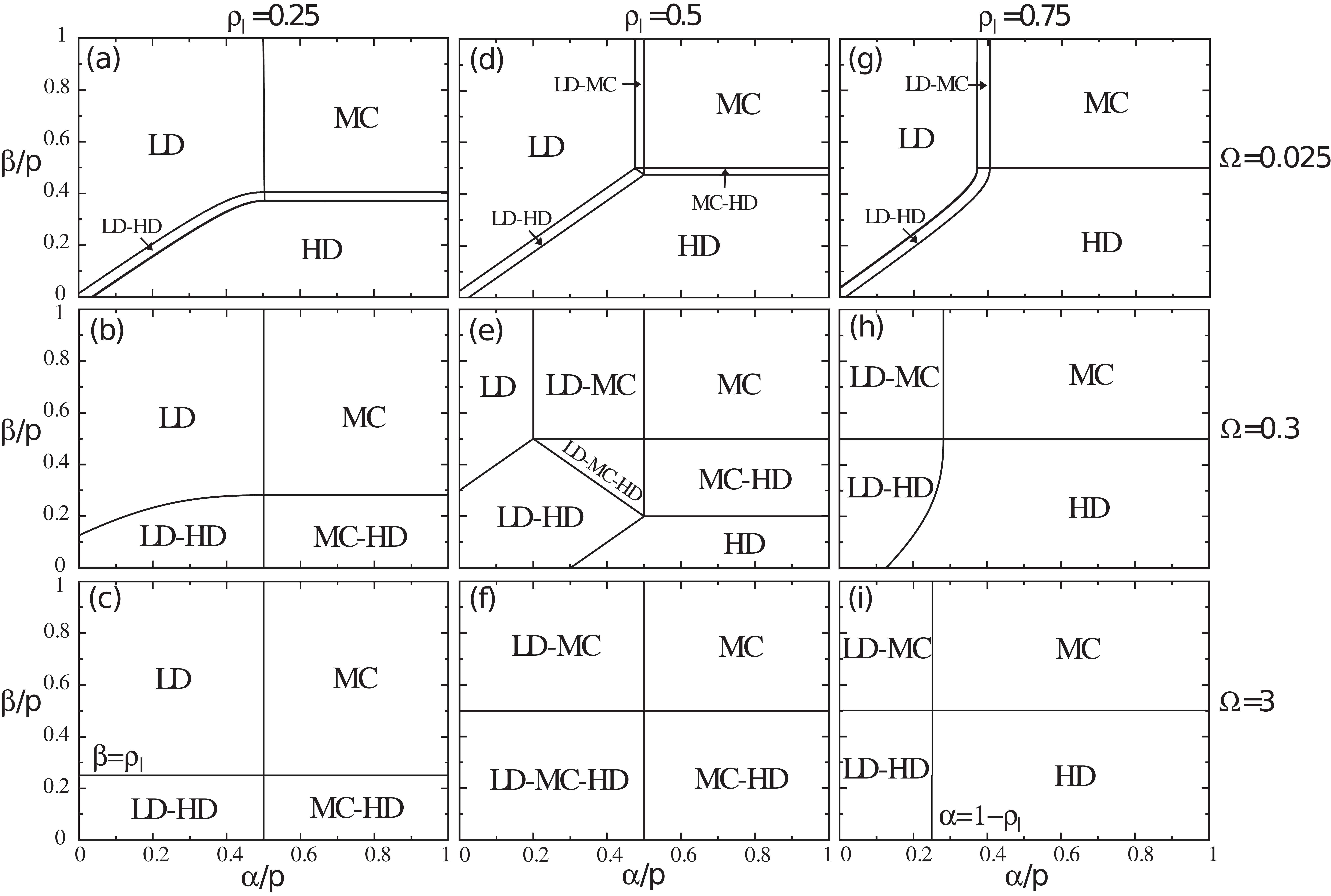}
\caption{
  The $(\alpha/p, \beta/p)$-phase diagram for  TASEP-LK on a single segment for the indicated values of $\Omega$ and $\rho_{ell}=K/(K+1)$. 
  For $\Omega\rightarrow 0$ we recover the TASEP phase diagram with the
  homogeneous LD, HD and MC phases.   For increasing values of $\Omega$ the
  heterogeneous LD-HD phase, which was restricted to a coexistence line for  $\Omega=0$, grows and plays a more prominent role.  The analytical expressions for the phase transitions for large $\Omega$ are also indicated.
}\label{fig:phaseDiagramLK}
\end{center}
\end{figure}

From the phase diagram for TASEP-LK (see figure \ref{fig:phaseDiagramLK}) we see that in addition to LD, HD and MC phases it also contains zones corresponding to the composite LD-HD, LD-MC, MC-HD and  LD-MC-HD phases.  
As $\Omega$ increases, the MC phase progressively dominates the phase diagram. 
This is not surprising, since here the MC phase is the equilibrium state corresponding to the homogeneous density  $\rho_\ell = \Omega_A/(\Omega_D+\Omega_A) = 1/2$ set by the reservoir.

\subsection{General case: $\Omega_A \neq \Omega_D$}
We now turn to the general case where $K\neq 1$.  Here we first
consider $K>1$, corresponding to a HD Langmuir phase ($\rho_l>1/2$).   The
case $K<1$ follows readily from the solution for $K>1$ by exploiting 
the particle-hole symmetry.  The two systems are related by the following transformation: 
\begin{equation}
  K\rightarrow 1/K, \
  (\alpha,\beta)\rightarrow (\beta,\alpha), \
  x\rightarrow 1-x \
  \mbox{and}  \
  \rho\rightarrow 1-\rho
  . 
\end{equation}
We define the rescaled density $\sigma[x]$ through
\begin{eqnarray}
 \sigma[x] \equiv \frac{K+1}{K-1}\left(2\rho[x]-1\right)-1,
\end{eqnarray}
such that the Langmuir density is given by $\sigma=0$. It is independent of $K$.
 
We define the two functions 
\begin{eqnarray}
Y_{\alpha}[x] &=& \left|\sigma[0]\right|\exp\left[2\Omega
\frac{(K+1)}{(K-1)}x + \sigma[0]\right],\\ 
Y_{\beta}[x] &=& \left|\sigma[1]\right|\exp\left[2\Omega
\frac{(K+1)}{(K-1)}(x-1) + \sigma[1]\right].
\end{eqnarray}
The boundary conditions, $\rho[0] = \alpha/p$ and $\rho[1] = 1-\beta/p$
determine  $\sigma[0]$ and $\sigma[1]$.  The solution for the profiles
$\sigma[x]$
are given in terms of the two branches $W_{-1}$ and $W_{0}$ of the real-valued
Lambert $W$ function.  We find the left boundary solution
\begin{eqnarray} 
 \sigma_{\alpha}[x] = W_{-1}\left[-Y_{\alpha}[x]\right]<-1,
\end{eqnarray}
corresponding to a density $\rho_{\alpha}<1/2$
and the right boundary solution
\begin{eqnarray}
 \sigma_{\beta}[x] = \left\{\begin{array}{ccc}W_0\left[Y_{\beta}[x]\right]>0&&
0\leq \beta
<p(1-\rho_\ell)\\0 &&\beta=p(1-\rho_\ell) \\
W_0\left[-Y_{\beta}[x]\right]<0&&p(1-\rho_\ell)<\beta\leq p/2 \\
W_0\left[-Y_{1/2}[x]\right]<0&&\beta>p/2
\end{array}\right.. \label{eq:rightBoundarySol}
\end{eqnarray}
Note that the left boundary solution is only defined for values of $x$
such that $Y_{\alpha}[x]<1/e$.  
The two solutions are matched at the position $x_w\in[0,1]$, for which the
current of both solutions are equal: $\rho_{\alpha}[x_w] =
1-\rho_\beta[x_w]$.  If $x_w\in[0,1]$ we are in the LD-HD phase (or the LD-MC
phase when $\beta>p/2$). Otherwise we
are in a LD, HD or MC phase, depending on the solution which dominates the
current. 
The MC phase corresponds to the boundary independent solution
$W_0\left[-Y_{1/2}[x]\right]$ and appears for $\beta>p/2$. Here (for $K>1$) MC has a density larger than one half.   All phases here generalize the equivalent homogeneous phases in TASEP.

We can now evaluate the solution in the various quadrants of the
$(\alpha/p,\beta/p)$ phase diagram: 
\begin{itemize}
 \item   
   $\alpha/p<1/2$ and $\beta/p<1/2$: in this case one can have either LD,
   LD-HD or HD phases, depending on the position of the domain wall $x_w$.
\item 
  $\alpha/p<1/2$ and $\beta/p>1/2$: one can have the LD, LD-MC or MC 
  phase.  Again, one has to calculate the position $x_w$ of the domain wall. 
  The right boundary solution is given by
  $W_0\left[-Y_{1/2}[x]\right]$,  independently of $\beta$. For
  $x_w<0$ we obtain the boundary independent MC phase.  
  We also remark that here
  the LD-MC phase has the particularity that the MC part is boundary
  (i.e.~$\beta$) independent.  
\item 
  $\alpha/p>1/2$ and $\beta/p<1/2$:
  the system is in the HD phase.  
\item 
  $\alpha/2>1/2$ and $\beta/p>1/2$: 
  the system is in the MC phase.
\end{itemize}

The phase diagrams in figure \ref{fig:phaseDiagramLK} are constructed as
follow: the transition between LD and LD-HD phases (and LD and LD-MC for
$\beta/p>1/2$) follows from the condition
$x_w=1$.  Analogously, the transition between the  LD-HD and HD (or
MC) phases follows from
the condition $x_w=0$. The transition between the HD and  MC phases is given
by $\beta/p=1/2$.  For
$\beta/p>1/2$ all transition lines are vertical independent of $\beta$. 

At low values of $\Omega$ the phase diagram involves four phases, 
i.e.~the LD, HD, MC, LD-HD and LD-MC
phase.  When increasing the exchange rate $\Omega$ the LD phase becomes
gradually smaller and eventually disappears at a critical value $\Omega_c$:
\begin{eqnarray}
\Omega_c = \frac{K-1}{2(K+1)}\left( -1 + \ln\left(\frac{K-1}{2K}\right) +
\frac{2K}{K-1}\right).  \label{eq:critUpper}
\end{eqnarray}

\subsection{The phase diagram of TASEP-LK at $\Omega\rightarrow \infty$}
The phase diagram of TASEP-LK is represented in figure
\ref{fig:phaseDiagramLK}.  In general we have no explicit analytical
expressions for the TASEP-LK phase diagram, but we do have the explicit
expression for the phase diagram at $\Omega\rightarrow\infty$.   Let us elaborate on
three different cases: 
\begin{itemize}
 \item 
   $K>1$: such that the Langmuir density $\rho_\ell>1/2$ corresponds 
   to a HD phase.
   \begin{itemize}
   \item 
     LD-HD: $\alpha/p<1-\rho_\ell$ and $\beta/p<1/2$ 
 \item 
     LD-MC: $\alpha/p<1-\rho_\ell$ and $\beta/p>1/2$ 
   \item 
     HD: $\alpha/p>1-\rho_\ell$ and $\beta/p<1/2$ 
   \item 
     M: $\alpha/p>1-\rho_\ell$  and $\beta/p>1/2$
   \end{itemize}

 \item  
   $K<1$, such that the Langmuir density $\rho_\ell<1/2$ corresponds 
   to a LD phase.  The
   phase diagram follows readily from that for $K>1$ using the 
   particle-hole symmetry transformations given above.

\item 
  $K=1$ is a special situation, as then the Langmuir density
  $\rho_\ell=1/2$ corresponds to half filling.  The Langmuir phase thus
  corresponds to the MC phase. For $K=1$ a part of the segment will reach this
  Langmuir phase (which for $K\neq 1$ is only the case in the limit $\Omega
  \rightarrow \infty$).  This leads to the LD-MC-HD, LD-MC and MC-HD phases. 
  Due to the linear density profiles at $K=1$ one can determine explicitly
  analytic
  expressions for the phase diagram \cite{parmeggiani2}, which is presented in
  figure \ref{fig:phaseDiagramLK}.  For values of $\Omega_c>1/2$ (which
  follows from equation (\ref{eq:critUpper})) the  phase diagram is given by the
  simple expression 
  \begin{itemize}
  \item 
    LD-HD or LD-MC-HD: $\alpha/p<1/2$, $\beta/p<1/2$
  \item 
    MC-HD:  $\alpha/p>1/2$ and $\beta/p<1/2$
  \item 
    LD-MC:  $\alpha/p<1/2$ and $\beta/p>1/2$
  \item 
    MC: $\alpha/p>1/2$  and $\beta/p>1/2$
  \end{itemize}
  The LD-HD phase is not present for $\Omega>1$.
\end{itemize}

\section{Universal expression for networks with infinite connectivity} \label{app:B}
We present here the analytical and universal expression for the 
current-density profile of infinitely connected graphs ($c \rightarrow \infty$). 
The first observation is that the approximate mean field equations (\ref{eq:simple}) are seen to become exact in the infinitely connected limit.  As discussed in section \ref{sec:tasep:infconnect}, the infinite connectivity limit amounts to considering all junctions blocked: $\rho_v=1$ for all junctions.   

This is reflected in the effective rates which scale with the average connectivity $c$ as  
$(\alpha^{\rm eff}, \beta^{\rm eff})\sim(\mathcal{O}(c^{-1}),
\mathcal{O}(c^{-1}))$, and therefore the expression for the current follows 
from the TASEP-LK single segment current  by setting $(\alpha/p,\beta/p) =
(0,0)$. For mean densities on the network $\rho>1/2$ we obtain (using $\rho = K/(1+K)$)
\begin{eqnarray}
 J/p &=&
\int^{x_w}_0\:\rho^{\infty}_{\alpha}\left[x\right]\left(1-\rho^{\infty}_{\alpha}
\left[x\right]\right)dx +
\int^1_{x_w}\rho^{\infty}_{\beta}\left[x\right]\left(1-\rho^{\infty}_{\beta}
\left[x\right] \right) dx 
\nonumber \\ 
&=& \rho -
\int^{x_w}_0\:\left(\rho^{\infty}_{\alpha}\left[x\right]\right)^2dx -
\int^1_{x_w}\left(\rho^{\infty}_{\beta}\left[x\right]\right)^2 dx 
\label{eq:currentInf}
\end{eqnarray}
with $x_w$ the domain wall position defined through the condition
$\rho^{\infty}_{\alpha}\left(x_w\right) =
1-\rho^{\infty}_{\beta}\left(x_w\right)$, 
and 
\begin{eqnarray}
 \sigma^{\infty}_{\alpha}\left[x\right] &=&
W_{-1}\left[-\frac{2\rho}{2\rho-1}\exp\left[\frac{2
\left(\Omega \:x-\rho\right)}{2\rho-1}\right]\right],  \\ 
\sigma^{\infty}_{\beta}\left[x\right] &=& 
W_0\left[\frac{2(1-\rho)}{2\rho-1}\exp\left[\frac{2
\left(\Omega(x-1)+(1-\rho)\right)}{2\rho-1}\right]\right].
\end{eqnarray}
We have $\sigma^{\infty}_{\alpha/\beta}[x] =
(2\rho^{\infty}_{\alpha/\beta}[x]-1)/(2\rho-1)-1$. 
For $\rho<1/2$ we can deduce the analoguous
expressions from particle-hole symmetry $(\rho\rightarrow 1-\rho)$, while for the special case $\rho=1/2$ 
the integrals in equation (\ref{eq:currentInf}) can be integrated explicitly to find the expression in 
 \ref{app:C1}.

\section{Half-filling for TASEP-LK on a network}\label{app:C}

\begin{figure}[t]
\begin{center}
\includegraphics[ width = 1\textwidth]{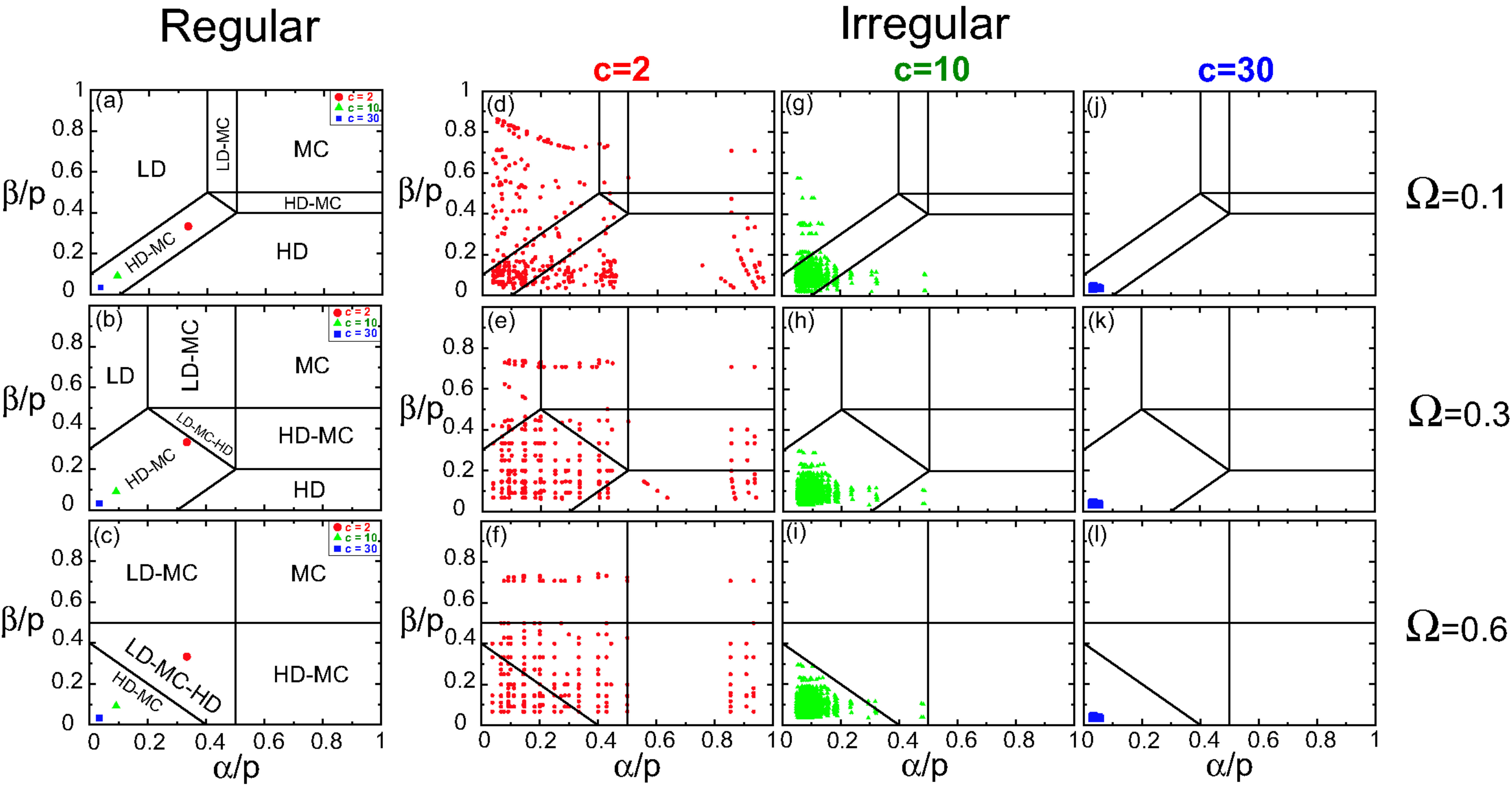}
\caption{Effective rate diagrams for TASEP-LK through 
    regular graphs (left) and irregular graphs (right), both at a
    total density of $\rho=1/2$, are presented for the given values of 
    $\Omega$ and $c$.  We have used the same graph instances as in figure
    \ref{fig:phaseDiagram}.  On regular networks the effective rates are equal
    for all segments, such that only one marker is plotted. 
}\label{fig:app2}
\end{center}
\end{figure}

The case of half-filling ($\rho=1/2$, corresponding to $K=1$) mathematically
simplifies TASEP-LK on a single segment since the density profiles are piecewise
linear functions (see \ref{app:LK}). The corresponding phase diagrams are known analytically, see figure \ref{fig:phaseDiagramLK} (d)-(f), and we exploit them for the effective rate diagrams for networks at
half filling (figure \ref{fig:app2}).    
Using these results we now derive several simple analytical
results  on the scale of the network. 

\paragraph{Infinite connectivity ($c \rightarrow \infty$) at
half-filling}\label{app:C1}
At infinite connectivity the effective rates cluster close to the origin,
see figure \ref{fig:app2}.  All segments decouple and become equivalent to
isolated open segments with $\alpha^{\rm eff}, \beta^{\rm eff} = (0,0)$.  We
recover the expressions in \ref{app:B}.  For $K=1$ we  can explicitly integrate
equation (\ref{eq:currentInf}) to find
\begin{eqnarray}
  J/p  = \left\{\begin{array}{cccc}\Omega/4-\Omega^2/12,
  & &\Omega<\Omega^\ast=1 &{\mbox{(LD-HD)}}\\ 1/4 -
  \Omega^{-1}/12,& &\Omega>\Omega^\ast=1 & {\mbox{(LD-MC-HD)}}\end{array}\right. 
  \label{eq:cInf}
\end{eqnarray}
We see that beyond some critical exchange parameter ($\Omega>\Omega^\ast=1$) a part of the segment attains the
Langmuir phase (i.e.~here the MC phase, since $K=1$ corresponds to half-filling).   Moreover, the current approaches its Langmuir value $J/p=1/4$ for
rather  small values of $\Omega$ already .



\paragraph{Regular networks at half-filling}
The special case of half-filling also allows to derive analytical expressions
for TASEP transport on regular networks.  For regular networks all segments
are equivalent and have the same effective rates given by equation
(\ref{eq:rhovT}) (see also figure \ref{fig:app2}).  As stated in the main text,
the current density profile can then be established from the formulas in
\ref{app:LK}. However, to find the average
current one still has to integrate the local expression $J^{\rm LK}[x]$ along the
segment $x\in[0,1]$. In general this is difficult, but the integration can be performed explicitely for half-filling. Then the current profile becomes quadratic, and we find 
\begin{equation}
  \fl \frac{J(\Omega)}{p} 
  =
  \left\{
  \begin{array}{lll}
    \frac{c}{(c+1)^2}  + \frac{\Omega}{4}\left(\frac{c-1}{c+1}\right)
    -\frac{\Omega^2}{12}
    &\qquad (\Omega<\Omega^\ast \mbox{ : LD-HD})
    \\
    \frac{1}{4} 
    - \frac{1}{\Omega} \, \frac{1}{12} \, \left(\frac{c-1}{c+1}\right)^3
    \left(1-\frac{2}{c+1}\right)^2
    &\qquad (\Omega>\Omega^\ast \mbox{ : LD-MC-HD})
  \end{array}
  \right.
\end{equation}
This shows that the current saturates gradually to its maximal value of
$1/4$ due to the appearance of a MC phase  in the middle of the segments, 
which is present beyond a threshold for the
exchange parameter, $\Omega>\Omega^\ast= (c-1)/(c+1)$. 
As $\Omega$ is increased
further the homogeneous Langmuir phase is attained asymptotically through growth
of the MC zone within the individual segments.  Note that this mechanism, for
which the Langmuir phase appears in the middle of the segment, is particular
to the $K=1$ case.

\ack
We acknowledge support
from ANR-09-BLAN-0395-02 and from the Scientific Council of the University of
Montpellier 2.

\section*{References}
\bibliographystyle{ieeetr}
\bibliography{bibliography}

\end{document}